\documentclass[
reprint,
superscriptaddress,
amsmath,amssymb,
aps,
pra,
floatfix,
]{revtex4-2}

\usepackage{graphicx}
\usepackage{dcolumn}
\usepackage{bm}
\usepackage{braket}
\usepackage{float}
\usepackage{xcolor}
\usepackage{hyperref}
\usepackage{cleveref}
\usepackage{physics}
\usepackage{cancel}
\usepackage{xcolor}
\usepackage{dsfont}
\usepackage{multirow}
\usepackage{babel}

\usepackage[mathlines]{lineno}

\def\[#1\]{\begin{align}#1\end{align}}

\def\ex[#1]{\exp\left(#1\right)}

\begin{document}

\title{The loss tolerance of cat breeding for fault-tolerant grid state generation}


\author{Olga Solodovnikova}
\email{olgasol@dtu.dk}
\author{Ulrik L. Andersen}
\email{ulrik.andersen@fysik.dtu.dk}
\author{Jonas S. Neergaard-Nielsen}
\email{jsne@fysik.dtu.dk}
\affiliation{Center for Macroscopic Quantum States (bigQ), Department of Physics, Technical University of Denmark, Building 307, Fysikvej, 2800 Kgs. Lyngby, Denmark}
\date{August 8, 2025}
\begin{abstract}

The development of a continuous-variable photonic quantum computer depends on the reliable preparation of high-quality Gottesman-Kitaev-Preskill states. The most promising GKP preparation scheme is the cat breeding protocol, which can generate GKP states deterministically given a source of squeezed cat states, using beam splitters, homodyne detectors and a feedforward displacement. However, analyzing the performance of the protocol under loss is cumbersome due to the exponential scaling of the system. By representing the Wigner function of the input states as a linear combination of Gaussians, we are able to quickly and accurately simulate several rounds of breeding with mixed input states. Using this novel method, we find that optical loss decreases the overall success probability of the protocol, and prohibits the preparation of a fault-tolerant GKP state when the loss exceeds 4\%. Our methodology is available as open-source code.

 
\end{abstract}
\maketitle
\section{Introduction}
The preparation of Gottesman-Kitaev-Preskill (GKP) states \cite{gottesman_encoding_2001} in traveling light is a crucial component in building a fault-tolerant continuous-variable (CV) photonic quantum computer \cite{menicucci_universal_2006,bourassa_blueprint_2021}. In the GKP error correction scheme, a qubit is encoded in a state of the infinite dimensional Hilbert space of an oscillator exhibiting translational symmetry in phase space. The wavefunction of the GKP states, also called grid states, has a periodic peak structure arranged on a grid in conjugate quadratures. Small displacement errors and Gaussian noise are mapped to logical Pauli errors, and a topological error correction code such as the surface code \cite{fowler_surface_2012} can be used to mitigate errors. This scheme is particularly appealing for optics because the required Clifford gates and stabilizer measurements can be implemented with standard linear optical components (beam splitters, phase shifters and homodyne detectors) in the measurement-based quantum computing model \cite{menicucci_fault-tolerant_2014}. Universality can be achieved via magic state distillation using a heterodyne detector \cite{baragiola_all-gaussian_2019}. 

While GKP states have been generated experimentally in superconducting microwave cavities \cite{campagne-ibarcq_quantum_2020,sivak_real-time_2023,lachance-quirion_autonomous_2024}, and the vibrational motion of trapped ions \cite{fluhmann_encoding_2019,de_neeve_error_2022}, their preparation in optics is challenging due to the lack of a strong non-linearity. Recently, Larsen et al. \cite{larsen_integrated_2025} heralded a four-peak logical GKP state 
using a low-loss integrated photonic chip. Although significant, this result is 
not yet useful for fault-tolerant measurement-based quantum computing \cite{larsen_fault-tolerant_2021,ostergaard_octo-rail_2025,aghaee_rad_scaling_2025}. Minimizing the optical losses, especially those caused by chip-to-fiber coupling, is a major hurdle. 

In optics, the most common non-Gaussian state preparation scheme involves using photon-number-resolving detectors (PNRDs) to probabilistically generate a non-Gaussian state from a multimode Gaussian state
\cite{su_conversion_2019,quesada_simulating_2019,tzitrin_progress_2020,eaton_measurement-based_2022,takase_gottesman-kitaev-preskill_2023,endo_non-gaussian_2023,endo_high-rate_2025}. This approach, which is similar to Gaussian Boson sampling (GBS) \cite{hamilton_gaussian_2017}, was the method used to prepare the GKP state in \cite{larsen_integrated_2025}. The Schrödinger cat breeding protocol \cite{vasconcelos_all-optical_2010, weigand_generating_2018, takase_generation_2024, aghaee_rad_scaling_2025} is another GKP state preparation scheme, which is deterministic in the asymptotic limit of several rounds of breeding, requiring a large supply of squeezed cat states (SCS). Unfortunately, we are limited by the success rate of the SCS preparation, as the schemes use small GBS devices to implement generalized photon subtraction \cite{ourjoumtsev_generating_2006, huang_optical_2015, takase_generation_2021, takase_generation_2024}, requiring projection onto high photon numbers to produce cats with large amplitudes. This demands a high level of squeezing and good photon-number-resolving capabilities. SCS can also be prepared from Fock states \cite{ourjoumtsev_generation_2007,eaton_non-gaussian_2019,winnel_deterministic_2024}, but this simply propagates the issue further back, as Fock state preparation schemes that use spontaneous parametric down conversion (SPDC) schemes face similar limitations \cite{tiedau_scalability_2019}. In cavity-QED systems, semi-deterministic cat state generation schemes have been developed \cite{hacker_deterministic_2019,hastrup_protocol_2022}. Regardless of the preparation scheme, the experimentally prepared SCS will be limited by photon loss and the non-unit quantum efficiency of detectors \cite{provaznik_benchmarking_2020}.
It is therefore relevant to re-examine the performance of the breeding protocol when the input states are noisy.


In this work, we analyze the breeding protocol when the ideal input states, the squeezed cat states, are subject to a loss channel. To this aim, we develop a fast simulation framework to model the multi-mode breeding circuit by using the linear combination of Gaussians (LCoG) representation \cite{bourassa_fast_2021}. Informed by our methodology, we argue that a feedforward displacement does not improve the quality of the prepared grid states in terms of the effective squeezing \cite{duivenvoorden_single-mode_2017}, a figure of merit used in setting fault-tolerance thresholds. We show that non-zero homodyne measurement outcomes can lead to incoherent phases on the Gaussian peaks, which cannot be corrected by an overall displacement operation. Our numerical results indicate that the breeding protocol is not fully deterministic, and its probability of success decreases significantly even with small amounts of loss. We find that the protocol fails to produce a fault-tolerant grid state when the input states experience 5\% loss or more.

The paper is structured as follows. In Sec.\ \ref{sec:preliminaries}, we cover the relevant phase-space methods and define GKP grid states. 
In Sec.\ \ref{sec:ideal_protocol}, the ideal (lossless) SCS breeding protocol is introduced, and the effect of the non-zero homodyne measurement outcomes is discussed. In Sec.\ \ref{sec:non-ideal_protocol}, loss is added to the input states and its effect on the quality of the generated grid states is shown through numerical simulations. In Sec.\ \ref{sec:simulation}, the simulation methods that were developed to perform the loss analysis and calculate the figures of mreit are presented. In Sec.\ \ref{sec:outlook}, we summarize our results and discuss their implications on the prospect of fault-tolerant quantum computing in optics.
\section{Preliminaries}\label{sec:preliminaries}
In this section, we introduce the background concepts in CV quantum optics. In Sec.\ \ref{subsec:phase_space_methods}, we cover phase-space methods used in the rest of the paper. In Sec.\ \ref{subsec:gkp_states}, we introduce GKP states and their finite energy approximations, as well as the effective squeezing \cite{duivenvoorden_single-mode_2017}, which is a figure of merit for the quality of a GKP state. In Sec.\ \ref{sec:SCS}, SCS are defined and their phase-space representation is described.

\subsection{Phase space methods}
\label{subsec:phase_space_methods}
A bosonic mode $i$ is a CV system described by its position $\hat{x}_i$ and momentum $\hat{p}_i$ quadrature operators, which satisfy the commutator relation $[\hat{x}_i,\hat{p}_j]=i\hbar\delta_{ij}$. These operators can be gathered in a vector mode-by-mode, $\hat{\vb*{q}}=(\hat{\vb*{q}}_i,\dots,\hat{\vb*{q}}_N)^T$, where $\hat{\vb*{q}}_i=(\hat{x}_i,\hat{p}_i)^T$. Multi-mode Gaussian states are fully characterized by their first and second moments: the displacement vector $\vb*{\mu}$ with elements $\mu_i=\langle \hat{q}_i \rangle$ and the covariance matrix $\vb*{\sigma}$ with elements $\sigma_{ij}=\frac{1}{2}\langle \{\hat{q}_i-\langle\hat{q}_i\rangle,\hat{q}_j-\langle\hat{q}_j\rangle\}\rangle$ \cite{weedbrook_gaussian_2012,serafini_quantum_2017,brask_gaussian_2022}, where the curly brackets indicate the anti-commutator. The Wigner-Weyl characteristic function is a quasi-probability distribution which fully characterizes a state $\hat{\rho}$,
\[
 \chi_{\hat{\rho}}(\vb*{\alpha})=\langle\hat{D}(\vb*{\alpha})\rangle = &\Tr[\hat{\rho}\hat{D}(\vb*{\alpha})],
\]
where $\vb*{\alpha}=(\vb*{\alpha}_1,\dots, \vb*{\alpha}_N)^T$ is a vector of displacements in each mode, which is divided into its real and imaginary parts, i.e. $\vb*{\alpha}_i=(\Re(\alpha_i),\Im(\alpha_i))^T$. The Wigner function is related to the characteristic function via a Fourier transform,
\[
W_{\hat{\rho}}(\vb*{q})=\frac{1}{(2\pi)^{N}}\int_{\mathds{R}^{2N}} \dd^{2N}{\vb*{\alpha}}\: e^{-i\vb*{q}^T\vb*{\Omega}\vb*{\alpha}}\chi_{\hat{\rho}}(\vb*{\alpha}),
\]
where $\vb*{\Omega}=\bigoplus_{i=1}^N\begin{pmatrix}
    0 & 1 \\ -1 & 0
\end{pmatrix}$ is the symplectic form and $N$ is the number of modes. 

In the following, we list the operators of common optical components which transform Gaussian states into Gaussian states. 
The displacement operator is
\[
\hat{D}(u+iv)=\exp\left[i\sqrt{\frac{2}{\hbar}}\left(v\hat{x}-u\hat{p}\right)\right],
\]
where a displacement in $x$ is
\[
\hat{X}(s)=\hat{D}\left( \frac{s}{\sqrt{ 2\hbar }} \right)=\exp\left[-\frac{is}{\hbar}\hat{p}\right],
\]
and transforms the quadrature operators as 
\[\hat{X}^{\dagger}(s)\hat{x}\hat{X}(s)=\hat{x}+s,\quad \hat{X}^{\dagger}(s)\hat{p}\hat{X}(s)=\hat{p}.
\]
A displacement in $p$ is given by
\[\hat{Z}(t)=\hat{D}\left( i\frac{t}{\sqrt{ 2\hbar }} \right)=\exp\left[\frac{it}{\hbar}\hat{x}\right],
\]
with the following transformations,
\[\hat{Z}^{\dagger}(t)\hat{x}\hat{Z}(t)=\hat{x},\quad \hat{Z}^{\dagger}(t)\hat{p}\hat{Z}(t)=\hat{p}+t.
\]
The squeezing operator is
\[
\hat{S}(z,\phi)=\exp\left[ \frac{\ln(z)}{2} (e^{-i\phi}\hat{a}^2-e^{i\phi}\hat{a}^{\dagger_{2}})\right],
\]
where $\hat{a}=\sqrt{2\hbar}^{-1}(\hat{x}+i\hat{p})$ and $\hat{a}^\dagger=\sqrt{2\hbar}^{-1}(\hat{x}-i\hat{p})$ are the canonical bosonic operators. For $\phi=0$ and squeezing strength $r=\ln(z)$, the operator squeezes the $x$-quadrature and anti-squeezes the $p$-quadrature,
\[
\hat{S}^{\dagger}(z)\hat{x}\hat{S} (z)=e^{-r}\hat{x}, \quad \hat{S}^{\dagger}(z)\hat{p}\hat{S}(z)=e^{r}\hat{p}.
\]
The rotation operator 
\[
\hat{R}(\phi)=\exp[i\phi\hat{a}^\dagger \hat{a}],
\]
transforms the quadrature operators as,
\[
\hat{R}^{\dagger}(\phi)\hat{x}
\hat{R}(\phi)=\cos(\phi)\hat{x}-\sin(\phi)\hat{p},\\ \hat{R}^{\dagger}(\phi)\hat{p}
\hat{R}(\phi)=\cos(\phi)\hat{p}+\sin(\phi)\hat{x}.
\]
And the beam splitter operation,
\[
\hat{B}(\theta)=\exp\left[\frac{i\theta}{\hbar}\left(\hat{p}_1\hat{x}_2 - \hat{x}_1\hat{p}_2\right)\right],
\]
with reflection amplitude $\sqrt{\eta}=\cos\theta$ and transmission amplitude $\sqrt{1-\eta}=\sin\theta$ as sketched in the diagram,
\begin{figure}[H]
    \centering
    \includegraphics[width=0.33\linewidth]{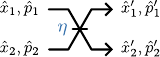}
    \label{fig:enter-label}
\end{figure}
\noindent transforms the quadratures as,
\[
\vb*{\hat{q}}_1'=\hat{B}^{\dagger}(\theta)\vb*{\hat{q}}_1
\hat{B}(\theta)=\sqrt{\eta}\hat{\vb*{q}}_1-\sqrt{1-\eta}\hat{\vb*{q}}_2,\\ 
\vb*{\hat{q}}_2'=\hat{B}^{\dagger}(\theta)\vb*{\hat{q}}_2
\hat{B}(\theta)=\sqrt{\eta}\hat{\vb*{q}}_2+\sqrt{1-\eta}\hat{\vb*{q}}_1.
\]
The quadrature transformations under Gaussian unitary operations can be gathered in the following equation, $\hat{U}^\dagger\hat{\vb*{q}}\hat{U}=\vb*{S}\hat{\vb*{q}}+\vb*{d}$ where $\vb*{S}$ is a symplectic matrix, satisfying $\vb*{S}\vb*{\Omega}\vb*{S}^T=\vb*{\Omega}$ and $\vb*{d}$ is a vector of displacements.

\subsection{Grid states}
\label{subsec:gkp_states}
In the GKP code, a two-dimensional logical subspace can be encoded in a single bosonic mode. The GKP codewords were designed in \cite{gottesman_encoding_2001} to be simultaneous $+1$ eigenstates of two commuting displacement operators $\hat{D}(\alpha)$ and $\hat{D}(\beta)$, where $\Im(\alpha\beta^*)=2\pi$ ensures the commutation condition \cite{hastrup_measurement-free_2021}. These two displacement operators and their powers form the stabilizer group, visualized as a periodic lattice in phase-space \cite{grimsmo_quantum_2021}. 
The logical Pauli operators are also displacement operators, $\bar{X}=\hat{D}(\alpha/2)$ and $\bar{Z}=\hat{D}(\beta/2)$, which must anti-commute. 

The two possible lattices one can construct in 2D are the rectangular and the triangular (hexagonal) lattice. For the rectangular lattice, the commutation condition implies $\alpha=2\pi/\beta^*$. A common choice is $\alpha=\sqrt{2\pi}$ and $\beta=i\sqrt{2\pi}$, which yields a square lattice, the most widely used GKP encoding. One choice for the triangular lattice is $\alpha=\sqrt{\frac{\pi}{2}}(\kappa_+ +i\kappa_-)$ and $\beta=\sqrt{\frac{\pi}{2}}(\kappa_-+i\kappa_+)$ with $\kappa_\pm = (\sqrt[4]{3})^{-1}\pm \sqrt[4]{3}$. The hexagonal lattice has the most optimal packing, and can correct larger displacement errors than the other encodings. The stabilizer commutation condition also implies that the area of the unit cell spanned by the two stabilizer displacement vectors is always $2\pi$ ($4\pi\hbar$ in phase-space) and each lattice can be transformed into another via a squeezing and/or rotation operations applied to both stabilizers. In the following, we shall primarily focus on the square-lattice encoding. 

\subsubsection{Ideal GKP states}
The GKP codewords in the computational basis of the square-lattice encoding are defined as a superposition of equally spaced peaks on a grid with spacing $2\sqrt{\pi\hbar}$ in $x$ and $\sqrt{\pi\hbar}$ in $p$ 
\cite{mensen_phase-space_2021},
\[
\ket{j_L}&=
\sum_{n\in\mathds{Z}_0}\ket{(2n+j)\sqrt{\pi\hbar}}_x, \\
&= 
\sum_{n\in\mathds{Z}_0}e^{ij\pi n} \ket{n\sqrt{\pi\hbar}}_p,
\]
where $j\in\{0,1\}$ indicates the logical qubit state. The logical bit flip operator is
\[
\hat{X}(\sqrt{\pi\hbar})\ket{0_L}=\ket{1_L},
\]
and the logical phase-flip operator satisfies
\[
  \hat{Z}(\sqrt{\pi\hbar})\ket{j_L}=(-1)^{j}\ket{j_L}.
\]
The conjugate basis is easily accessible with a Fourier transform, i.e. a $\frac{\pi}{2}$ rotation, 
\[
\ket{+_L}=\hat{R}\left(\frac{\pi}{2}\right)\ket{0_L}
\]
The codewords for the hexagonal-lattice encoding can be obtained by squeezing the square-lattice GKP codewords by $\sqrt[4]{3}$ along the $\pi/4$ axis,
\[
\ket{j_L}_\text{hex}=\hat{S}\left(\sqrt[4]{3},\frac{\pi}{4}\right)\ket{j_L}
\]
The symplectic matrix for the 
$\hat{S}\left(\sqrt[4]{3}, \frac{\pi}{4}\right)$
operation is 
\[
\vb*{S}_\text{hex} = \frac{1}{2}\begin{pmatrix}
\kappa_+  & \kappa_- \\ \kappa_- & \kappa_+
\end{pmatrix}
\]
An additional rotation of $\hat{R}(-\frac{\pi}{12})$ may be used to align the grid with the $x,p$ axes. 

Another interesting grid state is the qunaught or sensor state, introduced in \cite{duivenvoorden_single-mode_2017}, which is used to detect small phase-space displacements. It is defined as a superposition of peaks with spacing $\sqrt{2\pi\hbar}$ on a square grid in phase space,
\[
\ket{\varnothing}=  
\sum_{n}\ket{n\sqrt{ 2\pi\hbar }}_{x}=
 \sum_{n}\ket{n\sqrt{ 2\pi\hbar }}_{p}.
\]
The qunaught state can be obtained from the square-lattice encoding by squeezing the $\ket{0_L}$ state by $\sqrt{2}$ in $x$ (anti-squeezing by $\sqrt{2}$ in $p$),
\[
\ket{\varnothing}=\hat{S}(\sqrt{2})\ket{0_{L}}  
\]
This state is stabilized by two displacement operators with $\alpha=\sqrt{\pi}$ and $\beta=i\sqrt{\pi}$, however its code-space is one-dimensional, and therefore the state does not carry logical information. Two qunaught states can be combined on a balanced beam splitter to produce a Bell state in the square-lattice encoding, which is used for Knill-type error correction \cite{walshe_continuous-variable_2020}. 

Similarly, we can define hexagonal qunaught states, that when mixed on a balanced beam splitter result in a Bell state in the hexagonal encoding. First, we perform a squeezing operation on $\ket{0_L}$ to create a qunaught state, and then we squeeze it along the $\pi/4$ axis, as before:
\[
\ket{\varnothing}_\text{hex} = \hat{S}\left(\sqrt[4]{3}, \frac{\pi}{4}\right)\hat{S}(\sqrt{2})\ket{0_L}.
\]
 
\subsubsection{Approximate GKP states}
The wavefunction of an ideal GKP state is an infinite series of delta functions arranged on a particular lattice. Finite-energy GKP states can be modeled by replacing the delta functions by Gaussians of width $\Delta_x$ in $x$. The states are normalized by applying a Gaussian envelope of width $\Delta_p^{-1}$. In the $p$ quadrature, this corresponds to a series of Gaussian peaks of width $\Delta_p$ and overall envelope width $\Delta_x^{-1}$: 
\[
\ket{j_{\Delta_x,\Delta_p}}&=\sum_{n\in\mathds{Z}_0} e^{-\frac{1}{2}\Delta_p^2\left((2n+j)\sqrt{\pi\hbar}\right)^2}\hat{X}((2n+j)\sqrt{{\pi\hbar}})\ket{\Delta_x},\label{eq:GKP_approx_Delta_x}
\]
where $\ket{\Delta_x}_x=\hat{S}\left(\Delta_x^{-1}\right)\ket{0}$
and $\ket{0}$ is the vacuum. $\Delta_x=\Delta_p$ defines the symmetric finite energy GKP states.

Another way to approximate symmetric finite-energy GKP states is to apply a Fock damping operator on the ideal GKP states: $\ket{j_L^\epsilon}=e^{-\epsilon\hat{n}}\ket{j_L}$. The damping strength $\epsilon$ determines the width of the peaks and the Gaussian envelope. The variance of each peak (in both quadratures) is $\frac{\hbar}{2}\tanh(\epsilon)$, the center of each peak is scaled by $\sech(\epsilon)$, and the variance of the overall Gaussian envelope is $\frac{2}{\hbar}\tanh^{-1}(\epsilon)$ \cite{hastrup_analysis_2023}. When $\epsilon$ is small, $\ket{j_L^\epsilon}\simeq\ket{j_{\Delta_x,\Delta_p}}$ with $\Delta_x^2=\Delta_p^2=\tanh{\epsilon}$. The limit $\epsilon\to0$ recovers the ideal GKP states. 

\subsubsection{Effective squeezing}
The quality of approximate GKP states is often quantified by the squeezing of the Gaussian peaks relative to vacuum. This value can be calculated for arbitrary states from the expectation value of the GKP stabilizer in the opposite quadrature. For example, for the square-lattice encoding, the effective squeezing in $x$, $\Delta_x$, is calculated using the $p$-quadrature stabilizer, i.e. a $p$ grid displacement. The effective squeezing \cite{duivenvoorden_single-mode_2017} is defined as follows: 
\[
&\Delta_x = \sqrt{\frac{-2}{\abs{\beta}^2}\ln(\lvert\langle\hat{D}(\beta)\rangle\rvert)}\label{eq:eff_sqz_x},\\
&\Delta_p = \sqrt{\frac{-2}{\abs{\alpha}^2}\ln(\lvert\langle\hat{D}(\alpha)\rangle\rvert)}\label{eq:eff_sqz_p},
\]
where the stabilizer displacement values $\alpha$ and $\beta$ depend on the lattice, as listed in Table \ref{tab:stabilizers}. The squeezing can be expressed in decibel units via $\Delta_{\text{dB}}=-10\log_{10}(\Delta ^2)$. The $x$ and $p$ subscripts in the definition are specific to the square or rectangular lattice encoding. For the hexagonal encoding, the quadrature of the effective squeezing is the orthogonal quadrature of the displacement. Taking the absolute of the stabilizer expectation value ensures that the metric is invariant under displacements. The two figures of merit can be combined into a single quantity, 
the symmetric effective squeezing \cite{aghaee_rad_scaling_2025},
\[
\Delta_{\text{sym}}^2 = \frac{\Delta_x^2+\Delta_p^2}{2}.
\]
The current fault-tolerance threshold on the effective squeezing on the qunaught state has been improved from 10.1 dB \cite{larsen_fault-tolerant_2021,tzitrin_fault-tolerant_2021} to 9.75 dB in \cite{aghaee_rad_scaling_2025} thanks to improvements in the decoder. 

A closely related GKP quality measure is the nonlinear GKP squeezing introduced in \cite{marek_ground_2024}, which involves the expectation value of a linear combination of stabilizers. This figure of merit is not invariant under displacements, and it can therefore differentiate between logical states. Finally, the fidelity to an approximate GKP state in Eq.\ \eqref{eq:GKP_approx_Delta_x}, for given $\Delta_x$ and $\Delta_p$, can always be computed. However, it may not always capture the relevant features. For example, a squeezed vacuum state can have a high fidelity with a GKP state of only moderate effective squeezing.

\begin{table}[]
    \centering
    \caption{The two stabilizers displacements for the logical, sensor, hexagonal and hexagonal qunaught GKP states.}
    \begin{tabular}{lcc}\hline\hline
    State & $\alpha$ & $\beta$ \\
    \hline
    $\ket{j_L}$ & $\sqrt{2\pi}$  & $i\sqrt{2\pi}$ \\
    $\ket{\varnothing}$ & $\sqrt{\pi}$  & $i\sqrt{\pi}$ \\
    $\ket{j_L}_\text{hex}$ &  $\sqrt{\frac{\pi}{2}}(\kappa_++i\kappa_-)$ & $\sqrt{\frac{\pi}{2}}(\kappa_-+i\kappa_+)$ \\
    $\ket{\varnothing}_\text{hex}$ & $\frac{\sqrt{\pi}}{2}(\kappa_++i\kappa_-)$ & $\frac{\sqrt{\pi}}{2}(\kappa_-+i\kappa_+)$\\
    \hline\hline
    \end{tabular}

    \label{tab:stabilizers}
\end{table}

\subsection{Squeezed cat states (SCS)}
\label{sec:SCS}
Let $\ket{\psi_k}$ be a pure, squeezed cat with parity $k\in\{0,1\}$,
\[
\ket{\psi_k}=
\hat{S}(e^r)\left(\ket{\alpha}+(-1)^k \ket{-\alpha}  \right) .\label{eq:sqz_cat}
\]
Let the cat state be squeezed in the $x$-quadrature, $r>0$, and the amplitude of the cat be along the $x$-axis, $\alpha\in\mathbb{R}$. 
The wavefunction of the SCS in the position basis is a linear combination of two Gaussian peaks centered at $\pm\mu$ where $\mu=\sqrt{2\hbar}\alpha e^{-r}$ with variance $\sigma=\frac{\hbar}{2}e^{-2r}$,
\[
\psi_k(x)=G_{\mu,\sigma}(x)+(-1)^kG_{-\mu,\sigma}(x), \label{eq:sqz_cat_wavefun}
\]
where 
\[
G_{\mu,\sigma}(x)=(\pi \sigma)^{-\frac{1}{4}}\exp\left[ -\frac{(x-\mu)^2}{2\sigma} \right]\label{eq:Gaussian}.
\]
is a Gaussian function. As the density operator of the SCS is
\[
\hat{\rho}_k=\ketbra{\psi_k} = \hat{S}(r)\left(\ketbra{\alpha}+(-1)^k \ketbra{\alpha}{-\alpha} \right.\nonumber\\
\left.+(-1)^k \ketbra{-\alpha}{\alpha} + \ketbra{-\alpha}{-\alpha}\right)\hat{S}^\dagger(r),
\]
the Wigner function $W_{\hat{\rho}_{k}}(x,p)$ is a linear combination of four multivariate Gaussians \cite{bourassa_fast_2021},
\[
W_{\hat{\rho}_k}(\vb*{q})=G_{\vb*{\mu},\vb*{\sigma}}(\vb*{q})+(-1)^{k}e^{-2\alpha^2}G_{\vb*{\mu}_{z},\vb*{\sigma}}(\vb*{q}) \nonumber\\
+(-1)^{k}e^{-2\alpha^2}G_{-\vb*{\mu}_{z},\vb*{\sigma}}(\vb*{q})+G_{-\vb*{\mu},\vb*{\sigma}}(\vb*{q}) \label{eq:Wigner_sqz_cat}
\]
where 
\[
G_{\vb*{\mu},\vb*{\sigma}}(\vb*{q})=\frac{1}{\sqrt{  \det(2\pi\vb*{\sigma}) }}\exp\left[ -\frac{1}{2}(\vb*{q}-\vb*{\mu})^T \vb*{\sigma}^{-1}(\vb*{q}-\vb*{\mu}) \right],
\]
and $\vb*{q}=(x,p)^T$, $\vb*{\mu}=\sqrt{ 2\hbar }\begin{pmatrix}
\alpha e^{-r} \\
0
\end{pmatrix}$, $\vb*{\mu}_{z}=\sqrt{ 2\hbar }\begin{pmatrix}
0 \\
i\alpha e^{r}
\end{pmatrix}$, and $\vb*{\sigma}=\frac{\hbar}{2}\begin{pmatrix}
e^{-2r} & 0 \\
0 & e^{2r}
\end{pmatrix}$. The two Gaussians with complex means are complex conjugates of each other, and form the interference fringes at the center of the SCS. 

\subsubsection{Noisy squeezed cat states}
A Gaussian channel can be straightforwardly applied in the LCoG formalism by transforming each displacement vector and covariance matrix in the Wigner function in Eq.\ \eqref{eq:Wigner_sqz_cat} according to $\vb*{\mu}_k\mapsto \vb*{X}\vb*{\mu}_k $ and $\vb*{\sigma}\mapsto\vb*{X}\vb*{\sigma}\vb*{X}^T+\vb*{Y}$ \cite{weedbrook_gaussian_2012,bourassa_fast_2021}. For a photon loss channel with transmissivity $\eta$, $\vb*{X}=\sqrt{\eta}\mathds{1}_2$ and $\vb*{Y}=(1-\eta)\frac{\hbar}{2}\mathds{1}_2$. The density operator of a SCS that has undergone such a photon loss channel will be called $\hat{\rho}_{\text{cat}}^\eta$.
\section{Grid state preparation with ideal cat breeding}\label{sec:ideal_protocol}
In this section, we introduce the ideal SCS breeding protocol. In Sec.\ \ref{sec:pure_output_state}, we derive an expression for the wavefunction of the output state, which is a finite sum of equally spaced Gaussian peaks with a binomial distribution. Each peak is also weighed by a complex factor dependent on the homodyne measurement outcome. In Sec.\ \ref{sec:GKP_codewords}, we discuss strategies to generate GKP code words, which must have a particular grid spacing. We discuss how a feedforward squeezing operation can be used to align the grid to any GKP lattice. In Sec.\ \ref{sec:N=3}, we analyze the effect of the homodyne measurements on the weights of the peaks for the specific case of breeding three SCS. We find that the measurement-dependent weights can suppress the peaks, and certain measurement outcomes do not herald a viable grid state. 

\begin{figure}
    \centering
    \includegraphics[width=\linewidth]{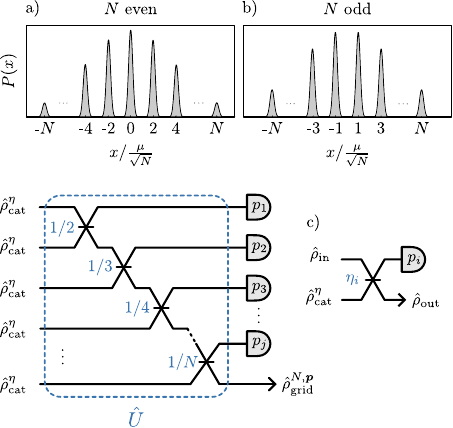}
    \caption{\textbf{The breeding circuit for grid state generation}. $N$ squeezed cat states $\hat\rho_\text{cat}^\eta$ are sent through a linear optical cascade, $\hat{U}$.  The ratios (blue) indicate the reflectivity $\eta_i$ of each beam splitter. The $p$ quadratures of all but the last mode are measured. A sketch of the marginal probability of the resulting grid state when breeding a) even and b) odd numbers of (equal parity) input states, post-selected on $\vb*{p}=(p_1,p_2,\dots,p_{N-1})^T=\vb*{0}$. A sub-circuit of the $i$'th breeding step is shown in c). For $i=1$, $\hat{\rho}_{\text{in}}=\hat{\rho}_\text{cat}^\eta$. The output state of step $i$ becomes the input state of the ($i+1$)'th breeding step. }
    \label{fig:cat_breeding_circuit}
\end{figure}

\subsection{Calculating the output grid state}
\label{sec:pure_output_state}
A schematic of the breeding circuit is shown in Fig.\ \ref{fig:cat_breeding_circuit}. $N$ SCS are sent through a beam splitter cascade, after which $N-1$ of the modes are measured with homodyne detection of the $p$-quadrature, heralding a grid state. The beam splitter cascade $\hat{U}$ is equivalent to the binary tree breeding circuit in the efficient breeding scheme described in \cite{vasconcelos_all-optical_2010}. 


\subsubsection*{Beam splitter cascade}
 Each beam splitter in the cascade transforms the quadratures linearly, 
 \[
 &\hat{B}_{i,i+1}^\dagger\hat{x}_i \hat{B}_{i,i+1}=\sqrt{\eta_i}\hat{x}_i-\sqrt{1-\eta_i}\hat{x}_{i+1} \nonumber \\
 &\hat{B}_{i,i+1}^\dagger \hat{x}_{i+1}\hat{B}_{i,i+1}=\sqrt{1-\eta_i}\hat{x}_{i}+\sqrt{\eta_i}\hat{x}_{i+1}
 \]
where $\eta_i$ is the reflectivity of the beam splitter (and similarly for $\hat{\vb*{p}}$). The beam splitter cascade, $\hat{U}=\prod_{i=1}^{N-1}\hat{B}_{i,i+1}$, transforms the quadrature operators as $\hat{U}^{\dagger} \hat{\vb*{x}}\hat{U}=\vb*{B}\hat{\vb*{x}}$, where the symplectic matrix $\vb*{B}$ is defined in Eq.\ \eqref{eq:BS_cascade} in Appendix \ref{app:bs_cascade}. The $x$-quadrature of the output mode is simply an equally weighted linear combination of all the input $x$-quadratures: $\hat{x}_{N}' = \frac{1}{\sqrt{ N }}\left( \sum_{i=1}^{N}\hat{x}_{i} \right)$. 
 
 Now, the beam splitter cascade is applied to a tensor product of $N$ squeezed cat states, $\hat{U}\bigotimes _{i=1}^{N}\ket{\psi_{k_i}}_i$. The wavefunction of the tensor product of $N$ squeezed cat states, $\ket{\Psi}=\bigotimes_{i=1}^N\ket{\psi_{k_i}}$, where $k_i$ is the parity of each SQS, can be written as a linear combination of $2^N$ multivariate Gaussians, 
\[
\Psi(\vb*{x})=\prod_{i=1}^{N}\psi_{k_{i}}(x_{i})=\sum_{n=1}^{2^N} \gamma(\vb*{k}_n)G_{\vb*{\mu_{n}},\vb*{\sigma}}(\vb*{x}) \label{eq:wavefun_Psi}
\]
where $\vb*{x}=(x_{1},x_{2},\dots,x_{N})^T$, and $\gamma(\vb*{k}_{n})\in\{-1,+1\}$ indicates the sign in front of each Gaussian, which is dependent on the parities of the SCS, $\vb*{k}_n\in\otimes_{i=j}^N\{0,k_j\}$ and $\gamma(\vb*{k}_n)=(-1)^{\sum_{i=1}^N k_{ni}}$ where $k_{ni}$ is the $i$'th element of $\vb*{k}_n$. The displacement vector is also unique for each term, $\vb*{\mu}_n\in\{\mu,-\mu\}^{N}$. $\vb*{\sigma}$ is the $N\times N$ matrix of covariances, which is diagonal: $\vb*{\sigma}=\frac{\hbar}{2}e^{-2r}\mathds{1}_N$. We can apply the beam splitter cascade unitary by simply transforming $\vb*{\mu}$ and $\vb*{\sigma}$ in the wavefunction in Eq.\ \eqref{eq:wavefun_Psi}. In particular, since $\vb*{\sigma}$ is proportional to the identity matrix, beam splitters have no effect on the covariance matrix, $\vb*{\sigma}\mapsto\vb*{B}\vb*{\sigma}\vb*{B}^T=\vb*{\sigma}$. The vector of means, on the other hand, "sees" the entanglement, $\vb*{\mu}_n\mapsto\vb*{B}\vb*{\mu}_n$. Since $\vb*{\sigma}$ is still proportional to the identity, we can factor the multivariate Gaussians into products of $N$ single-variate Gaussians after applying the beam splitters,
\[
\Psi(\vb*{B}^{-1}\vb*{x})=\sum_{n=1}^{2^N}\gamma(\vb*{k}_{n})\prod_{i=1}^NG_{\mu_{ni}',\sigma}(x_{i}),\label{eq:wavefun_UPsi}
\]
where $\mu_{ni}'=\sum_{j}B_{ij}\mu_{nj}$, $n$ is the sum index, $i$ is the mode index.

\subsubsection*{Homodyne measurements}
The homodyne measurement projects $N-1$ modes onto the momentum eigenstates $\ket{\vb*{p}}_p$, where $\vb*{p}=(p_{1},p_{2},\dots,p_{N-1})^T$ is a vector of measurement results. The projection results in the factor $\braket{\vb*{p}}{\tilde{\vb*{x}}}=e^{-i\vb*{p}^T \tilde{\vb*{x}}/\hbar}$, where $\tilde{\vb*{x}}=(x_{1},x_{2},\dots,x_{N-1})^T$, in the multivariate wavefunction,
\[
\bra{\vb*{p}}_{p}\hat{U}\bigotimes_{i=1}^{N} \ket{\psi_{k_i}} = \int  \dd^N{{\vb*{x}}} \:\Psi(\vb*{B}^{-1}\vb*{x}) e^{-i \vb*{p}^T \tilde{\vb*{x}} /\hbar} \ket{x_{N}} \nonumber \\
=\int \dd{x_{N}}\: \sum_{n=1}^{2^N} \gamma(\vb*{k}_{n}) G_{\mu_{nN}',\sigma}(x_{N}) \nonumber\\
\times \underbrace{\left[ \int \dd^{N-1}{\tilde{\vb*{x}}}\: e^{-i \vb*{p}^T \tilde{\vb*{x}}/\hbar}\prod_{i=1}^{N-1}G_{\mu_{ni}',\sigma}(x_{i}) \right]}_{f_n(\vb*{p})} \ket{x_{N}}\label{eq:wavefun_homo_intermediate}
\]
In the second line of Eq.\ \eqref{eq:wavefun_homo_intermediate}, we inserted the expression for the wavefunction Eq.\ \eqref{eq:wavefun_UPsi}, and factored out the $N$'th Gaussian, which is not involved in the integration over $\tilde{\vb*{x}}$. The $N-1$ integrals can be evaluated analytically,
\[
f_{n}(\vb*{p}) &= \prod_{i=1}^{N-1} (\pi \sigma)^{1/4}e^{-ip_{i}\mu_{ni}'/\hbar}\exp\left[ -\frac{p_i^2\sigma}{2} \right] \\
&\propto e^{-i\tilde{\vb*{p}}^T\vb*{B}\vb*{\mu}_n/\hbar}\nonumber
\]
where $\tilde{\vb*{p}}=\mqty(\vb*{p} \\ 0)$.
The output state $\ket{\psi_{\text{grid}}}$ of the breeding circuit is a linear combination of Gaussians, each weighted by a factor dependent on the measurement outcome $\vb*{p}$:
\[
\psi_{\text{grid}}(x_{N})\propto \sum_{n=1}^{2^N}\gamma(\vb*{k}_{n})f_{n}(\vb*{p})G_{\mu_{nN}',\sigma}(x_{N})\label{eq:psi_grid_long}
\]
\subsubsection*{Simplifying the output grid state}
The wavefunction of the output state in Eq.\ \eqref{eq:psi_grid_long} can be simplified by noting that the beam splitter cascade equally weighs the means of each quadrature in the final mode: $\mu_{nN}'=\frac{1}{\sqrt{ N }}\sum_{i=1}^{N}\mu_{ni}$. Due to the symmetry of the cats, the means will have $N+1$ possible unique values, with their occurrence given by a binomial factor. Therefore, the sum with $2^N$ terms can be recast into a sum over $N+1$ terms with binomial coefficients,
\[
\psi_{\text{grid}}(x_{N})\propto \sum_{n\in\mathcal{N}}\begin{pmatrix}
N  \\
N- \abs{n}
\end{pmatrix}  \tilde{f}(\vb*{k_n},\vb*{p}) G_{ \frac{n\mu}{\sqrt{ N }},\sigma}(x_{N}) \label{eq:psi_grid}
\]
where $\mathcal{N}= \{-N,\dots,-2,0,2,\dots,N\}$ for $N$ even and $\mathcal{N}= \{-N,\dots,-3,-1,1,3,\dots,N\}$ for $N$ odd. The marginal distributions of Eq.\ \eqref{eq:psi_grid} for $\vb*{p}=\vb*{0}$ and $\vb*{k}_n=\vb*{0}$ are sketched in Fig.\ \ref{fig:cat_breeding_circuit}a and \ref{fig:cat_breeding_circuit}b. The location of the $N+1$ peaks for even and odd numbers of input states is at even or odd values of $x$ modulo $\frac{\mu}{\sqrt{N}}$. Each breeding step increases the number of peaks by one. $\tilde{f}(\vb*{k}_n, \vb*{p})$ is the combined measurement- and parity-dependent factor for each unique peak. For some values of $\vb*{p}$, $\tilde{f}(\vb*{k}_n,\vb*{p})$ can suppress certain peaks, which can cause the breeding protocol to fail at producing a grid state that is useful for quantum error correction. The effect of the homodyne measurement result on the peak suppression is studied in detail in section \ref{sec:N=3} when breeding $N=3$ squeezed cats.


\subsection{Preparing GKP codewords}
\label{sec:GKP_codewords}

The quality of the prepared grid state can be evaluated using the effective squeezing $\Delta_x$ from Eq.\ \eqref{eq:eff_sqz_x}, and $\Delta_p$ from Eq.\ \eqref{eq:eff_sqz_p}. Most important is that the peaks are narrow, have a large envelope, and properly aligned with the GKP grid. 

For the grid state in Eq.\ \eqref{eq:psi_grid}, each peak has variance equal to $\Delta_x = e^{-r}$, i.e. the initial variance of SCS. Meanwhile, $\Delta_p$ is related to the envelope of Eq.\ \eqref{eq:psi_grid}, which is binomial. When $N$ is large, the binomial envelope approximates well a Gaussian envelope. 
However, in a practical implementation of the breeding protocol, $N$ will not be in this limit, because non-Gaussian states are notoriously difficult to prepare. The outer peaks of the prepared grid state in Eq.\ \eqref{eq:psi_grid} will therefore be more suppressed than in the finite energy approximation of GKP states in Eq.\ \eqref{eq:GKP_approx_Delta_x}. 

In order to align the grid state with a specific GKP lattice, the initial peak locations of the cats must be multiplied by $\sqrt{N}$ to counter the shrinking effect due to the beam splitter cascade. Specifically, for the logical grid spacing of $2\sqrt{\pi\hbar}$ in $x$, $\mu=\sqrt{N\pi\hbar}$. For the qunaught state with spacing $\sqrt{2\pi\hbar}$, $\mu=\sqrt{\frac{N\pi\hbar}{2}}$. 

Instead of fixing $\mu$ at the beginning, which requires SCS with large amplitudes, one could instead breed identical cats with smaller $\mu$, and use an active squeezing operation on the output state to align it with a desired GKP grid. For example, $N$ copies of squeezed cats with peak location $\mu$ can be bred, and the squeezing operation $\hat{S}\left(\frac{\mu}{\sqrt{N}\mu_T}\right)$ can be applied to obtain a grid with spacing $\mu_T=\sqrt{\pi\hbar}$ for a logical state or $\mu_T=\sqrt{\frac{\pi\hbar}{2}}$ for the qunaught state. If a hexagonal grid is desired, we can follow up with a $\hat{S}(\sqrt[4]{3},\frac{\pi}{4})$ operation. However, since the first 
operation is anti-squeezing in $x$ if $\mu<\sqrt{N}\mu_T$, the variance of the peaks in $x$ will increase, which lowers $\Delta_x$. There is therefore a trade-off to be considered if smaller cats, which require fewer photon numbers to generate, are used. Aghaee et al. \cite{aghaee_rad_scaling_2025} also suggests using variable beam splitters in the breeding circuit to breed non-identical cats, that is cats with different initial $\mu$'s, followed by one final active squeezing operation on the output. Such operations are examples of deterministic Gaussian CP maps examined by \cite{hahn_deterministic_2022}. 

When targeting a specific codeword, i.e. $\ket{0_L}$ or $\ket{1_L}$ or the qunaught state, a corrective displacement equal to half the grid spacing in either $x$ or $p$ direction may need to be applied, depending on the parity of $N$ and the parity of the input cats $\vb*{k}$. Since the effective squeezing is invariant under displacement, this figure of merit concerns itself only with alignment on  particular code-space lattice, and we assume that this feedforward operation can be performed at a later step, or in post-processing. 

\subsection{Breeding three cats} \label{sec:N=3}
We now turn our attention back to the expression for the heralded grid state in Eq. \eqref{eq:psi_grid}. In order to understand the effect of the measurement-dependent factors on the weights, we explicitly calculate the output state for $N=3$.
\begin{table}[]
    \centering
    \begin{ruledtabular}
    \begin{tabular}{c|c|c|c}
       $n$  & $\vb*{\mu}_n/\mu$ & $\vb*{B}\vb*{\mu}_n / \mu$ & $f_n(\vb*{p})$ \\
       \colrule
       1 & $(1,1,1)$ & $(0,0,\sqrt{ 3 })$ & 1 \\
       2 & $(-1,1,1)$ & $\left( -\sqrt{ 2 },-\sqrt{ \frac{2}{3} }, \frac{1}{\sqrt{ 3 }} \right)$ & $e^{i\mu\left(\sqrt{ 2 } p_{1}+\sqrt{ \frac{2}{3} } p_{2}\right)/\hbar}$ \\
       3 & $(1,-1,1)$ & $\left( \sqrt{ 2 },-\sqrt{ \frac{2}{3} }, \frac{1}{\sqrt{ 3 }} \right)$ & $e^{-i \mu\left( \sqrt{ 2 }p_{1}-\sqrt{ \frac{2}{3} }p_{2}  \right)/\hbar}$ \\
       4 & $(1,1,-1)$ & $\left( 0,2\sqrt{ \frac{2}{3} }, \frac{1}{\sqrt{ 3 }} \right)$ & $e^{-i \mu 2\sqrt{ \frac{2}{3} }p_{2}}/\hbar$ \\
       5 & $(-1,-1,1)$ & $\left( 0,-2\sqrt{ \frac{2}{3} }, -\frac{1}{\sqrt{ 3 }} \right)$ & $e^{i \mu 2\sqrt{ \frac{2}{3} }p_{2}}/\hbar$ \\
       6 & $(1,-1,-1)$ & $\left( \sqrt{ 2 },\sqrt{ \frac{2}{3} }, -\frac{1}{\sqrt{ 3 }} \right)$ & $e^{-i \mu\left(\sqrt{ 2 }p_{1}+\sqrt{\frac{2}{3}} p_{2}\right)/\hbar}$ \\
       7 & $(-1,1,-1)$ & $\left( -\sqrt{ 2 },\sqrt{ \frac{2}{3} }, -\frac{1}{\sqrt{ 3 }} \right)$ & $e^{i \mu\left(\sqrt{ 2 }p_{1}-\sqrt{\frac{2}{3}} p_{2}\right)/\hbar}$ \\
       8 & $(-1,-1,-1)$ & $\left(0, 0, -\sqrt{ 3 } \right)$ & 1\\ 
    \end{tabular}
    \end{ruledtabular}
    \caption{The mean vectors and measurement-dependent phase in the wavefunction evolution during the breeding protocol for $N=3$ input states.}
    \label{tab:N=3}
\end{table}
\begin{figure*}
    \centering
    \includegraphics[width=\linewidth]{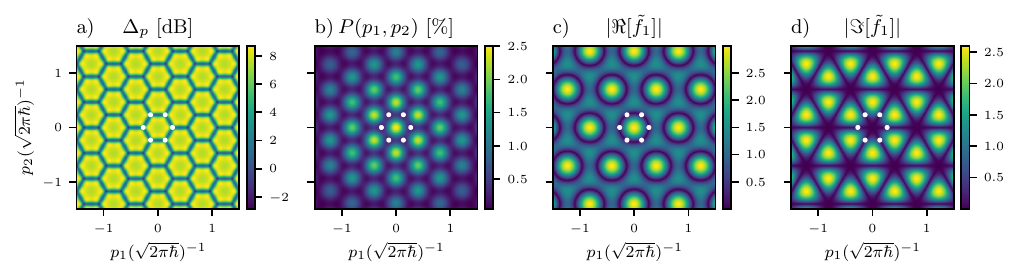}
    \caption{\textbf{The effect of the homodyne measurement outcomes when breeding three SCS}. The effective squeezing $\Delta_p$ a) and the measurement probability b) of the grid state produced when breeding three ideal cat states with $-12$ dB of squeezing with $\mu=\sqrt{\frac{3\pi\hbar}{2}}$ set to target the grid of a qunaught state. The output state is post-selected on homodyne outcomes $(p_1,p_2)$.  
    The real c) and imaginary parts d) of the measurement-dependent factor $\tilde{f}_{1}(p_1,p_2)$ in Eq.\ \eqref{eq:N=3phase} of the inner peaks of the output grid state. The white points lie on the intersection of a circle with radius $d = \frac{\pi\hbar}{4\mu}\sqrt{\frac{7}{2}}$ in c) and the six lines of rotational symmetry radiating from the central hexagon in d), which form the hexagonal pattern of low effective squeezing a). }
    \label{fig:N=3}
\end{figure*}

\begin{figure*}
    \centering
    \includegraphics[width=\linewidth]{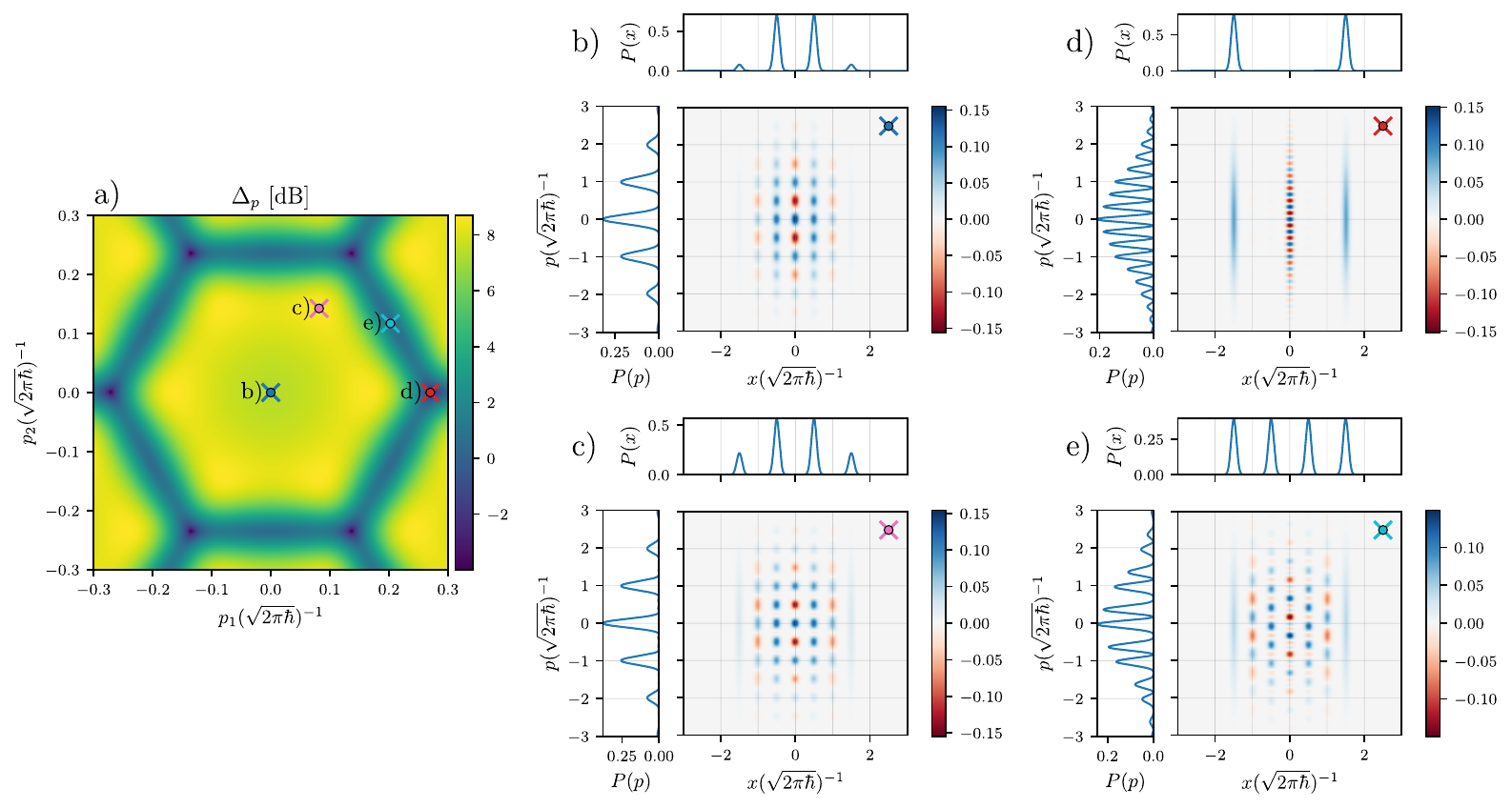}
    \caption{\textbf{Post-selected output states}. A zoom in of the central hexagon in the effective squeezing of the output grid state when breeding three cats a). The Wigner functions and marginal distributions of the output states post selected on $(p_1,p_2)$ coordinates b)-e) marked by the colored crosses in a). The effective squeezing $\Delta_p$ at each coordinate is $7.37$ dB b), $8.70$ dB c), $-3.53$ dB d), and  $0.57$ dB e). The effective squeezing is $\Delta_x=12$ dB for b)-e).}
    \label{fig:N=3points}
\end{figure*} 

There are a total of $2^3=8$ combinations of multi-mode mean vectors in the multivariate wavefunction in Eq. \eqref{eq:wavefun_Psi}, as listed in the first column of Table \ref{tab:N=3}.  After applying the linear optical cascade, we see that the last mode has four unique means $\pm \sqrt{3}\mu$ and $\pm\frac{1}{\sqrt{3}}\mu$ in the second column of Table \ref{tab:N=3}. The output state of the breeding protocol can therefore be written as a linear combination of four Gaussians,
\[
\psi_{\text{grid}}(x_3) &\propto G_{\sqrt{ 3 }\mu,\sigma}(x_{3})+3\tilde{f}_{1}(p_{1},p_{2})G_{\frac{1}{\sqrt{ 3 }\mu},\sigma}(x_3)\nonumber\\
&+3 \tilde{f}_{-1}(p_{1},p_{2})G_{-\frac{1}{\sqrt{ 3 }\mu},\sigma}(x_{3})+G_{-\sqrt{ 3 }\mu,\sigma}(x_{3}),
\]
where the overall weight factor $\tilde{f}_{1}(p_1,p_2)$ is found by summing the individual weight factors $f_n(p_1,p_2)$ of rows $n=2,3,4$ in Table \ref{tab:N=3}, and $\tilde{f}_{-1}(p_1,p_2)$ is found by summing the weight factors from rows $n=5,6,7$. Below, we give an explicit expression for the weights when each cat has parity $0$:
\[
\tilde{f}_1(p_1,p_2)&=\tilde{f}_{-1}^*(p_1,p_2), \nonumber\\
&=2e^{i \mu \sqrt{ \frac{2}{3} }p_{2}/\hbar}\cos(\mu \sqrt{ 2} p_{1}/\hbar)+e^{-i \mu 2 \sqrt{ \frac{2}{3} }p_{2}/\hbar}.\label{eq:N=3phase}
\] 
For different cat parities, the weights in each row of Table \ref{tab:N=3} would potentially have different signs, resulting in different values of $\tilde{f}_{\pm 1}(p_1,p_2)$.

In Fig. \ref{fig:N=3}a, the effect of the homodyne measurement result $\vb*{p}=(p_1,p_2)$ on the effective squeezing $\Delta_p$ is shown when breeding cats with $\mu=\frac{\sqrt{3\pi\hbar}}{2}$. The effective squeezing exhibits a hexagonal pattern, where dark regions at the edges indicate a severe failure of the breeding protocol. Fig.\ \ref{fig:N=3points}d and \ref{fig:N=3points}e show the Wigner function for the post-selected states on such points. Fig.\ \ref{fig:N=3points}d shows a complete failure in increasing the number of peaks. Rather, a larger cat appears to have been produced (breeding with $x$-measurements can also be used to create cats with large amplitudes \cite{laghaout_amplification_2013,sychev_enlargement_2017}). In Fig.\ \ref{fig:N=3points}e, the inner peaks in the $x$ quadrature are suppressed due to weights on the individual peaks, an effect that is especially noticeable in the $p$-quadrature, in which the peaks appear to split. It is also clear from their Wigner functions that applying displacements to these states will not improve their GKP-quality.

The hexagonal pattern in the effective squeezing arises from the minima of the weight function on the inner peaks, $\tilde{f}_1(p_1,p_2)$, which is plotted in Fig.\ \ref{fig:N=3}c, and Fig.\ \ref{fig:N=3}d. The minima of the real part form a pattern of circles, while those of the imaginary part has minima forming a pattern of equilateral triangles. Together, this shapes the hexagonal pattern in $\Delta_p$, where the vertices of the hexagons lie on the circumference of the circle and on the six edges of the triangles. 

Another peculiar feature is that the best GKP candidate is not post-selected on $\vb*{p} = (0,0)$. Inside each hexagon (Fig.\ \ref{fig:N=3points}a) is a smaller, brighter inner hexagon whose vertices, although less likely, appear to create a wider envelope in $x$, as shown in the Wigner function in Fig.\ \ref{fig:N=3points}c, resulting in a better $\Delta_p$ value.

The regions of poor effective squeezing are, fortunately, located near the minimal regions of the probability distribution of the homodyne measurement results in Fig.\ \ref{fig:N=3}b, as highlighted by the white points. The overall success probability of the breeding protocol should still be high, and asymptotically approach unity as the number of cats is increased \cite{weigand_generating_2018}. We define the success of the breeding protocol as one where the effective squeezing is at least as good as when post-selected on $p=0$,
\[
\Delta_p(\vb*{p})\leq\Delta_p(\vb*{p}=0).
\]
The resulting success probability is computed numerically (for simulation methods, see Sec.\ \ref{sec:simulation}) and plotted in Fig.\ \ref{fig:sampling_eta} (solid, dark blue line). For each $N$, we use squeezed cats with $\mu=\frac{\sqrt{N\pi\hbar}}{2}$ and squeezing values of $-12$ dB and $-15$ dB . Under ideal conditions, the success probability appears to climb from around 80\% towards unity. The probability of preparing a fault-tolerant qunaught state with symmetric effective squeezing of at least 9.75 dB is already at least 80\% with four ideal cat states. This success rate is effectively deterministic, compared to GKP preparation rates of GBS devices. However, the introduction of loss on the input states, as explored in Sec.\ \ref{sec:non-ideal_protocol} has a significant effect on the success probability. 


\section{Grid state preparation with lossy input states}\label{sec:non-ideal_protocol}

\begin{figure}
    \centering
    \includegraphics[width=\linewidth]{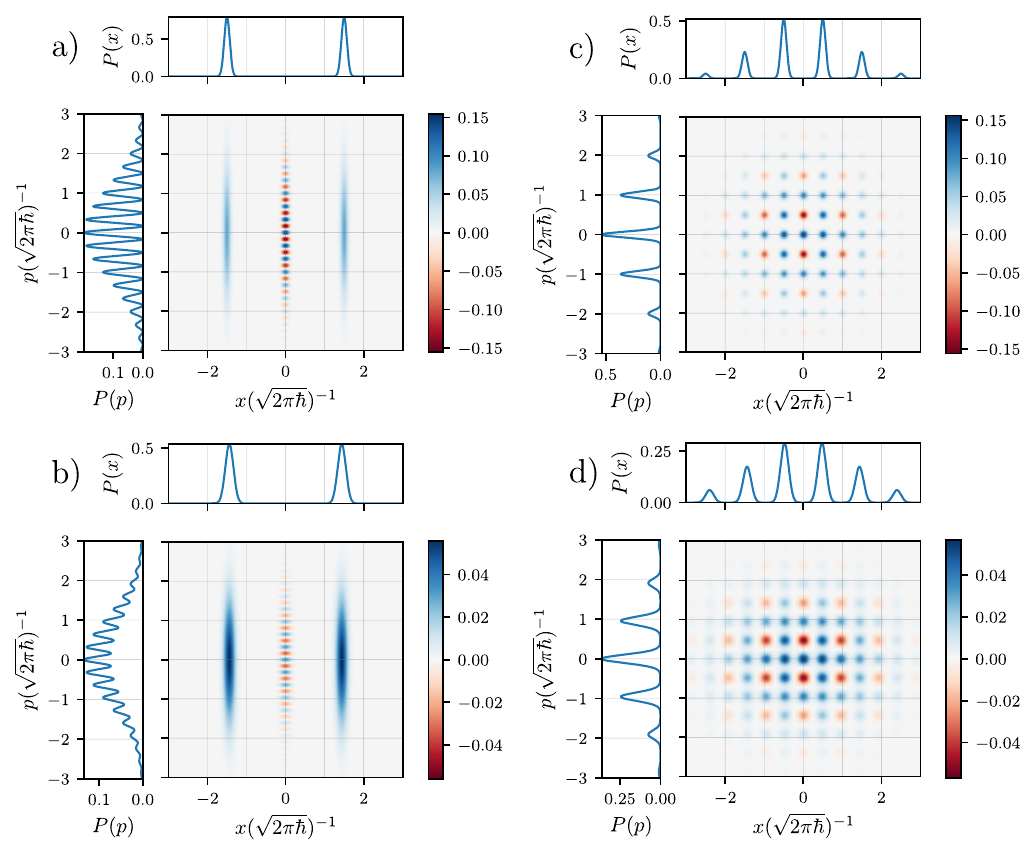}
    \caption{\textbf{Effect of the loss channel}. The Wigner function of the $r=-12$ dB input cat state after a loss channel with transmissivity of $\eta = 1$ a) and $\eta=0.92$ b). The Wigner function of the output state after breeding $N=9$ copies of the input state on the left and post-selecting on $\vb*{p}=0$ is shown in c) and d). For c), $\Delta_x=12.00$ dB, $\Delta_p=11.73$ dB, and $\Delta_s=11.87$ dB. For d) $\Delta_x=7.52$ dB, $\Delta_p=7.04$ dB and $\Delta_s=7.27$ dB. } 
    \label{fig:wigner_loss}
\end{figure}

\begin{figure*}
    \centering
    \includegraphics[width=0.95\linewidth]
    {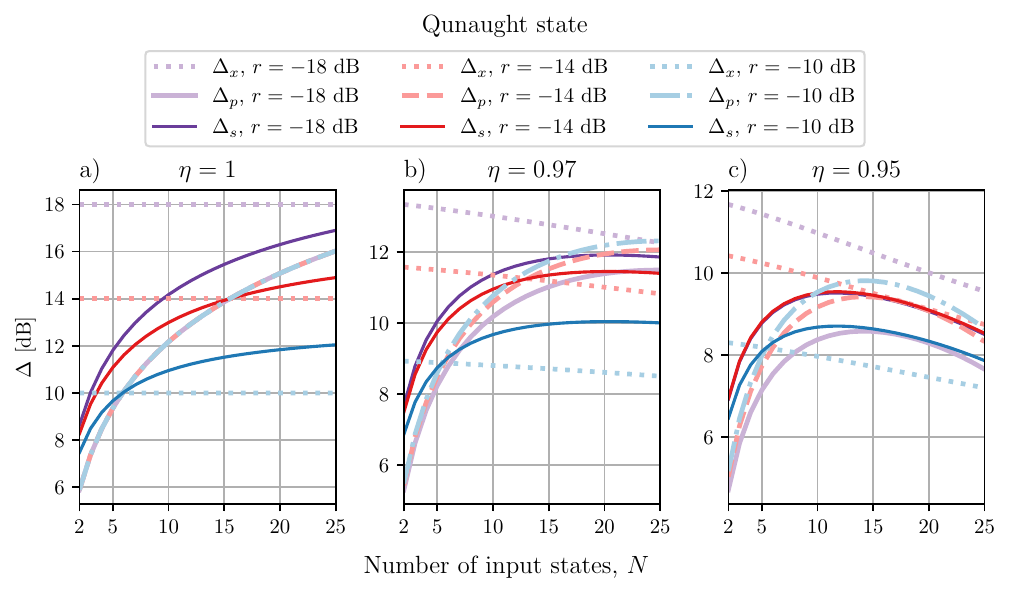}
    \caption{\textbf{The effect of loss on the effective squeezing}. The evolution of the effective squeezing in both quadratures $\Delta_x$ and $\Delta_p$ (faded colors) when targeting the qunaught state for 10, 14, 18 dB of cat squeezing as more input states are used with given loss channel transmissivities $\eta$. The grid state is post-selected on $\vb*{p}=0$. The symmetric effective squeezing $\Delta_s$ is also shown as a bright line. At $\eta=0.95$, the purple line is right below the red line.}
    \label{fig:effective_sqz_p=0_qunaught}
\end{figure*}

\begin{figure*}
    \centering
    \includegraphics[width=0.95\linewidth]
    {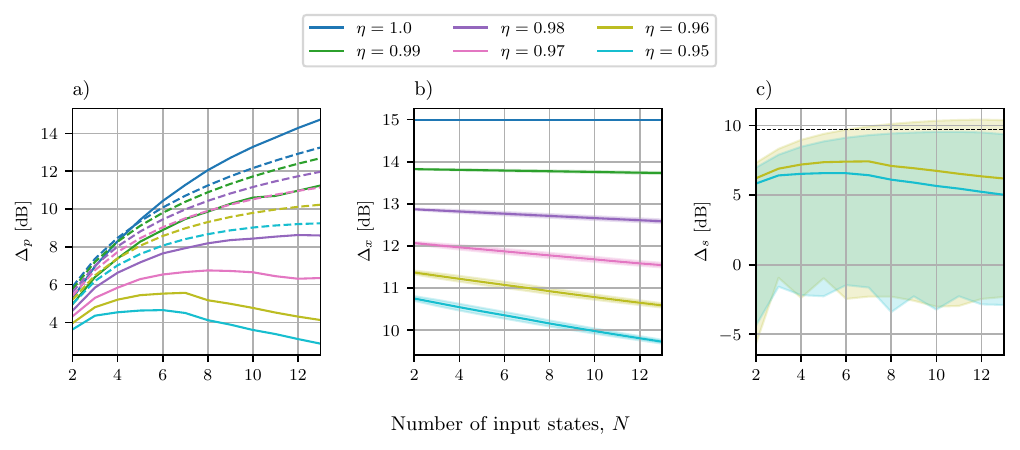}
    \caption{\textbf{The effect of loss on the effective squeezing, averaged over the homodyne measurement outcomes}. The average effective squeezing (solid lines) of the output state after breeding $N$ cats with $-15$ dB squeezing and a transmissivity channel of $\eta$. a) The effective squeezing in $p$, $\Delta_p$. The dashed lines are $\Delta_p$ of the state heralded at $\vb*{p}=\vb*{0}$. b) The effective squeezing in $x$, $\Delta_x$. The shaded region is $\pm$ one standard deviation of $\Delta_x$. c) The symmetric effective squeezing $\Delta_s$. The shaded region is between the maximum and minimum symmetric effective squeezing value of the sampled states. The dashed black line is at the fault tolerance threshold of $9.75$ dB. } \label{fig:sampling_-15_dB}
\end{figure*}

In an experimental setting, loss and noise are inevitably present to some extent. In this section, we show how optical losses in the setup affects the performance and the success rate of the breeding protocol by studying, which has implications on the ability to generate fault-tolerant GKP states. 

\begin{figure}
    \centering
    \includegraphics[width=\linewidth]{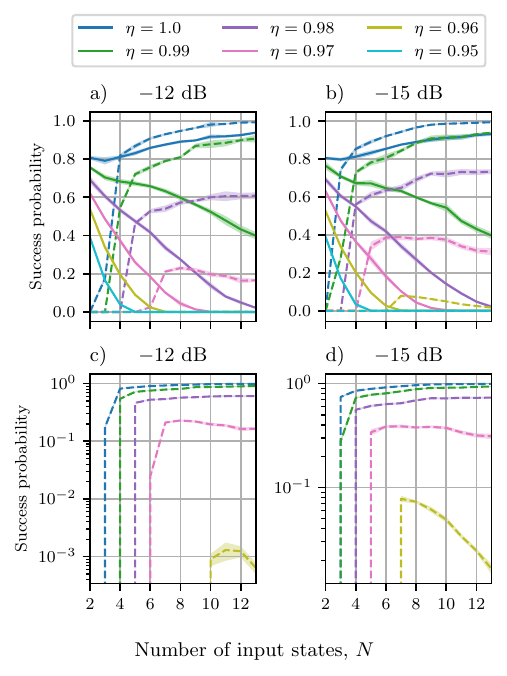}
    \caption{\textbf{The success probability of the breeding protocol}. The probability of $\Delta_p(\vb*{p})\geq\Delta_p(\vb*{p}=0)$ (solid line) is shown in a)-b). The probability of breeding a fault-tolerant qunaught state with symmetric effective squeezing $\Delta_s(\vb*{p})\geq 9.75$ dB (dashed line) is shown for different values of the loss channel transmissivity $\eta$ on the input squeezed cats. The shaded regions are over $\pm$ one standard deviation away from the average.}
    \label{fig:sampling_eta}
\end{figure}

We model the experimental imperfections by applying a photon loss channel of transmissivity $\eta$ on the squeezed cats. Notably, the loss channel scales the means by $\sqrt{\eta}$, so input states intended to be used for many rounds of breeding with large amplitudes, $\mu\propto\sqrt{N}$,  are affected by loss more severely than for smaller $N$. In Fig.\ \ref{fig:wigner_loss}a and \ref{fig:wigner_loss}b, the Wigner functions of the input squeezed cats with $\mu=\sqrt{\frac{9\pi\hbar}{2}}$ is shown before and after a moderate loss channel of $\eta=0.92$, which was chosen for visualization purposes. The Gaussian convolution increases the variance of the peaks in $x$ and greatly affects the interference fringes in $p$. Fig.\ \ref{fig:wigner_loss}c and Fig.\ \ref{fig:wigner_loss}d show the output state (post-selected on $\vb*{p}=0$) when nine copies of the squeezed cat are bred. In addition to the variance of each peak increasing when breeding the lossy state, the grid of the output state has also shrunk inwards slightly in both directions, overall causing a decrease in the effective squeezing in both quadratures. For a visualization of the effect of different homodyne measurement outcomes when breeding lossy states, see Fig.\ \ref{fig:N=3_loss} in Appendix \ref{app:sm_figures}. 

 In principle, the shrinking of the grid can be accompanied by shrinking the stabilizer displacements of the code-space, resulting in stabilizers that no longer commute \cite{aghaee_rad_scaling_2025}. When measuring stabilizers, this shrinking can be accounted for during the homodyne detector binning, which led to better fault-tolerance threshold values \cite{tzitrin_fault-tolerant_2021}. However, in this work we always compare with the perfect qunaught lattice, for which the fault-tolerance threshold is defined, i.e. we do not shrink the stabilizer displacements when evaluating Eq.\ \eqref{eq:eff_sqz_x} and \eqref{eq:eff_sqz_p}.

 Since the loss scales the means by $\sqrt{\eta}$, it can in general be mitigated by starting with a larger SCS whose amplitude is scaled by $\sqrt{\eta}^{-1}$. While this improves $\Delta_x$, it is at the expense of lowering $\Delta_p$, which becomes significantly worse. For example, pre-scaling the cat state in Fig.\ \ref{fig:wigner_loss} before applying the loss results in a grid state with $\Delta_x=8.60$ dB and $\Delta_p=5.28$ dB, giving a symmetric effective squeezing of $\Delta_s=6.63$ dB, which is worse than without scaling. Therefore, we do not consider mitigating the loss this way. 

 In the following, we investigate the quality of the generated grid states (targeting a qunaught state) when loss channels with transmissivity $\eta\in[0.95,1]$ have been applied on the SCS.  First, we consider the grid state heralded on $\vb*{p}=\vb*{0}$ in Fig.\ \ref{fig:effective_sqz_p=0_qunaught}. As the loss is increased, the effective squeezing in both quadratures begins to decrease with more rounds of breeding, an effect which is exacerbated in $p$ when the squeezing is high. A similar effect is seen in the preparation of logical states in the square-lattice encoding, see Fig.\ \ref{fig:effective_sqz_p=0_logical} in the Appendix. When the cat parities are equal, the effective squeezing at $\vb*{p}=0$ is a good indication of the overall performance of the breeding protocol, as this is the most probable outcome. In the presence of loss, the probability of the homodyne measurements will be convolved with a Gaussian, reducing the probability of the $\vb*{p}=0$ outcome and increasing the likelihood of undesirable outcomes.

 We investigate the average performance of breeding by drawing samples from the homodyne measurement distribution, and calculate the effective squeezing of the heralded states. The implementation details are discussed in Sec.\ \ref{sec:simulation}. In Fig.\ \ref{fig:sampling_-15_dB}, we plot the average effective squeezing. With the introduction of loss, the average $\Delta_p$ becomes lower than $\Delta_p$ at $\vb*{p}=0$. In Fig.\ \ref{fig:sampling_-15_dB}c, the spread of the symmetric effective squeezing is shown for $\eta=0.96$ and $\eta=0.95$. $\Delta_s$ manages to surpass the fault tolerance threshold for $\eta=0.96$, but not for $\eta=0.95$, indicating that generating a fault-tolerant qunaught state is not possible when the loss is $5\%$ or more with current state-of-the-art squeezing (-15 dB \cite{vahlbruch_detection_2016}) in the seed cats. The success probability of the breeding protocol is shown in Fig.\ \ref{fig:sampling_eta}. Increasing the squeezing from $-12$ to $-15$ dB appears to benefit the probability of fault-tolerance. Increasing the number of breeding rounds results in diminishing returns in the success probability when the loss is 2\% or more.

\section{Simulation methods}\label{sec:simulation}
In this section we introduce the simulation framework used to generate the results in Sec.\ \ref{sec:ideal_protocol} and \ref{sec:non-ideal_protocol}. Using the LCoG formalism, we calculate the Wigner function of the conditional output state of the breeding circuit in Fig.\ \ref{fig:cat_breeding_circuit} by tracking only the weights, displacement vectors and the covariance matrix. The homodyne measurement outcomes can either be chosen, or they can be sampled based on their probability distribution. In order to investigate the performance of the breeding protocol on average, we sample the homodyne outcomes and calculate the effective squeezing of the ensemble. When computing the heralded output states, the number of Gaussians in the Wigner function can be reduced after each breeding iteration, allowing us to simulate several rounds of breeding in a fast framework.

\subsection{Obtaining the mixed output grid state}\label{sec:mixed_output_state}

The simulation of the entire breeding process can be divided into $N-1$ iterations of the sub-circuit in Fig.\ \ref{fig:cat_breeding_circuit}d. The input states $\hat{\rho}_\text{cat}^\eta$ (which are in general mixed) can be straightforwardly represented in the linear combination of Gaussians formalism \cite{bourassa_fast_2021}. This avoids the prohibitive memory requirements of simulating density matrices in the Fock basis. Initially, the Wigner function of a single-mode SCS can be written as a sum of Gaussians in Eq.\ \eqref{eq:Wigner_sqz_cat}. More generally,
\[
W_{\hat{\rho}}(\vb*{q})=\sum_{k\in\mathcal{N}} c_k G_{\vb*{\mu}_{k},\vb*{\sigma}}(\vb*{q}),
\]
where $\mathcal{N}$ is a set of indices. The state $\hat{\rho}$ is therefore fully characterized by the weights $c_k$, means $\vb*{\mu}_k$ and a single covariance matrix $\vb*{\sigma}$. Furthermore, its evolution throughout the entire breeding sub-circuit can be represented by transforming the individual weights, means and the covariance matrix, i.e. at no point is it required to compute the actual Wigner function, except for visualization purposes. 

We make a small modification; the exponential arguments of the weights, $w_k = \ln(c_k)$ are tracked instead of the weights themselves,
\[
W_{\hat{\rho}}(\vb*{q})=\sum_{k\in\mathcal{N}} e^{w_{k}} G_{\vb*{\mu}_{k},\vb*{\sigma}}(\vb*{q}).\label{eq:Wig_sum_of_G_log}
\]
This helps mitigate rounding errors that can occur when some of the log-weights are on the order of $w_k<\mathcal{O}(-10^{2})$. See Appendix \ref{app:numerical_stability} for a comment about the numerical stability. 

Now, we proceed with the calculation of an iteration of the breeding sub-circuit  in Fig.\ \ref{fig:cat_breeding_circuit}d, which consists of the following steps: 
\begin{enumerate}
    \item Compute the weights, means and covariance matrix in the Wigner function of the tensor product $\hat{\rho}_{\text{in}}\otimes\hat{\rho}_{\text{cat}}$.
    \item  Apply the beam splitter (Gaussian operation).
    \item Pick or sample the homodyne measurement outcome $p_i$ of the top mode. 
    \item Compute the measurement-induced weights, means and covariance matrix of output state $\hat{\rho}_{\text{out}}$ post-selected on outcome $p_i$.
    \item  Reduce the number of Gaussians in the Wigner function of the output state.
\end{enumerate}
The implementation of each step in the LCoG formalism is described in detail below. 
\subsubsection*{Tensor product of two non-Gaussian states}
The tensor product of two states in the LCoG formalism is obtained simply by multiplying the Wigner functions of each state, resulting in
\[
W_{\hat{\rho}_{\text{in}}\otimes\hat{\rho}_{\text{cat}}}(\vb*{q})=\sum_{k \in\mathcal{N}}\sum_{l \in\mathcal{M}}e^{w_{k}+w_{l}}G_{\vb*{\mu}_{k} \oplus \vb*{\mu}_{l},\vb*{\sigma}_{1} \oplus \vb*{\sigma}_{2}}(\vb*{q}) 
\]
where $\vb*{q}=(\vb*{q}_i,\vb*{q}_{i+1})^T$, $\vb*{\mu}_{k}\oplus \vb*{\mu}_l=\mqty(\vb*{\mu}_{k} \\ \vb*{\mu}_{l})$ and $\vb*{\sigma}_{1}\oplus \vb*{\sigma}_{2}=\mqty(\vb*{\sigma}_{1} & \vb*{0} \\ \vb*{0} & \vb*{\sigma}_{2})$. The log-weights are simply added. We can define a new set of indices $j\in\mathcal{J}$, where the number of Gaussians in $\mathcal{J}=(\mathcal{N},\mathcal{M)}$ is equal to $4 n_i$, where $i$ is the breeding iteration number and $n_i$ is the number of Gaussians in the input state $(n_1=4)$, because each new cat input comes with 4 Gaussians. Naively, the number of Gaussians scales as $4^N$, where $N-1$ is the number of breeding rounds.

\subsubsection*{Gaussian operation}
The beam splitter is applied on each mean by performing the matrix operation, $\vb*{\mu}_{j}\mapsto\vb*{B} \vb*{\mu}_{j}$ and a single transformation of the covariance matrix $\vb*{\sigma}\mapsto\vb*{B}\vb*{\sigma}\vb*{B}^T$, where $\vb*{B}$ is a $4\times4$ symplectic matrix. The weights remain unchanged. 

\subsubsection*{Sampling the $p$-quadrature}
We can either pick the homodyne measurement outcome $m$, or we can drawn it from its probability distribution, which is given by
\[
P(m)&=\sum_{j\in\mathcal{J}}e^{w_{j}}G_{a_{jp},A_{p}}(m)\nonumber\\
&=\frac{1}{\sqrt{2\pi A_p}}\sum_{j}e^{w_{j}-\frac{1}{2}A_{p}^{-1}\left(a_{jp}-m\right)^2} \nonumber\\
&=\sum_{j}e^{w_j + \frac{1}{2}A_{p}^{-1}\beta_{j}^2-iA_{p}^{-1}\beta_{j}(\alpha_{j}-m)}G_{\alpha_{j},A_{p}}(m)\label{eq:homodyne_probability}
\]
where $a_{jp}$ is the $p$-element of the $j$'th mean vector in the first mode, 
$\vb*{\mu}_{j}=(\vb*{\mu}_{j1}, \vb*{\mu}_{j2})^T=(a_{jx},a_{jp},b_{jx},b_{jp})^T$. And $A_p$ is the variance in the $p$-quadrature of the first mode, $\vb*{\sigma}=\begin{pmatrix}
\vb*{\sigma}_{11} & \vb*{\sigma}_{12} \\
\vb*{\sigma}_{12}^T & \vb*{\sigma}_{22}
\end{pmatrix}
$, and 
$\vb*{\sigma}_{11}=\mqty(A_{x} & A_{xp} \\ A_{xp} & A_{p})$. The means are complex, $a_{jp}=\alpha_{j}+i\beta_{j}$, with $\alpha_j,\beta_j\in\mathds{R}$ and the expression for $P(m)$ can be rearranged to extract the complex phase contribution in each Gaussian due to the imaginary part of the mean in the last line of Eq.\ \eqref{eq:homodyne_probability}. When the summation over $j$ is computed, $P(m)$ becomes real, because the terms in the sum can be grouped by pairs that are complex conjugates of each other.

In order to pick a sample $p_i$ from the probability distribution $P(m)$, we use a rejection sampling technique for distributions that are LCoGs with complex means as described in  \cite{bourassa_fast_2021}. First, we define an upper bounding function $g(m)\geq P(m)$, by taking the absolute value of the coefficients, 
\[
g(m) = \sum_{j} e^{w_j+\frac{1}{2}A_{p}^{-1}\beta_{j}^2}G_{\alpha_{j},A_p}(m)
\]
This enables us to first sample from a discrete distribution of peaks with probability equal to the coefficients, $P(j)=e^{w_j+\frac{1}{2}A_{p}^{-1}\beta_{j}^2}/\mathcal{M}$, where $\mathcal{M}=\sum_je^{w_j+\frac{1}{2}A_{p}^{-1}\beta_{j}^2}$. Once we have picked a peak $j$, we sample $p_i$ from a continuous Gaussian distribution $p_i\in\mathcal{N}(\alpha_j,A_p)$.  The sample is accepted if $P(p_i) \geq y$, where $y$ is picked uniformly between the interval $y\in[0,g(p_i)]$. If the sample is rejected, the entire process is restarted. At minimum, three sums of exponential functions must be computed to approve a sample; $\mathcal{M}$, $P(p_i)$, and $g(p_i)$, which is performed numerically using the \texttt{logsumexp} function from SciPy \cite{virtanen_scipy_2020}.

When breeding squeezed (lossy) cat states using phase-less beam splitters, $\alpha_j=0$ for all $j$, and the upper-bounding function can be sampled from the normal distribution $p_i\in\mathcal{N}(0,A_p)$, followed by the rejection/acceptance procedure as before. Once a candidate sample is accepted, we post-select the intermediate output state on the $p_i$

\subsubsection*{Compute the post-selected state}
The output state of a given homodyne measurement outcome $p_i$ can now be computed using the transformation rules of partial Gaussian measurements \cite{bourassa_fast_2021} applied on each of the Gaussians in the Wigner function in Eq.\ \eqref{eq:Wig_sum_of_G_log}, yielding
\[
W_{\hat{\rho}_{\text{out}}}(\vb*{q}_{i+1})=\sum_{j\in\mathcal{J}}e^{\tilde{w}_j}G_{\tilde{\vb*{\mu}}_{j},\tilde{\vb*{\sigma}}}(\vb*{q}_{i+1}),
\]
where
\[
&\tilde{\vb*{\sigma}}=\vb*{\sigma}_{22}-A_{p}^{-1}\vb*{\sigma}_{12}^{T}\vb*{\Pi}\vb*{\sigma}_{12},\\
&\tilde{\vb*{\mu}}_j = \vb*{\mu}_{j2}+A_{p}^{-1}\vb*{\sigma}_{12}^{T}\vb*{\Pi}(\vb*{m}-\vb*{\mu}_{j1}),\\
&\tilde{w}_j=w_{j}-\frac{1}{2}A_{p}^{-1}\left(a_{jp}-p_i\right)^2
\]
where $\vb*{m}=(0,p_i)^T$ and $\vb*{\Pi}=\mqty(0&0 \\ 0&1)$. When each cat has equal squeezing and the loss is uniform, $\vb*{\sigma}_{12}=0_{2,2}$, and there is no overall measurement-induced displacement. In this case, the effect of the measurement appears solely on the weights $\tilde{w}_j$ of the Gaussian components. 

\subsubsection*{Reducing the number of Gaussians in the sum}
After each iteration of the breeding sub-circuit, the number of Gaussians in the output state grows by a factor of 4, leading to a total number of $4^N$ Gaussians at the end of the protocol. This quickly becomes intractable to keep track of in memory. Similar to the ideal protocol, Gaussians with equal means can be grouped together, and the Wigner function of the output state can be re-expressed using only $(i+2)^2$ Gaussians, where $i$ is the breeding step number, resulting in $(N+1)^2$ Gaussians at the end of the breeding protocol,
\[
W_{\hat{\rho}_{\text{out}}}(\vb*{q}_{i+1})=\sum_{j'\in\mathcal{J}'} G_{\tilde{\vb*{\mu}}_{j'},\vb*{\sigma}}(\vb*{q}_{i+1}) \sum_{k\in\mathcal{K}_{j'}} e^{\tilde{w}_{k}},
\]
where $\mathcal{J}'$ is a set numbering the unique means $\tilde{\vb*{\mu}}_{j'}$, while $\mathcal{K}_{j'}$ is a set numbering the weights belonging to each unique mean, labeled by $j'$. The reduction procedure involves finding the unique means, and calculating a sum over exponential functions for each $j$'. This enables us to simulate the breeding of a very large number of cats that are mixed because we significantly delay the growth in the number of Gaussians. Note, that this strategy is tailored for the breeding of identical cats, i.e. cats with equal means. Hence, if we breed cats with non-uniform loss, a full reduction is no longer applicable, but a speedup could still be provided in certain cases.

\subsubsection*{Computing the effective squeezing of the output state}

After iterating the breeding sub-circuit the required number of times, either via sampling, or by post-selecting on a measurement outcome pattern, $\vb*{p}$, we arrive at compact expression for the Wigner function of the output state, in the form of a LCoG. Now, we would like to evaluate its effective squeezing via Eq.\ \eqref{eq:eff_sqz_x}, \eqref{eq:eff_sqz_p}, which is related to the characteristic function of the state $\chi(\vb*{\alpha})=\mathrm{Tr}[\hat{D}(\vb*{\alpha})\hat{\rho}]$.  

Similar to the Wigner function, the characteristic function of a Gaussian state is fully characterized by its covariance matrix $\vb*{\sigma}$, displacement vector $\vb*{\mu}$ and the symplectic form $\vb*{\Omega}$ \cite{weedbrook_gaussian_2012, serafini_quantum_2017},
\[
\chi_{\hat{\rho}}(\vb*{\alpha}) = \exp\left[-\frac{1}{2}\vb*{\alpha}^T\vb*{\Omega}\vb*{\sigma}\vb*{\Omega}^T\vb*{\alpha} +i\vb*{\mu}^T\vb*{\Omega \alpha}\right],
\]
where
$\vb*{\alpha} =(\alpha_x,\alpha_p)^T$ and $\alpha = \alpha_x + i\alpha_p$. 
If the Wigner function is a linear combination of Gaussians, the characteristic function will also be a linear combination of Gaussians (here with identical $\vb*{\sigma}$),
\[
\chi_{\hat{\rho}_\text{grid}}(\vb*{\alpha}) = \sum_{j\in\mathcal{J}'}e^{w_{j}}\exp[-\frac{1}{2}\vb*{\alpha}^T \vb*{\Omega}\vb*{\sigma}\vb*{\Omega}^T\vb*{\alpha}+i\vb*{\mu}_{j}^T\vb*{\Omega}\vb*{\alpha}]
\]
This is a sum over exponential functions, which is handled by \texttt{logsumexp} from SciPy \cite{virtanen_scipy_2020}.

\subsection{Simulation details}

The results shown in Fig.\ \ref{fig:sampling_-15_dB} and \ref{fig:sampling_eta} were generated by averaging over 10,000 homodyne measurement samples for $N\in[2,8]$, and 5,000 samples for $N\in[9,13]$ collected on a MacBook Pro. To reduce the error bars for $\eta=0.96$ and $\eta=0.95$, 50,000 samples were collected for $N\in[8,13]$ using the Niflheim supercomputing cluster at DTU Physics.

\section{Summary and Outlook}\label{sec:outlook}

In this work, we have developed numerical tools using the LCoG framework \cite{bourassa_fast_2021}, which enables the accurate and fast simulation of grid state preparation via cat breeding. By dividing the breeding circuit into sub-circuits, it is possible to reduce the number of Gaussians at each breeding step, allowing the simulation of many rounds of breeding with a minimal number of Gaussians. By representing the Wigner function, it became possible to investigate the effect of optical loss and non-zero homodyne measurement outcomes on the performance of the cat breeding protocol. We found that the presence of 5\% loss or more on the seed states inhibits the preparation of a fault-tolerant GKP state. The presence of loss also decreases the success probability of the grid state generation, which in light of the low SCS generation probability from GBS devices, is a discouraging result. However, multiplexing the GBS devices and selecting the best candidates as suggested by \cite{takase_generation_2024,aghaee_rad_scaling_2025} will help boost the success rate. While in our analysis, we only consider the loss on one optical path, i.e. the static beam splitters are assumed to be lossless, and the homodyne detectors have unit efficiency, Aghaee et al. \cite{aghaee_rad_scaling_2025} consider multiple loss paths and find loss thresholds that are even more restrictive, e.g. around 1\%. The engineering such a pure system is a enormous challenge. 

The methodology described in this work can be used to apply arbitrary Gaussian CPTP maps on the input states, and study their effect on the output state of the breeding circuit. Additionally, the breeding of any state that can be written as a LCoG can be simulated. For example, arbitrary superpositions of Fock states can be written as a superposition of coherent states \cite{marshall_simulation_2023}, which fit neatly within the LCoG formalism. As suggested by \cite{aghaee_rad_scaling_2025}, three-peak approximate GKP states from three-mode GBS devices can also be used in combination with cat states as seed states. The addition of approximate GKP states using Eq.\ \eqref{eq:GKP_approx_Delta_x} as seed states to the breeding protocol is a straightforward application of our methodology. In \cite{zheng_gaussian_2023}, the breeding of binomial states (an approximate three-peak GKP state) for the generation of a qunaught state was studied by alternating between $x$ and $p$ measurements at each step of the protocol. In their analysis, the authors also discovered a variation in the quality of the output state that was dependent on the homodyne measurement outcome. When studying cat breeding, we found that alternating between $x$ and $p$ measurements during the breeding of ideal cats produced a grid state with higher effective squeezing. However, as soon as loss was added, the scheme immediately became inferior to measuring $p$. When generalizing the input states, the method of reducing the number of Gaussians described here may no longer have good performance, since it is tailored to states of the form Eq.\ \eqref{eq:GKP_approx_Delta_x}. Therefore, other reduction methods may have to be considered, e.g. using the stellar rank \cite{chabaud_stellar_2020} developed by us in \cite{solodovnikova_fast_2025}. In future work, the connection between the catability \cite{brauer_catability_2025}, which is a new figure of merit for the "catness" of a state, and the suitability of approximate cat states for breeding grid states could be studied.

\section{Code and Data availability}
The code is written in Python and available at \url{https://github.com/qpit/breeding}. The results of the homodyne measurement sampling can also be found in the repository. 

\section{Acknowledgments}
OS thanks Eli Bourassa, Rafael  Alexander and Timo Hillmann for valuable discussions about cat breeding and the log-weights method in the linear combinations of Gaussians framework. OS also thanks Niklas Budinger for reading and providing valuable feedback to this manuscript.

The authors acknowledge support from the Danish National Research Foundation (bigQ, DNRF0142), the European Union's Horizon Europe CLUSTEC project (no.\ 101080173), the European Research Council ClusterQ project (no.\ 101055224), and the Novo Nordisk Foundation Data Science Research Infrastructure 2022 grant (no.\ NNF22OC0078009).

The authors would also like to acknowledge the use of several open source libraries \cite{harris_array_2020,virtanen_scipy_2020,hunter_matplotlib_2007,kluyver_jupyter_2016} .

\bibliography{references} 

\begin{thebibliography}{61}%
\makeatletter
\providecommand \@ifxundefined [1]{%
 \@ifx{#1\undefined}
}%
\providecommand \@ifnum [1]{%
 \ifnum #1\expandafter \@firstoftwo
 \else \expandafter \@secondoftwo
 \fi
}%
\providecommand \@ifx [1]{%
 \ifx #1\expandafter \@firstoftwo
 \else \expandafter \@secondoftwo
 \fi
}%
\providecommand \natexlab [1]{#1}%
\providecommand \enquote  [1]{``#1''}%
\providecommand \bibnamefont  [1]{#1}%
\providecommand \bibfnamefont [1]{#1}%
\providecommand \citenamefont [1]{#1}%
\providecommand \href@noop [0]{\@secondoftwo}%
\providecommand \href [0]{\begingroup \@sanitize@url \@href}%
\providecommand \@href[1]{\@@startlink{#1}\@@href}%
\providecommand \@@href[1]{\endgroup#1\@@endlink}%
\providecommand \@sanitize@url [0]{\catcode `\\12\catcode `\$12\catcode `\&12\catcode `\#12\catcode `\^12\catcode `\_12\catcode `\%12\relax}%
\providecommand \@@startlink[1]{}%
\providecommand \@@endlink[0]{}%
\providecommand \url  [0]{\begingroup\@sanitize@url \@url }%
\providecommand \@url [1]{\endgroup\@href {#1}{\urlprefix }}%
\providecommand \urlprefix  [0]{URL }%
\providecommand \Eprint [0]{\href }%
\providecommand \doibase [0]{https://doi.org/}%
\providecommand \selectlanguage [0]{\@gobble}%
\providecommand \bibinfo  [0]{\@secondoftwo}%
\providecommand \bibfield  [0]{\@secondoftwo}%
\providecommand \translation [1]{[#1]}%
\providecommand \BibitemOpen [0]{}%
\providecommand \bibitemStop [0]{}%
\providecommand \bibitemNoStop [0]{.\EOS\space}%
\providecommand \EOS [0]{\spacefactor3000\relax}%
\providecommand \BibitemShut  [1]{\csname bibitem#1\endcsname}%
\let\auto@bib@innerbib\@empty
\bibitem [{\citenamefont {Gottesman}\ \emph {et~al.}(2001)\citenamefont {Gottesman}, \citenamefont {Kitaev},\ and\ \citenamefont {Preskill}}]{gottesman_encoding_2001}%
  \BibitemOpen
  \bibfield  {author} {\bibinfo {author} {\bibfnamefont {D.}~\bibnamefont {Gottesman}}, \bibinfo {author} {\bibfnamefont {A.}~\bibnamefont {Kitaev}},\ and\ \bibinfo {author} {\bibfnamefont {J.}~\bibnamefont {Preskill}},\ }\bibfield  {title} {\bibinfo {title} {Encoding a qubit in an oscillator},\ }\href {https://doi.org/10.1103/PhysRevA.64.012310} {\bibfield  {journal} {\bibinfo  {journal} {Physical Review A}\ }\textbf {\bibinfo {volume} {64}},\ \bibinfo {pages} {012310} (\bibinfo {year} {2001})}\BibitemShut {NoStop}%
\bibitem [{\citenamefont {Menicucci}\ \emph {et~al.}(2006)\citenamefont {Menicucci}, \citenamefont {{van Loock}}, \citenamefont {Gu}, \citenamefont {Weedbrook}, \citenamefont {Ralph},\ and\ \citenamefont {Nielsen}}]{menicucci_universal_2006}%
  \BibitemOpen
  \bibfield  {author} {\bibinfo {author} {\bibfnamefont {N.~C.}\ \bibnamefont {Menicucci}}, \bibinfo {author} {\bibfnamefont {P.}~\bibnamefont {{van Loock}}}, \bibinfo {author} {\bibfnamefont {M.}~\bibnamefont {Gu}}, \bibinfo {author} {\bibfnamefont {C.}~\bibnamefont {Weedbrook}}, \bibinfo {author} {\bibfnamefont {T.~C.}\ \bibnamefont {Ralph}},\ and\ \bibinfo {author} {\bibfnamefont {M.~A.}\ \bibnamefont {Nielsen}},\ }\bibfield  {title} {\bibinfo {title} {Universal {{Quantum Computation}} with {{Continuous-Variable Cluster States}}},\ }\href {https://doi.org/10.1103/PhysRevLett.97.110501} {\bibfield  {journal} {\bibinfo  {journal} {Physical Review Letters}\ }\textbf {\bibinfo {volume} {97}},\ \bibinfo {pages} {110501} (\bibinfo {year} {2006})}\BibitemShut {NoStop}%
\bibitem [{\citenamefont {Bourassa}\ \emph {et~al.}(2021{\natexlab{a}})\citenamefont {Bourassa}, \citenamefont {Alexander}, \citenamefont {Vasmer}, \citenamefont {Patil}, \citenamefont {Tzitrin}, \citenamefont {Matsuura}, \citenamefont {Su}, \citenamefont {Baragiola}, \citenamefont {Guha}, \citenamefont {Dauphinais}, \citenamefont {Sabapathy}, \citenamefont {Menicucci},\ and\ \citenamefont {Dhand}}]{bourassa_blueprint_2021}%
  \BibitemOpen
  \bibfield  {author} {\bibinfo {author} {\bibfnamefont {J.~E.}\ \bibnamefont {Bourassa}}, \bibinfo {author} {\bibfnamefont {R.~N.}\ \bibnamefont {Alexander}}, \bibinfo {author} {\bibfnamefont {M.}~\bibnamefont {Vasmer}}, \bibinfo {author} {\bibfnamefont {A.}~\bibnamefont {Patil}}, \bibinfo {author} {\bibfnamefont {I.}~\bibnamefont {Tzitrin}}, \bibinfo {author} {\bibfnamefont {T.}~\bibnamefont {Matsuura}}, \bibinfo {author} {\bibfnamefont {D.}~\bibnamefont {Su}}, \bibinfo {author} {\bibfnamefont {B.~Q.}\ \bibnamefont {Baragiola}}, \bibinfo {author} {\bibfnamefont {S.}~\bibnamefont {Guha}}, \bibinfo {author} {\bibfnamefont {G.}~\bibnamefont {Dauphinais}}, \bibinfo {author} {\bibfnamefont {K.~K.}\ \bibnamefont {Sabapathy}}, \bibinfo {author} {\bibfnamefont {N.~C.}\ \bibnamefont {Menicucci}},\ and\ \bibinfo {author} {\bibfnamefont {I.}~\bibnamefont {Dhand}},\ }\bibfield  {title} {\bibinfo {title} {Blueprint for a {{Scalable Photonic Fault-Tolerant Quantum Computer}}},\ }\href
  {https://doi.org/10.22331/q-2021-02-04-392} {\bibfield  {journal} {\bibinfo  {journal} {Quantum}\ }\textbf {\bibinfo {volume} {5}},\ \bibinfo {pages} {392} (\bibinfo {year} {2021}{\natexlab{a}})}\BibitemShut {NoStop}%
\bibitem [{\citenamefont {Fowler}\ \emph {et~al.}(2012)\citenamefont {Fowler}, \citenamefont {Mariantoni}, \citenamefont {Martinis},\ and\ \citenamefont {Cleland}}]{fowler_surface_2012}%
  \BibitemOpen
  \bibfield  {author} {\bibinfo {author} {\bibfnamefont {A.~G.}\ \bibnamefont {Fowler}}, \bibinfo {author} {\bibfnamefont {M.}~\bibnamefont {Mariantoni}}, \bibinfo {author} {\bibfnamefont {J.~M.}\ \bibnamefont {Martinis}},\ and\ \bibinfo {author} {\bibfnamefont {A.~N.}\ \bibnamefont {Cleland}},\ }\bibfield  {title} {\bibinfo {title} {Surface codes: {{Towards}} practical large-scale quantum computation},\ }\href {https://doi.org/10.1103/PhysRevA.86.032324} {\bibfield  {journal} {\bibinfo  {journal} {Physical Review A}\ }\textbf {\bibinfo {volume} {86}},\ \bibinfo {pages} {032324} (\bibinfo {year} {2012})}\BibitemShut {NoStop}%
\bibitem [{\citenamefont {Menicucci}(2014)}]{menicucci_fault-tolerant_2014}%
  \BibitemOpen
  \bibfield  {author} {\bibinfo {author} {\bibfnamefont {N.~C.}\ \bibnamefont {Menicucci}},\ }\bibfield  {title} {\bibinfo {title} {Fault-{{Tolerant Measurement-Based Quantum Computing}} with {{Continuous-Variable Cluster States}}},\ }\href {https://doi.org/10.1103/PhysRevLett.112.120504} {\bibfield  {journal} {\bibinfo  {journal} {Physical Review Letters}\ }\textbf {\bibinfo {volume} {112}},\ \bibinfo {pages} {120504} (\bibinfo {year} {2014})}\BibitemShut {NoStop}%
\bibitem [{\citenamefont {Baragiola}\ \emph {et~al.}(2019)\citenamefont {Baragiola}, \citenamefont {Pantaleoni}, \citenamefont {Alexander}, \citenamefont {Karanjai},\ and\ \citenamefont {Menicucci}}]{baragiola_all-gaussian_2019}%
  \BibitemOpen
  \bibfield  {author} {\bibinfo {author} {\bibfnamefont {B.~Q.}\ \bibnamefont {Baragiola}}, \bibinfo {author} {\bibfnamefont {G.}~\bibnamefont {Pantaleoni}}, \bibinfo {author} {\bibfnamefont {R.~N.}\ \bibnamefont {Alexander}}, \bibinfo {author} {\bibfnamefont {A.}~\bibnamefont {Karanjai}},\ and\ \bibinfo {author} {\bibfnamefont {N.~C.}\ \bibnamefont {Menicucci}},\ }\bibfield  {title} {\bibinfo {title} {All-{{Gaussian}} universality and fault tolerance with the {{Gottesman-Kitaev-Preskill}} code},\ }\href {https://doi.org/10.1103/PhysRevLett.123.200502} {\bibfield  {journal} {\bibinfo  {journal} {Physical Review Letters}\ }\textbf {\bibinfo {volume} {123}},\ \bibinfo {pages} {200502} (\bibinfo {year} {2019})}\BibitemShut {NoStop}%
\bibitem [{\citenamefont {{Campagne-Ibarcq}}\ \emph {et~al.}(2020)\citenamefont {{Campagne-Ibarcq}}, \citenamefont {Eickbusch}, \citenamefont {Touzard}, \citenamefont {{Zalys-Geller}}, \citenamefont {Frattini}, \citenamefont {Sivak}, \citenamefont {Reinhold}, \citenamefont {Puri}, \citenamefont {Shankar}, \citenamefont {Schoelkopf}, \citenamefont {Frunzio}, \citenamefont {Mirrahimi},\ and\ \citenamefont {Devoret}}]{campagne-ibarcq_quantum_2020}%
  \BibitemOpen
  \bibfield  {author} {\bibinfo {author} {\bibfnamefont {P.}~\bibnamefont {{Campagne-Ibarcq}}}, \bibinfo {author} {\bibfnamefont {A.}~\bibnamefont {Eickbusch}}, \bibinfo {author} {\bibfnamefont {S.}~\bibnamefont {Touzard}}, \bibinfo {author} {\bibfnamefont {E.}~\bibnamefont {{Zalys-Geller}}}, \bibinfo {author} {\bibfnamefont {N.~E.}\ \bibnamefont {Frattini}}, \bibinfo {author} {\bibfnamefont {V.~V.}\ \bibnamefont {Sivak}}, \bibinfo {author} {\bibfnamefont {P.}~\bibnamefont {Reinhold}}, \bibinfo {author} {\bibfnamefont {S.}~\bibnamefont {Puri}}, \bibinfo {author} {\bibfnamefont {S.}~\bibnamefont {Shankar}}, \bibinfo {author} {\bibfnamefont {R.~J.}\ \bibnamefont {Schoelkopf}}, \bibinfo {author} {\bibfnamefont {L.}~\bibnamefont {Frunzio}}, \bibinfo {author} {\bibfnamefont {M.}~\bibnamefont {Mirrahimi}},\ and\ \bibinfo {author} {\bibfnamefont {M.~H.}\ \bibnamefont {Devoret}},\ }\bibfield  {title} {\bibinfo {title} {Quantum error correction of a qubit encoded in grid states of an oscillator},\ }\href
  {https://doi.org/10.1038/s41586-020-2603-3} {\bibfield  {journal} {\bibinfo  {journal} {Nature}\ }\textbf {\bibinfo {volume} {584}},\ \bibinfo {pages} {368} (\bibinfo {year} {2020})}\BibitemShut {NoStop}%
\bibitem [{\citenamefont {Sivak}\ \emph {et~al.}(2023)\citenamefont {Sivak}, \citenamefont {Eickbusch}, \citenamefont {Royer}, \citenamefont {Singh}, \citenamefont {Tsioutsios}, \citenamefont {Ganjam}, \citenamefont {Miano}, \citenamefont {Brock}, \citenamefont {Ding}, \citenamefont {Frunzio}, \citenamefont {Girvin}, \citenamefont {Schoelkopf},\ and\ \citenamefont {Devoret}}]{sivak_real-time_2023}%
  \BibitemOpen
  \bibfield  {author} {\bibinfo {author} {\bibfnamefont {V.~V.}\ \bibnamefont {Sivak}}, \bibinfo {author} {\bibfnamefont {A.}~\bibnamefont {Eickbusch}}, \bibinfo {author} {\bibfnamefont {B.}~\bibnamefont {Royer}}, \bibinfo {author} {\bibfnamefont {S.}~\bibnamefont {Singh}}, \bibinfo {author} {\bibfnamefont {I.}~\bibnamefont {Tsioutsios}}, \bibinfo {author} {\bibfnamefont {S.}~\bibnamefont {Ganjam}}, \bibinfo {author} {\bibfnamefont {A.}~\bibnamefont {Miano}}, \bibinfo {author} {\bibfnamefont {B.~L.}\ \bibnamefont {Brock}}, \bibinfo {author} {\bibfnamefont {A.~Z.}\ \bibnamefont {Ding}}, \bibinfo {author} {\bibfnamefont {L.}~\bibnamefont {Frunzio}}, \bibinfo {author} {\bibfnamefont {S.~M.}\ \bibnamefont {Girvin}}, \bibinfo {author} {\bibfnamefont {R.~J.}\ \bibnamefont {Schoelkopf}},\ and\ \bibinfo {author} {\bibfnamefont {M.~H.}\ \bibnamefont {Devoret}},\ }\bibfield  {title} {\bibinfo {title} {Real-time quantum error correction beyond break-even},\ }\href {https://doi.org/10.1038/s41586-023-05782-6} {\bibfield
  {journal} {\bibinfo  {journal} {Nature}\ }\textbf {\bibinfo {volume} {616}},\ \bibinfo {pages} {50} (\bibinfo {year} {2023})}\BibitemShut {NoStop}%
\bibitem [{\citenamefont {{Lachance-Quirion}}\ \emph {et~al.}(2024)\citenamefont {{Lachance-Quirion}}, \citenamefont {Lemonde}, \citenamefont {Simoneau}, \citenamefont {{St-Jean}}, \citenamefont {Lemieux}, \citenamefont {Turcotte}, \citenamefont {Wright}, \citenamefont {Lacroix}, \citenamefont {{Fr{\'e}chette-Viens}}, \citenamefont {Shillito}, \citenamefont {Hopfmueller}, \citenamefont {Tremblay}, \citenamefont {Frattini}, \citenamefont {Camirand~Lemyre},\ and\ \citenamefont {{St-Jean}}}]{lachance-quirion_autonomous_2024}%
  \BibitemOpen
  \bibfield  {author} {\bibinfo {author} {\bibfnamefont {D.}~\bibnamefont {{Lachance-Quirion}}}, \bibinfo {author} {\bibfnamefont {M.-A.}\ \bibnamefont {Lemonde}}, \bibinfo {author} {\bibfnamefont {J.~O.}\ \bibnamefont {Simoneau}}, \bibinfo {author} {\bibfnamefont {L.}~\bibnamefont {{St-Jean}}}, \bibinfo {author} {\bibfnamefont {P.}~\bibnamefont {Lemieux}}, \bibinfo {author} {\bibfnamefont {S.}~\bibnamefont {Turcotte}}, \bibinfo {author} {\bibfnamefont {W.}~\bibnamefont {Wright}}, \bibinfo {author} {\bibfnamefont {A.}~\bibnamefont {Lacroix}}, \bibinfo {author} {\bibfnamefont {J.}~\bibnamefont {{Fr{\'e}chette-Viens}}}, \bibinfo {author} {\bibfnamefont {R.}~\bibnamefont {Shillito}}, \bibinfo {author} {\bibfnamefont {F.}~\bibnamefont {Hopfmueller}}, \bibinfo {author} {\bibfnamefont {M.}~\bibnamefont {Tremblay}}, \bibinfo {author} {\bibfnamefont {N.~E.}\ \bibnamefont {Frattini}}, \bibinfo {author} {\bibfnamefont {J.}~\bibnamefont {Camirand~Lemyre}},\ and\ \bibinfo {author} {\bibfnamefont {P.}~\bibnamefont
  {{St-Jean}}},\ }\bibfield  {title} {\bibinfo {title} {Autonomous {{Quantum Error Correction}} of {{Gottesman-Kitaev-Preskill States}}},\ }\href {https://doi.org/10.1103/PhysRevLett.132.150607} {\bibfield  {journal} {\bibinfo  {journal} {Physical Review Letters}\ }\textbf {\bibinfo {volume} {132}},\ \bibinfo {pages} {150607} (\bibinfo {year} {2024})}\BibitemShut {NoStop}%
\bibitem [{\citenamefont {Fl{\"u}hmann}\ \emph {et~al.}(2019)\citenamefont {Fl{\"u}hmann}, \citenamefont {Nguyen}, \citenamefont {Marinelli}, \citenamefont {Negnevitsky}, \citenamefont {Mehta},\ and\ \citenamefont {Home}}]{fluhmann_encoding_2019}%
  \BibitemOpen
  \bibfield  {author} {\bibinfo {author} {\bibfnamefont {C.}~\bibnamefont {Fl{\"u}hmann}}, \bibinfo {author} {\bibfnamefont {T.~L.}\ \bibnamefont {Nguyen}}, \bibinfo {author} {\bibfnamefont {M.}~\bibnamefont {Marinelli}}, \bibinfo {author} {\bibfnamefont {V.}~\bibnamefont {Negnevitsky}}, \bibinfo {author} {\bibfnamefont {K.}~\bibnamefont {Mehta}},\ and\ \bibinfo {author} {\bibfnamefont {J.~P.}\ \bibnamefont {Home}},\ }\bibfield  {title} {\bibinfo {title} {Encoding a qubit in a trapped-ion mechanical oscillator},\ }\href {https://doi.org/10.1038/s41586-019-0960-6} {\bibfield  {journal} {\bibinfo  {journal} {Nature}\ }\textbf {\bibinfo {volume} {566}},\ \bibinfo {pages} {513} (\bibinfo {year} {2019})}\BibitemShut {NoStop}%
\bibitem [{\citenamefont {De~Neeve}\ \emph {et~al.}(2022)\citenamefont {De~Neeve}, \citenamefont {Nguyen}, \citenamefont {Behrle},\ and\ \citenamefont {Home}}]{de_neeve_error_2022}%
  \BibitemOpen
  \bibfield  {author} {\bibinfo {author} {\bibfnamefont {B.}~\bibnamefont {De~Neeve}}, \bibinfo {author} {\bibfnamefont {T.-L.}\ \bibnamefont {Nguyen}}, \bibinfo {author} {\bibfnamefont {T.}~\bibnamefont {Behrle}},\ and\ \bibinfo {author} {\bibfnamefont {J.~P.}\ \bibnamefont {Home}},\ }\bibfield  {title} {\bibinfo {title} {Error correction of a logical grid state qubit by dissipative pumping},\ }\href {https://doi.org/10.1038/s41567-021-01487-7} {\bibfield  {journal} {\bibinfo  {journal} {Nature Physics}\ }\textbf {\bibinfo {volume} {18}},\ \bibinfo {pages} {296} (\bibinfo {year} {2022})}\BibitemShut {NoStop}%
\bibitem [{\citenamefont {Larsen}\ \emph {et~al.}(2025)\citenamefont {Larsen}, \citenamefont {Bourassa}, \citenamefont {Kocsis}, \citenamefont {Tasker}, \citenamefont {Chadwick}, \citenamefont {{Gonz{\'a}lez-Arciniegas}}, \citenamefont {Hastrup}, \citenamefont {{Lopetegui-Gonz{\'a}lez}}, \citenamefont {Miatto}, \citenamefont {Motamedi}, \citenamefont {Noro}, \citenamefont {Roeland}, \citenamefont {Baby}, \citenamefont {Chen}, \citenamefont {Contu}, \citenamefont {Di~Luch}, \citenamefont {Drago}, \citenamefont {Giesbrecht}, \citenamefont {Grainge}, \citenamefont {Krasnokutska}, \citenamefont {Menotti}, \citenamefont {Morrison}, \citenamefont {Puviraj}, \citenamefont {Rezaei~Shad}, \citenamefont {Hussain}, \citenamefont {McMahon}, \citenamefont {Ortmann}, \citenamefont {Collins}, \citenamefont {Ma}, \citenamefont {Phillips}, \citenamefont {Seymour}, \citenamefont {Tang}, \citenamefont {Yang}, \citenamefont {Vernon}, \citenamefont {Alexander},\ and\ \citenamefont {Mahler}}]{larsen_integrated_2025}%
  \BibitemOpen
  \bibfield  {author} {\bibinfo {author} {\bibfnamefont {M.~V.}\ \bibnamefont {Larsen}}, \bibinfo {author} {\bibfnamefont {J.~E.}\ \bibnamefont {Bourassa}}, \bibinfo {author} {\bibfnamefont {S.}~\bibnamefont {Kocsis}}, \bibinfo {author} {\bibfnamefont {J.~F.}\ \bibnamefont {Tasker}}, \bibinfo {author} {\bibfnamefont {R.~S.}\ \bibnamefont {Chadwick}}, \bibinfo {author} {\bibfnamefont {C.}~\bibnamefont {{Gonz{\'a}lez-Arciniegas}}}, \bibinfo {author} {\bibfnamefont {J.}~\bibnamefont {Hastrup}}, \bibinfo {author} {\bibfnamefont {C.~E.}\ \bibnamefont {{Lopetegui-Gonz{\'a}lez}}}, \bibinfo {author} {\bibfnamefont {F.~M.}\ \bibnamefont {Miatto}}, \bibinfo {author} {\bibfnamefont {A.}~\bibnamefont {Motamedi}}, \bibinfo {author} {\bibfnamefont {R.}~\bibnamefont {Noro}}, \bibinfo {author} {\bibfnamefont {G.}~\bibnamefont {Roeland}}, \bibinfo {author} {\bibfnamefont {R.}~\bibnamefont {Baby}}, \bibinfo {author} {\bibfnamefont {H.}~\bibnamefont {Chen}}, \bibinfo {author} {\bibfnamefont {P.}~\bibnamefont {Contu}}, \bibinfo
  {author} {\bibfnamefont {I.}~\bibnamefont {Di~Luch}}, \bibinfo {author} {\bibfnamefont {C.}~\bibnamefont {Drago}}, \bibinfo {author} {\bibfnamefont {M.}~\bibnamefont {Giesbrecht}}, \bibinfo {author} {\bibfnamefont {T.}~\bibnamefont {Grainge}}, \bibinfo {author} {\bibfnamefont {I.}~\bibnamefont {Krasnokutska}}, \bibinfo {author} {\bibfnamefont {M.}~\bibnamefont {Menotti}}, \bibinfo {author} {\bibfnamefont {B.}~\bibnamefont {Morrison}}, \bibinfo {author} {\bibfnamefont {C.}~\bibnamefont {Puviraj}}, \bibinfo {author} {\bibfnamefont {K.}~\bibnamefont {Rezaei~Shad}}, \bibinfo {author} {\bibfnamefont {B.}~\bibnamefont {Hussain}}, \bibinfo {author} {\bibfnamefont {J.}~\bibnamefont {McMahon}}, \bibinfo {author} {\bibfnamefont {J.~E.}\ \bibnamefont {Ortmann}}, \bibinfo {author} {\bibfnamefont {M.~J.}\ \bibnamefont {Collins}}, \bibinfo {author} {\bibfnamefont {C.}~\bibnamefont {Ma}}, \bibinfo {author} {\bibfnamefont {D.~S.}\ \bibnamefont {Phillips}}, \bibinfo {author} {\bibfnamefont {M.}~\bibnamefont {Seymour}},
  \bibinfo {author} {\bibfnamefont {Q.~Y.}\ \bibnamefont {Tang}}, \bibinfo {author} {\bibfnamefont {B.}~\bibnamefont {Yang}}, \bibinfo {author} {\bibfnamefont {Z.}~\bibnamefont {Vernon}}, \bibinfo {author} {\bibfnamefont {R.~N.}\ \bibnamefont {Alexander}},\ and\ \bibinfo {author} {\bibfnamefont {D.~H.}\ \bibnamefont {Mahler}},\ }\bibfield  {title} {\bibinfo {title} {Integrated photonic source of {{Gottesman}}--{{Kitaev}}--{{Preskill}} qubits},\ }\href {https://doi.org/10.1038/s41586-025-09044-5} {\bibfield  {journal} {\bibinfo  {journal} {Nature}\ }\textbf {\bibinfo {volume} {642}},\ \bibinfo {pages} {587} (\bibinfo {year} {2025})}\BibitemShut {NoStop}%
\bibitem [{\citenamefont {Larsen}\ \emph {et~al.}(2021)\citenamefont {Larsen}, \citenamefont {Chamberland}, \citenamefont {Noh}, \citenamefont {{Neergaard-Nielsen}},\ and\ \citenamefont {Andersen}}]{larsen_fault-tolerant_2021}%
  \BibitemOpen
  \bibfield  {author} {\bibinfo {author} {\bibfnamefont {M.~V.}\ \bibnamefont {Larsen}}, \bibinfo {author} {\bibfnamefont {C.}~\bibnamefont {Chamberland}}, \bibinfo {author} {\bibfnamefont {K.}~\bibnamefont {Noh}}, \bibinfo {author} {\bibfnamefont {J.~S.}\ \bibnamefont {{Neergaard-Nielsen}}},\ and\ \bibinfo {author} {\bibfnamefont {U.~L.}\ \bibnamefont {Andersen}},\ }\bibfield  {title} {\bibinfo {title} {Fault-{{Tolerant Continuous-Variable Measurement-based Quantum Computation Architecture}}},\ }\href {https://doi.org/10.1103/PRXQuantum.2.030325} {\bibfield  {journal} {\bibinfo  {journal} {PRX Quantum}\ }\textbf {\bibinfo {volume} {2}},\ \bibinfo {pages} {030325} (\bibinfo {year} {2021})}\BibitemShut {NoStop}%
\bibitem [{\citenamefont {{\O}stergaard}\ \emph {et~al.}(2025)\citenamefont {{\O}stergaard}, \citenamefont {Budinger}, \citenamefont {Larsen}, \citenamefont {van Loock}, \citenamefont {{Neergaard-Nielsen}},\ and\ \citenamefont {Andersen}}]{ostergaard_octo-rail_2025}%
  \BibitemOpen
  \bibfield  {author} {\bibinfo {author} {\bibfnamefont {E.~E.~B.}\ \bibnamefont {{\O}stergaard}}, \bibinfo {author} {\bibfnamefont {N.}~\bibnamefont {Budinger}}, \bibinfo {author} {\bibfnamefont {M.~V.}\ \bibnamefont {Larsen}}, \bibinfo {author} {\bibfnamefont {P.}~\bibnamefont {van Loock}}, \bibinfo {author} {\bibfnamefont {J.~S.}\ \bibnamefont {{Neergaard-Nielsen}}},\ and\ \bibinfo {author} {\bibfnamefont {U.~L.}\ \bibnamefont {Andersen}},\ }\href {https://doi.org/10.48550/arXiv.2502.19393} {\bibinfo {title} {The {{Octo-Rail Lattice}}: A four-dimensional cluster state design}} (\bibinfo {year} {2025}),\ \Eprint {https://arxiv.org/abs/2502.19393} {arXiv:2502.19393 [quant-ph]} \BibitemShut {NoStop}%
\bibitem [{\citenamefont {Aghaee~Rad}\ \emph {et~al.}(2025)\citenamefont {Aghaee~Rad}, \citenamefont {Ainsworth}, \citenamefont {Alexander}, \citenamefont {Altieri}, \citenamefont {Askarani}, \citenamefont {Baby}, \citenamefont {Banchi}, \citenamefont {Baragiola}, \citenamefont {Bourassa}, \citenamefont {Chadwick}, \citenamefont {Charania}, \citenamefont {Chen}, \citenamefont {Collins}, \citenamefont {Contu}, \citenamefont {D'Arcy}, \citenamefont {Dauphinais}, \citenamefont {De~Prins}, \citenamefont {Deschenes}, \citenamefont {Di~Luch}, \citenamefont {Duque}, \citenamefont {Edke}, \citenamefont {Fayer}, \citenamefont {Ferracin}, \citenamefont {Ferretti}, \citenamefont {Gefaell}, \citenamefont {Glancy}, \citenamefont {{Gonz{\'a}lez-Arciniegas}}, \citenamefont {Grainge}, \citenamefont {Han}, \citenamefont {Hastrup}, \citenamefont {Helt}, \citenamefont {Hillmann}, \citenamefont {Hundal}, \citenamefont {Izumi}, \citenamefont {Jaeken}, \citenamefont {Jonas}, \citenamefont {Kocsis}, \citenamefont {Krasnokutska},
  \citenamefont {Larsen}, \citenamefont {Laskowski}, \citenamefont {Laudenbach}, \citenamefont {Lavoie}, \citenamefont {Li}, \citenamefont {Lomonte}, \citenamefont {Lopetegui}, \citenamefont {Luey}, \citenamefont {Lund}, \citenamefont {Ma}, \citenamefont {Madsen}, \citenamefont {Mahler}, \citenamefont {Mantilla~Calder{\'o}n}, \citenamefont {Menotti}, \citenamefont {Miatto}, \citenamefont {Morrison}, \citenamefont {Nadkarni}, \citenamefont {Nakamura}, \citenamefont {Neuhaus}, \citenamefont {Niu}, \citenamefont {Noro}, \citenamefont {Papirov}, \citenamefont {Pesah}, \citenamefont {Phillips}, \citenamefont {Plick}, \citenamefont {Rogalsky}, \citenamefont {Rortais}, \citenamefont {{Sabines-Chesterking}}, \citenamefont {{Safavi-Bayat}}, \citenamefont {Sazhaev}, \citenamefont {Seymour}, \citenamefont {Rezaei~Shad}, \citenamefont {Silverman}, \citenamefont {Srinivasan}, \citenamefont {Stephan}, \citenamefont {Tang}, \citenamefont {Tasker}, \citenamefont {Teo}, \citenamefont {Then}, \citenamefont {Tremblay},
  \citenamefont {Tzitrin}, \citenamefont {Vaidya}, \citenamefont {Vasmer}, \citenamefont {Vernon}, \citenamefont {Villalobos}, \citenamefont {Walshe}, \citenamefont {Weil}, \citenamefont {Xin}, \citenamefont {Yan}, \citenamefont {Yao}, \citenamefont {Zamani~Abnili},\ and\ \citenamefont {Zhang}}]{aghaee_rad_scaling_2025}%
  \BibitemOpen
  \bibfield  {author} {\bibinfo {author} {\bibfnamefont {H.}~\bibnamefont {Aghaee~Rad}}, \bibinfo {author} {\bibfnamefont {T.}~\bibnamefont {Ainsworth}}, \bibinfo {author} {\bibfnamefont {R.~N.}\ \bibnamefont {Alexander}}, \bibinfo {author} {\bibfnamefont {B.}~\bibnamefont {Altieri}}, \bibinfo {author} {\bibfnamefont {M.~F.}\ \bibnamefont {Askarani}}, \bibinfo {author} {\bibfnamefont {R.}~\bibnamefont {Baby}}, \bibinfo {author} {\bibfnamefont {L.}~\bibnamefont {Banchi}}, \bibinfo {author} {\bibfnamefont {B.~Q.}\ \bibnamefont {Baragiola}}, \bibinfo {author} {\bibfnamefont {J.~E.}\ \bibnamefont {Bourassa}}, \bibinfo {author} {\bibfnamefont {R.~S.}\ \bibnamefont {Chadwick}}, \bibinfo {author} {\bibfnamefont {I.}~\bibnamefont {Charania}}, \bibinfo {author} {\bibfnamefont {H.}~\bibnamefont {Chen}}, \bibinfo {author} {\bibfnamefont {M.~J.}\ \bibnamefont {Collins}}, \bibinfo {author} {\bibfnamefont {P.}~\bibnamefont {Contu}}, \bibinfo {author} {\bibfnamefont {N.}~\bibnamefont {D'Arcy}}, \bibinfo {author}
  {\bibfnamefont {G.}~\bibnamefont {Dauphinais}}, \bibinfo {author} {\bibfnamefont {R.}~\bibnamefont {De~Prins}}, \bibinfo {author} {\bibfnamefont {D.}~\bibnamefont {Deschenes}}, \bibinfo {author} {\bibfnamefont {I.}~\bibnamefont {Di~Luch}}, \bibinfo {author} {\bibfnamefont {S.}~\bibnamefont {Duque}}, \bibinfo {author} {\bibfnamefont {P.}~\bibnamefont {Edke}}, \bibinfo {author} {\bibfnamefont {S.~E.}\ \bibnamefont {Fayer}}, \bibinfo {author} {\bibfnamefont {S.}~\bibnamefont {Ferracin}}, \bibinfo {author} {\bibfnamefont {H.}~\bibnamefont {Ferretti}}, \bibinfo {author} {\bibfnamefont {J.}~\bibnamefont {Gefaell}}, \bibinfo {author} {\bibfnamefont {S.}~\bibnamefont {Glancy}}, \bibinfo {author} {\bibfnamefont {C.}~\bibnamefont {{Gonz{\'a}lez-Arciniegas}}}, \bibinfo {author} {\bibfnamefont {T.}~\bibnamefont {Grainge}}, \bibinfo {author} {\bibfnamefont {Z.}~\bibnamefont {Han}}, \bibinfo {author} {\bibfnamefont {J.}~\bibnamefont {Hastrup}}, \bibinfo {author} {\bibfnamefont {L.~G.}\ \bibnamefont {Helt}}, \bibinfo
  {author} {\bibfnamefont {T.}~\bibnamefont {Hillmann}}, \bibinfo {author} {\bibfnamefont {J.}~\bibnamefont {Hundal}}, \bibinfo {author} {\bibfnamefont {S.}~\bibnamefont {Izumi}}, \bibinfo {author} {\bibfnamefont {T.}~\bibnamefont {Jaeken}}, \bibinfo {author} {\bibfnamefont {M.}~\bibnamefont {Jonas}}, \bibinfo {author} {\bibfnamefont {S.}~\bibnamefont {Kocsis}}, \bibinfo {author} {\bibfnamefont {I.}~\bibnamefont {Krasnokutska}}, \bibinfo {author} {\bibfnamefont {M.~V.}\ \bibnamefont {Larsen}}, \bibinfo {author} {\bibfnamefont {P.}~\bibnamefont {Laskowski}}, \bibinfo {author} {\bibfnamefont {F.}~\bibnamefont {Laudenbach}}, \bibinfo {author} {\bibfnamefont {J.}~\bibnamefont {Lavoie}}, \bibinfo {author} {\bibfnamefont {M.}~\bibnamefont {Li}}, \bibinfo {author} {\bibfnamefont {E.}~\bibnamefont {Lomonte}}, \bibinfo {author} {\bibfnamefont {C.~E.}\ \bibnamefont {Lopetegui}}, \bibinfo {author} {\bibfnamefont {B.}~\bibnamefont {Luey}}, \bibinfo {author} {\bibfnamefont {A.~P.}\ \bibnamefont {Lund}}, \bibinfo {author}
  {\bibfnamefont {C.}~\bibnamefont {Ma}}, \bibinfo {author} {\bibfnamefont {L.~S.}\ \bibnamefont {Madsen}}, \bibinfo {author} {\bibfnamefont {D.~H.}\ \bibnamefont {Mahler}}, \bibinfo {author} {\bibfnamefont {L.}~\bibnamefont {Mantilla~Calder{\'o}n}}, \bibinfo {author} {\bibfnamefont {M.}~\bibnamefont {Menotti}}, \bibinfo {author} {\bibfnamefont {F.~M.}\ \bibnamefont {Miatto}}, \bibinfo {author} {\bibfnamefont {B.}~\bibnamefont {Morrison}}, \bibinfo {author} {\bibfnamefont {P.~J.}\ \bibnamefont {Nadkarni}}, \bibinfo {author} {\bibfnamefont {T.}~\bibnamefont {Nakamura}}, \bibinfo {author} {\bibfnamefont {L.}~\bibnamefont {Neuhaus}}, \bibinfo {author} {\bibfnamefont {Z.}~\bibnamefont {Niu}}, \bibinfo {author} {\bibfnamefont {R.}~\bibnamefont {Noro}}, \bibinfo {author} {\bibfnamefont {K.}~\bibnamefont {Papirov}}, \bibinfo {author} {\bibfnamefont {A.}~\bibnamefont {Pesah}}, \bibinfo {author} {\bibfnamefont {D.~S.}\ \bibnamefont {Phillips}}, \bibinfo {author} {\bibfnamefont {W.~N.}\ \bibnamefont {Plick}}, \bibinfo
  {author} {\bibfnamefont {T.}~\bibnamefont {Rogalsky}}, \bibinfo {author} {\bibfnamefont {F.}~\bibnamefont {Rortais}}, \bibinfo {author} {\bibfnamefont {J.}~\bibnamefont {{Sabines-Chesterking}}}, \bibinfo {author} {\bibfnamefont {S.}~\bibnamefont {{Safavi-Bayat}}}, \bibinfo {author} {\bibfnamefont {E.}~\bibnamefont {Sazhaev}}, \bibinfo {author} {\bibfnamefont {M.}~\bibnamefont {Seymour}}, \bibinfo {author} {\bibfnamefont {K.}~\bibnamefont {Rezaei~Shad}}, \bibinfo {author} {\bibfnamefont {M.}~\bibnamefont {Silverman}}, \bibinfo {author} {\bibfnamefont {S.~A.}\ \bibnamefont {Srinivasan}}, \bibinfo {author} {\bibfnamefont {M.}~\bibnamefont {Stephan}}, \bibinfo {author} {\bibfnamefont {Q.~Y.}\ \bibnamefont {Tang}}, \bibinfo {author} {\bibfnamefont {J.~F.}\ \bibnamefont {Tasker}}, \bibinfo {author} {\bibfnamefont {Y.~S.}\ \bibnamefont {Teo}}, \bibinfo {author} {\bibfnamefont {R.~B.}\ \bibnamefont {Then}}, \bibinfo {author} {\bibfnamefont {J.~E.}\ \bibnamefont {Tremblay}}, \bibinfo {author} {\bibfnamefont
  {I.}~\bibnamefont {Tzitrin}}, \bibinfo {author} {\bibfnamefont {V.~D.}\ \bibnamefont {Vaidya}}, \bibinfo {author} {\bibfnamefont {M.}~\bibnamefont {Vasmer}}, \bibinfo {author} {\bibfnamefont {Z.}~\bibnamefont {Vernon}}, \bibinfo {author} {\bibfnamefont {L.~F. S. S.~M.}\ \bibnamefont {Villalobos}}, \bibinfo {author} {\bibfnamefont {B.~W.}\ \bibnamefont {Walshe}}, \bibinfo {author} {\bibfnamefont {R.}~\bibnamefont {Weil}}, \bibinfo {author} {\bibfnamefont {X.}~\bibnamefont {Xin}}, \bibinfo {author} {\bibfnamefont {X.}~\bibnamefont {Yan}}, \bibinfo {author} {\bibfnamefont {Y.}~\bibnamefont {Yao}}, \bibinfo {author} {\bibfnamefont {M.}~\bibnamefont {Zamani~Abnili}},\ and\ \bibinfo {author} {\bibfnamefont {Y.}~\bibnamefont {Zhang}},\ }\bibfield  {title} {\bibinfo {title} {Scaling and networking a modular photonic quantum computer},\ }\href {https://doi.org/10.1038/s41586-024-08406-9} {\bibfield  {journal} {\bibinfo  {journal} {Nature}\ }\textbf {\bibinfo {volume} {638}},\ \bibinfo {pages} {912} (\bibinfo {year}
  {2025})}\BibitemShut {NoStop}%
\bibitem [{\citenamefont {Su}\ \emph {et~al.}(2019)\citenamefont {Su}, \citenamefont {Myers},\ and\ \citenamefont {Sabapathy}}]{su_conversion_2019}%
  \BibitemOpen
  \bibfield  {author} {\bibinfo {author} {\bibfnamefont {D.}~\bibnamefont {Su}}, \bibinfo {author} {\bibfnamefont {C.~R.}\ \bibnamefont {Myers}},\ and\ \bibinfo {author} {\bibfnamefont {K.~K.}\ \bibnamefont {Sabapathy}},\ }\bibfield  {title} {\bibinfo {title} {Conversion of {{Gaussian}} states to non-{{Gaussian}} states using photon-number-resolving detectors},\ }\href {https://doi.org/10.1103/PhysRevA.100.052301} {\bibfield  {journal} {\bibinfo  {journal} {Physical Review A}\ }\textbf {\bibinfo {volume} {100}},\ \bibinfo {pages} {052301} (\bibinfo {year} {2019})}\BibitemShut {NoStop}%
\bibitem [{\citenamefont {Quesada}\ \emph {et~al.}(2019)\citenamefont {Quesada}, \citenamefont {Helt}, \citenamefont {Izaac}, \citenamefont {Arrazola}, \citenamefont {Shahrokhshahi}, \citenamefont {Myers},\ and\ \citenamefont {Sabapathy}}]{quesada_simulating_2019}%
  \BibitemOpen
  \bibfield  {author} {\bibinfo {author} {\bibfnamefont {N.}~\bibnamefont {Quesada}}, \bibinfo {author} {\bibfnamefont {L.~G.}\ \bibnamefont {Helt}}, \bibinfo {author} {\bibfnamefont {J.}~\bibnamefont {Izaac}}, \bibinfo {author} {\bibfnamefont {J.~M.}\ \bibnamefont {Arrazola}}, \bibinfo {author} {\bibfnamefont {R.}~\bibnamefont {Shahrokhshahi}}, \bibinfo {author} {\bibfnamefont {C.~R.}\ \bibnamefont {Myers}},\ and\ \bibinfo {author} {\bibfnamefont {K.~K.}\ \bibnamefont {Sabapathy}},\ }\bibfield  {title} {\bibinfo {title} {Simulating realistic non-{{Gaussian}} state preparation},\ }\href {https://doi.org/10.1103/PhysRevA.100.022341} {\bibfield  {journal} {\bibinfo  {journal} {Physical Review A}\ }\textbf {\bibinfo {volume} {100}},\ \bibinfo {pages} {022341} (\bibinfo {year} {2019})}\BibitemShut {NoStop}%
\bibitem [{\citenamefont {Tzitrin}\ \emph {et~al.}(2020)\citenamefont {Tzitrin}, \citenamefont {Bourassa}, \citenamefont {Menicucci},\ and\ \citenamefont {Sabapathy}}]{tzitrin_progress_2020}%
  \BibitemOpen
  \bibfield  {author} {\bibinfo {author} {\bibfnamefont {I.}~\bibnamefont {Tzitrin}}, \bibinfo {author} {\bibfnamefont {J.~E.}\ \bibnamefont {Bourassa}}, \bibinfo {author} {\bibfnamefont {N.~C.}\ \bibnamefont {Menicucci}},\ and\ \bibinfo {author} {\bibfnamefont {K.~K.}\ \bibnamefont {Sabapathy}},\ }\bibfield  {title} {\bibinfo {title} {Progress towards practical qubit computation using approximate {{Gottesman-Kitaev-Preskill}} codes},\ }\href {https://doi.org/10.1103/PhysRevA.101.032315} {\bibfield  {journal} {\bibinfo  {journal} {Physical Review A}\ }\textbf {\bibinfo {volume} {101}},\ \bibinfo {pages} {032315} (\bibinfo {year} {2020})}\BibitemShut {NoStop}%
\bibitem [{\citenamefont {Eaton}\ \emph {et~al.}(2022)\citenamefont {Eaton}, \citenamefont {{Gonz{\'a}lez-Arciniegas}}, \citenamefont {Alexander}, \citenamefont {Menicucci},\ and\ \citenamefont {Pfister}}]{eaton_measurement-based_2022}%
  \BibitemOpen
  \bibfield  {author} {\bibinfo {author} {\bibfnamefont {M.}~\bibnamefont {Eaton}}, \bibinfo {author} {\bibfnamefont {C.}~\bibnamefont {{Gonz{\'a}lez-Arciniegas}}}, \bibinfo {author} {\bibfnamefont {R.~N.}\ \bibnamefont {Alexander}}, \bibinfo {author} {\bibfnamefont {N.~C.}\ \bibnamefont {Menicucci}},\ and\ \bibinfo {author} {\bibfnamefont {O.}~\bibnamefont {Pfister}},\ }\bibfield  {title} {\bibinfo {title} {Measurement-based generation and preservation of cat and grid states within a continuous-variable cluster state},\ }\href {https://doi.org/10.22331/q-2022-07-20-769} {\bibfield  {journal} {\bibinfo  {journal} {Quantum}\ }\textbf {\bibinfo {volume} {6}},\ \bibinfo {pages} {769} (\bibinfo {year} {2022})}\BibitemShut {NoStop}%
\bibitem [{\citenamefont {Takase}\ \emph {et~al.}(2023)\citenamefont {Takase}, \citenamefont {Fukui}, \citenamefont {Kawasaki}, \citenamefont {Asavanant}, \citenamefont {Endo}, \citenamefont {Yoshikawa}, \citenamefont {Van~Loock},\ and\ \citenamefont {Furusawa}}]{takase_gottesman-kitaev-preskill_2023}%
  \BibitemOpen
  \bibfield  {author} {\bibinfo {author} {\bibfnamefont {K.}~\bibnamefont {Takase}}, \bibinfo {author} {\bibfnamefont {K.}~\bibnamefont {Fukui}}, \bibinfo {author} {\bibfnamefont {A.}~\bibnamefont {Kawasaki}}, \bibinfo {author} {\bibfnamefont {W.}~\bibnamefont {Asavanant}}, \bibinfo {author} {\bibfnamefont {M.}~\bibnamefont {Endo}}, \bibinfo {author} {\bibfnamefont {J.-i.}\ \bibnamefont {Yoshikawa}}, \bibinfo {author} {\bibfnamefont {P.}~\bibnamefont {Van~Loock}},\ and\ \bibinfo {author} {\bibfnamefont {A.}~\bibnamefont {Furusawa}},\ }\bibfield  {title} {\bibinfo {title} {Gottesman-{{Kitaev-Preskill}} qubit synthesizer for propagating light},\ }\href {https://doi.org/10.1038/s41534-023-00772-y} {\bibfield  {journal} {\bibinfo  {journal} {npj Quantum Information}\ }\textbf {\bibinfo {volume} {9}},\ \bibinfo {pages} {98} (\bibinfo {year} {2023})}\BibitemShut {NoStop}%
\bibitem [{\citenamefont {Endo}\ \emph {et~al.}(2023)\citenamefont {Endo}, \citenamefont {He}, \citenamefont {Sonoyama}, \citenamefont {Takahashi}, \citenamefont {Kashiwazaki}, \citenamefont {Umeki}, \citenamefont {Takasu}, \citenamefont {Hattori}, \citenamefont {Fukuda}, \citenamefont {Fukui}, \citenamefont {Takase}, \citenamefont {Asavanant}, \citenamefont {Marek}, \citenamefont {Filip},\ and\ \citenamefont {Furusawa}}]{endo_non-gaussian_2023}%
  \BibitemOpen
  \bibfield  {author} {\bibinfo {author} {\bibfnamefont {M.}~\bibnamefont {Endo}}, \bibinfo {author} {\bibfnamefont {R.}~\bibnamefont {He}}, \bibinfo {author} {\bibfnamefont {T.}~\bibnamefont {Sonoyama}}, \bibinfo {author} {\bibfnamefont {K.}~\bibnamefont {Takahashi}}, \bibinfo {author} {\bibfnamefont {T.}~\bibnamefont {Kashiwazaki}}, \bibinfo {author} {\bibfnamefont {T.}~\bibnamefont {Umeki}}, \bibinfo {author} {\bibfnamefont {S.}~\bibnamefont {Takasu}}, \bibinfo {author} {\bibfnamefont {K.}~\bibnamefont {Hattori}}, \bibinfo {author} {\bibfnamefont {D.}~\bibnamefont {Fukuda}}, \bibinfo {author} {\bibfnamefont {K.}~\bibnamefont {Fukui}}, \bibinfo {author} {\bibfnamefont {K.}~\bibnamefont {Takase}}, \bibinfo {author} {\bibfnamefont {W.}~\bibnamefont {Asavanant}}, \bibinfo {author} {\bibfnamefont {P.}~\bibnamefont {Marek}}, \bibinfo {author} {\bibfnamefont {R.}~\bibnamefont {Filip}},\ and\ \bibinfo {author} {\bibfnamefont {A.}~\bibnamefont {Furusawa}},\ }\bibfield  {title} {\bibinfo {title} {Non-{{Gaussian}}
  quantum state generation by multi-photon subtraction at the telecommunication wavelength},\ }\href {https://doi.org/10.1364/oe.486270} {\bibfield  {journal} {\bibinfo  {journal} {Optics Express}\ }\textbf {\bibinfo {volume} {31}},\ \bibinfo {pages} {12865} (\bibinfo {year} {2023})}\BibitemShut {NoStop}%
\bibitem [{\citenamefont {Endo}\ \emph {et~al.}(2025)\citenamefont {Endo}, \citenamefont {Nomura}, \citenamefont {Sonoyama}, \citenamefont {Takahashi}, \citenamefont {Takasu}, \citenamefont {Fukuda}, \citenamefont {Kashiwazaki}, \citenamefont {Inoue}, \citenamefont {Umeki}, \citenamefont {Nehra}, \citenamefont {Marek}, \citenamefont {Filip}, \citenamefont {Takase}, \citenamefont {Asavanant},\ and\ \citenamefont {Furusawa}}]{endo_high-rate_2025}%
  \BibitemOpen
  \bibfield  {author} {\bibinfo {author} {\bibfnamefont {M.}~\bibnamefont {Endo}}, \bibinfo {author} {\bibfnamefont {T.}~\bibnamefont {Nomura}}, \bibinfo {author} {\bibfnamefont {T.}~\bibnamefont {Sonoyama}}, \bibinfo {author} {\bibfnamefont {K.}~\bibnamefont {Takahashi}}, \bibinfo {author} {\bibfnamefont {S.}~\bibnamefont {Takasu}}, \bibinfo {author} {\bibfnamefont {D.}~\bibnamefont {Fukuda}}, \bibinfo {author} {\bibfnamefont {T.}~\bibnamefont {Kashiwazaki}}, \bibinfo {author} {\bibfnamefont {A.}~\bibnamefont {Inoue}}, \bibinfo {author} {\bibfnamefont {T.}~\bibnamefont {Umeki}}, \bibinfo {author} {\bibfnamefont {R.}~\bibnamefont {Nehra}}, \bibinfo {author} {\bibfnamefont {P.}~\bibnamefont {Marek}}, \bibinfo {author} {\bibfnamefont {R.}~\bibnamefont {Filip}}, \bibinfo {author} {\bibfnamefont {K.}~\bibnamefont {Takase}}, \bibinfo {author} {\bibfnamefont {W.}~\bibnamefont {Asavanant}},\ and\ \bibinfo {author} {\bibfnamefont {A.}~\bibnamefont {Furusawa}},\ }\href {https://doi.org/10.48550/arXiv.2502.08952}
  {\bibinfo {title} {High-{{Rate Four Photon Subtraction}} from {{Squeezed Vacuum}}: {{Preparing Cat State}} for {{Optical Quantum Computation}}}} (\bibinfo {year} {2025}),\ \Eprint {https://arxiv.org/abs/2502.08952} {arXiv:2502.08952 [quant-ph]} \BibitemShut {NoStop}%
\bibitem [{\citenamefont {Hamilton}\ \emph {et~al.}(2017)\citenamefont {Hamilton}, \citenamefont {Kruse}, \citenamefont {Sansoni}, \citenamefont {Barkhofen}, \citenamefont {Silberhorn},\ and\ \citenamefont {Jex}}]{hamilton_gaussian_2017}%
  \BibitemOpen
  \bibfield  {author} {\bibinfo {author} {\bibfnamefont {C.~S.}\ \bibnamefont {Hamilton}}, \bibinfo {author} {\bibfnamefont {R.}~\bibnamefont {Kruse}}, \bibinfo {author} {\bibfnamefont {L.}~\bibnamefont {Sansoni}}, \bibinfo {author} {\bibfnamefont {S.}~\bibnamefont {Barkhofen}}, \bibinfo {author} {\bibfnamefont {C.}~\bibnamefont {Silberhorn}},\ and\ \bibinfo {author} {\bibfnamefont {I.}~\bibnamefont {Jex}},\ }\bibfield  {title} {\bibinfo {title} {Gaussian {{Boson Sampling}}},\ }\href {https://doi.org/10.1103/PhysRevLett.119.170501} {\bibfield  {journal} {\bibinfo  {journal} {Physical Review Letters}\ }\textbf {\bibinfo {volume} {119}},\ \bibinfo {pages} {170501} (\bibinfo {year} {2017})}\BibitemShut {NoStop}%
\bibitem [{\citenamefont {Vasconcelos}\ \emph {et~al.}(2010)\citenamefont {Vasconcelos}, \citenamefont {Sanz},\ and\ \citenamefont {Glancy}}]{vasconcelos_all-optical_2010}%
  \BibitemOpen
  \bibfield  {author} {\bibinfo {author} {\bibfnamefont {H.~M.}\ \bibnamefont {Vasconcelos}}, \bibinfo {author} {\bibfnamefont {L.}~\bibnamefont {Sanz}},\ and\ \bibinfo {author} {\bibfnamefont {S.}~\bibnamefont {Glancy}},\ }\bibfield  {title} {\bibinfo {title} {All-optical generation of states for ``{{Encoding}} a qubit in an oscillator''},\ }\href {https://doi.org/10.1364/OL.35.003261} {\bibfield  {journal} {\bibinfo  {journal} {Optics Letters}\ }\textbf {\bibinfo {volume} {35}},\ \bibinfo {pages} {3261} (\bibinfo {year} {2010})}\BibitemShut {NoStop}%
\bibitem [{\citenamefont {Weigand}\ and\ \citenamefont {Terhal}(2018)}]{weigand_generating_2018}%
  \BibitemOpen
  \bibfield  {author} {\bibinfo {author} {\bibfnamefont {D.~J.}\ \bibnamefont {Weigand}}\ and\ \bibinfo {author} {\bibfnamefont {B.~M.}\ \bibnamefont {Terhal}},\ }\bibfield  {title} {\bibinfo {title} {Generating grid states from {{Schr{\"o}dinger-cat}} states without postselection},\ }\href {https://doi.org/10.1103/PhysRevA.97.022341} {\bibfield  {journal} {\bibinfo  {journal} {Physical Review A}\ }\textbf {\bibinfo {volume} {97}},\ \bibinfo {pages} {022341} (\bibinfo {year} {2018})}\BibitemShut {NoStop}%
\bibitem [{\citenamefont {Takase}\ \emph {et~al.}(2024)\citenamefont {Takase}, \citenamefont {Hanamura}, \citenamefont {Nagayoshi}, \citenamefont {Bourassa}, \citenamefont {Alexander}, \citenamefont {Kawasaki}, \citenamefont {Asavanant}, \citenamefont {Endo},\ and\ \citenamefont {Furusawa}}]{takase_generation_2024}%
  \BibitemOpen
  \bibfield  {author} {\bibinfo {author} {\bibfnamefont {K.}~\bibnamefont {Takase}}, \bibinfo {author} {\bibfnamefont {F.}~\bibnamefont {Hanamura}}, \bibinfo {author} {\bibfnamefont {H.}~\bibnamefont {Nagayoshi}}, \bibinfo {author} {\bibfnamefont {J.~E.}\ \bibnamefont {Bourassa}}, \bibinfo {author} {\bibfnamefont {R.~N.}\ \bibnamefont {Alexander}}, \bibinfo {author} {\bibfnamefont {A.}~\bibnamefont {Kawasaki}}, \bibinfo {author} {\bibfnamefont {W.}~\bibnamefont {Asavanant}}, \bibinfo {author} {\bibfnamefont {M.}~\bibnamefont {Endo}},\ and\ \bibinfo {author} {\bibfnamefont {A.}~\bibnamefont {Furusawa}},\ }\bibfield  {title} {\bibinfo {title} {Generation of flying logical qubits using generalized photon subtraction with adaptive {{Gaussian}} operations},\ }\href {https://doi.org/10.1103/PhysRevA.110.012436} {\bibfield  {journal} {\bibinfo  {journal} {Physical Review A}\ }\textbf {\bibinfo {volume} {110}},\ \bibinfo {pages} {012436} (\bibinfo {year} {2024})}\BibitemShut {NoStop}%
\bibitem [{\citenamefont {Ourjoumtsev}\ \emph {et~al.}(2006)\citenamefont {Ourjoumtsev}, \citenamefont {{Tualle-Brouri}}, \citenamefont {Laurat},\ and\ \citenamefont {Grangier}}]{ourjoumtsev_generating_2006}%
  \BibitemOpen
  \bibfield  {author} {\bibinfo {author} {\bibfnamefont {A.}~\bibnamefont {Ourjoumtsev}}, \bibinfo {author} {\bibfnamefont {R.}~\bibnamefont {{Tualle-Brouri}}}, \bibinfo {author} {\bibfnamefont {J.}~\bibnamefont {Laurat}},\ and\ \bibinfo {author} {\bibfnamefont {P.}~\bibnamefont {Grangier}},\ }\bibfield  {title} {\bibinfo {title} {Generating {{Optical Schr{\"o}dinger Kittens}} for {{Quantum Information Processing}}},\ }\href {https://doi.org/10.1126/science.1122858} {\bibfield  {journal} {\bibinfo  {journal} {Science}\ }\textbf {\bibinfo {volume} {312}},\ \bibinfo {pages} {83} (\bibinfo {year} {2006})}\BibitemShut {NoStop}%
\bibitem [{\citenamefont {Huang}\ \emph {et~al.}(2015)\citenamefont {Huang}, \citenamefont {Le~Jeannic}, \citenamefont {Ruaudel}, \citenamefont {Verma}, \citenamefont {Shaw}, \citenamefont {Marsili}, \citenamefont {Nam}, \citenamefont {Wu}, \citenamefont {Zeng}, \citenamefont {Jeong}, \citenamefont {Filip}, \citenamefont {Morin},\ and\ \citenamefont {Laurat}}]{huang_optical_2015}%
  \BibitemOpen
  \bibfield  {author} {\bibinfo {author} {\bibfnamefont {K.}~\bibnamefont {Huang}}, \bibinfo {author} {\bibfnamefont {H.}~\bibnamefont {Le~Jeannic}}, \bibinfo {author} {\bibfnamefont {J.}~\bibnamefont {Ruaudel}}, \bibinfo {author} {\bibfnamefont {V.~B.}\ \bibnamefont {Verma}}, \bibinfo {author} {\bibfnamefont {M.~D.}\ \bibnamefont {Shaw}}, \bibinfo {author} {\bibfnamefont {F.}~\bibnamefont {Marsili}}, \bibinfo {author} {\bibfnamefont {S.~W.}\ \bibnamefont {Nam}}, \bibinfo {author} {\bibfnamefont {E.}~\bibnamefont {Wu}}, \bibinfo {author} {\bibfnamefont {H.}~\bibnamefont {Zeng}}, \bibinfo {author} {\bibfnamefont {Y.-C.}\ \bibnamefont {Jeong}}, \bibinfo {author} {\bibfnamefont {R.}~\bibnamefont {Filip}}, \bibinfo {author} {\bibfnamefont {O.}~\bibnamefont {Morin}},\ and\ \bibinfo {author} {\bibfnamefont {J.}~\bibnamefont {Laurat}},\ }\bibfield  {title} {\bibinfo {title} {Optical {{Synthesis}} of {{Large-Amplitude Squeezed Coherent-State Superpositions}} with {{Minimal Resources}}},\ }\href
  {https://doi.org/10.1103/PhysRevLett.115.023602} {\bibfield  {journal} {\bibinfo  {journal} {Physical Review Letters}\ }\textbf {\bibinfo {volume} {115}},\ \bibinfo {pages} {023602} (\bibinfo {year} {2015})}\BibitemShut {NoStop}%
\bibitem [{\citenamefont {Takase}\ \emph {et~al.}(2021)\citenamefont {Takase}, \citenamefont {Yoshikawa}, \citenamefont {Asavanant}, \citenamefont {Endo},\ and\ \citenamefont {Furusawa}}]{takase_generation_2021}%
  \BibitemOpen
  \bibfield  {author} {\bibinfo {author} {\bibfnamefont {K.}~\bibnamefont {Takase}}, \bibinfo {author} {\bibfnamefont {J.-i.}\ \bibnamefont {Yoshikawa}}, \bibinfo {author} {\bibfnamefont {W.}~\bibnamefont {Asavanant}}, \bibinfo {author} {\bibfnamefont {M.}~\bibnamefont {Endo}},\ and\ \bibinfo {author} {\bibfnamefont {A.}~\bibnamefont {Furusawa}},\ }\bibfield  {title} {\bibinfo {title} {Generation of optical {{Schr{\"o}dinger}} cat states by generalized photon subtraction},\ }\href {https://doi.org/10.1103/PhysRevA.103.013710} {\bibfield  {journal} {\bibinfo  {journal} {Physical Review A}\ }\textbf {\bibinfo {volume} {103}},\ \bibinfo {pages} {013710} (\bibinfo {year} {2021})}\BibitemShut {NoStop}%
\bibitem [{\citenamefont {Ourjoumtsev}\ \emph {et~al.}(2007)\citenamefont {Ourjoumtsev}, \citenamefont {Jeong}, \citenamefont {{Tualle-Brouri}},\ and\ \citenamefont {Grangier}}]{ourjoumtsev_generation_2007}%
  \BibitemOpen
  \bibfield  {author} {\bibinfo {author} {\bibfnamefont {A.}~\bibnamefont {Ourjoumtsev}}, \bibinfo {author} {\bibfnamefont {H.}~\bibnamefont {Jeong}}, \bibinfo {author} {\bibfnamefont {R.}~\bibnamefont {{Tualle-Brouri}}},\ and\ \bibinfo {author} {\bibfnamefont {P.}~\bibnamefont {Grangier}},\ }\bibfield  {title} {\bibinfo {title} {Generation of optical `{{Schr{\"o}dinger}} cats' from photon number states},\ }\href {https://doi.org/10.1038/nature06054} {\bibfield  {journal} {\bibinfo  {journal} {Nature}\ }\textbf {\bibinfo {volume} {448}},\ \bibinfo {pages} {784} (\bibinfo {year} {2007})}\BibitemShut {NoStop}%
\bibitem [{\citenamefont {Eaton}\ \emph {et~al.}(2019)\citenamefont {Eaton}, \citenamefont {Nehra},\ and\ \citenamefont {Pfister}}]{eaton_non-gaussian_2019}%
  \BibitemOpen
  \bibfield  {author} {\bibinfo {author} {\bibfnamefont {M.}~\bibnamefont {Eaton}}, \bibinfo {author} {\bibfnamefont {R.}~\bibnamefont {Nehra}},\ and\ \bibinfo {author} {\bibfnamefont {O.}~\bibnamefont {Pfister}},\ }\bibfield  {title} {\bibinfo {title} {Non-{{Gaussian}} and {{Gottesman}}--{{Kitaev}}--{{Preskill}} state preparation by photon catalysis},\ }\href {https://doi.org/10.1088/1367-2630/ab5330} {\bibfield  {journal} {\bibinfo  {journal} {New Journal of Physics}\ }\textbf {\bibinfo {volume} {21}},\ \bibinfo {pages} {113034} (\bibinfo {year} {2019})}\BibitemShut {NoStop}%
\bibitem [{\citenamefont {Winnel}\ \emph {et~al.}(2024)\citenamefont {Winnel}, \citenamefont {Guanzon}, \citenamefont {Singh},\ and\ \citenamefont {Ralph}}]{winnel_deterministic_2024}%
  \BibitemOpen
  \bibfield  {author} {\bibinfo {author} {\bibfnamefont {M.~S.}\ \bibnamefont {Winnel}}, \bibinfo {author} {\bibfnamefont {J.~J.}\ \bibnamefont {Guanzon}}, \bibinfo {author} {\bibfnamefont {D.}~\bibnamefont {Singh}},\ and\ \bibinfo {author} {\bibfnamefont {T.~C.}\ \bibnamefont {Ralph}},\ }\bibfield  {title} {\bibinfo {title} {Deterministic {{Preparation}} of {{Optical Squeezed Cat}} and {{Gottesman-Kitaev-Preskill States}}},\ }\href {https://doi.org/10.1103/PhysRevLett.132.230602} {\bibfield  {journal} {\bibinfo  {journal} {Physical Review Letters}\ }\textbf {\bibinfo {volume} {132}},\ \bibinfo {pages} {230602} (\bibinfo {year} {2024})}\BibitemShut {NoStop}%
\bibitem [{\citenamefont {Tiedau}\ \emph {et~al.}(2019)\citenamefont {Tiedau}, \citenamefont {Bartley}, \citenamefont {Harder}, \citenamefont {Lita}, \citenamefont {Nam}, \citenamefont {Gerrits},\ and\ \citenamefont {Silberhorn}}]{tiedau_scalability_2019}%
  \BibitemOpen
  \bibfield  {author} {\bibinfo {author} {\bibfnamefont {J.}~\bibnamefont {Tiedau}}, \bibinfo {author} {\bibfnamefont {T.~J.}\ \bibnamefont {Bartley}}, \bibinfo {author} {\bibfnamefont {G.}~\bibnamefont {Harder}}, \bibinfo {author} {\bibfnamefont {A.~E.}\ \bibnamefont {Lita}}, \bibinfo {author} {\bibfnamefont {S.~W.}\ \bibnamefont {Nam}}, \bibinfo {author} {\bibfnamefont {T.}~\bibnamefont {Gerrits}},\ and\ \bibinfo {author} {\bibfnamefont {C.}~\bibnamefont {Silberhorn}},\ }\bibfield  {title} {\bibinfo {title} {Scalability of parametric down-conversion for generating higher-order {{Fock}} states},\ }\href {https://doi.org/10.1103/PhysRevA.100.041802} {\bibfield  {journal} {\bibinfo  {journal} {Physical Review A}\ }\textbf {\bibinfo {volume} {100}},\ \bibinfo {pages} {041802} (\bibinfo {year} {2019})}\BibitemShut {NoStop}%
\bibitem [{\citenamefont {Hacker}\ \emph {et~al.}(2019)\citenamefont {Hacker}, \citenamefont {Welte}, \citenamefont {Daiss}, \citenamefont {Shaukat}, \citenamefont {Ritter}, \citenamefont {Li},\ and\ \citenamefont {Rempe}}]{hacker_deterministic_2019}%
  \BibitemOpen
  \bibfield  {author} {\bibinfo {author} {\bibfnamefont {B.}~\bibnamefont {Hacker}}, \bibinfo {author} {\bibfnamefont {S.}~\bibnamefont {Welte}}, \bibinfo {author} {\bibfnamefont {S.}~\bibnamefont {Daiss}}, \bibinfo {author} {\bibfnamefont {A.}~\bibnamefont {Shaukat}}, \bibinfo {author} {\bibfnamefont {S.}~\bibnamefont {Ritter}}, \bibinfo {author} {\bibfnamefont {L.}~\bibnamefont {Li}},\ and\ \bibinfo {author} {\bibfnamefont {G.}~\bibnamefont {Rempe}},\ }\bibfield  {title} {\bibinfo {title} {Deterministic creation of entangled atom--light {{Schr{\"o}dinger-cat}} states},\ }\href {https://doi.org/10.1038/s41566-018-0339-5} {\bibfield  {journal} {\bibinfo  {journal} {Nature Photonics}\ }\textbf {\bibinfo {volume} {13}},\ \bibinfo {pages} {110} (\bibinfo {year} {2019})}\BibitemShut {NoStop}%
\bibitem [{\citenamefont {Hastrup}\ and\ \citenamefont {Andersen}(2022)}]{hastrup_protocol_2022}%
  \BibitemOpen
  \bibfield  {author} {\bibinfo {author} {\bibfnamefont {J.}~\bibnamefont {Hastrup}}\ and\ \bibinfo {author} {\bibfnamefont {U.~L.}\ \bibnamefont {Andersen}},\ }\bibfield  {title} {\bibinfo {title} {Protocol for {{Generating Optical Gottesman-Kitaev-Preskill States}} with {{Cavity QED}}},\ }\href {https://doi.org/10.1103/PhysRevLett.128.170503} {\bibfield  {journal} {\bibinfo  {journal} {Physical Review Letters}\ }\textbf {\bibinfo {volume} {128}},\ \bibinfo {pages} {170503} (\bibinfo {year} {2022})}\BibitemShut {NoStop}%
\bibitem [{\citenamefont {Provazn{\'i}k}\ \emph {et~al.}(2020)\citenamefont {Provazn{\'i}k}, \citenamefont {Lachman}, \citenamefont {Filip},\ and\ \citenamefont {Marek}}]{provaznik_benchmarking_2020}%
  \BibitemOpen
  \bibfield  {author} {\bibinfo {author} {\bibfnamefont {J.}~\bibnamefont {Provazn{\'i}k}}, \bibinfo {author} {\bibfnamefont {L.}~\bibnamefont {Lachman}}, \bibinfo {author} {\bibfnamefont {R.}~\bibnamefont {Filip}},\ and\ \bibinfo {author} {\bibfnamefont {P.}~\bibnamefont {Marek}},\ }\bibfield  {title} {\bibinfo {title} {Benchmarking photon number resolving detectors},\ }\href {https://doi.org/10.1364/OE.389619} {\bibfield  {journal} {\bibinfo  {journal} {Optics Express}\ }\textbf {\bibinfo {volume} {28}},\ \bibinfo {pages} {14839} (\bibinfo {year} {2020})}\BibitemShut {NoStop}%
\bibitem [{\citenamefont {Bourassa}\ \emph {et~al.}(2021{\natexlab{b}})\citenamefont {Bourassa}, \citenamefont {Quesada}, \citenamefont {Tzitrin}, \citenamefont {Sz{\'a}va}, \citenamefont {Isacsson}, \citenamefont {Izaac}, \citenamefont {Sabapathy}, \citenamefont {Dauphinais},\ and\ \citenamefont {Dhand}}]{bourassa_fast_2021}%
  \BibitemOpen
  \bibfield  {author} {\bibinfo {author} {\bibfnamefont {J.~E.}\ \bibnamefont {Bourassa}}, \bibinfo {author} {\bibfnamefont {N.}~\bibnamefont {Quesada}}, \bibinfo {author} {\bibfnamefont {I.}~\bibnamefont {Tzitrin}}, \bibinfo {author} {\bibfnamefont {A.}~\bibnamefont {Sz{\'a}va}}, \bibinfo {author} {\bibfnamefont {T.}~\bibnamefont {Isacsson}}, \bibinfo {author} {\bibfnamefont {J.}~\bibnamefont {Izaac}}, \bibinfo {author} {\bibfnamefont {K.~K.}\ \bibnamefont {Sabapathy}}, \bibinfo {author} {\bibfnamefont {G.}~\bibnamefont {Dauphinais}},\ and\ \bibinfo {author} {\bibfnamefont {I.}~\bibnamefont {Dhand}},\ }\bibfield  {title} {\bibinfo {title} {Fast {{Simulation}} of {{Bosonic Qubits}} via {{Gaussian Functions}} in {{Phase Space}}},\ }\href {https://doi.org/10.1103/PRXQuantum.2.040315} {\bibfield  {journal} {\bibinfo  {journal} {PRX Quantum}\ }\textbf {\bibinfo {volume} {2}},\ \bibinfo {pages} {040315} (\bibinfo {year} {2021}{\natexlab{b}})}\BibitemShut {NoStop}%
\bibitem [{\citenamefont {Duivenvoorden}\ \emph {et~al.}(2017)\citenamefont {Duivenvoorden}, \citenamefont {Terhal},\ and\ \citenamefont {Weigand}}]{duivenvoorden_single-mode_2017}%
  \BibitemOpen
  \bibfield  {author} {\bibinfo {author} {\bibfnamefont {K.}~\bibnamefont {Duivenvoorden}}, \bibinfo {author} {\bibfnamefont {B.~M.}\ \bibnamefont {Terhal}},\ and\ \bibinfo {author} {\bibfnamefont {D.}~\bibnamefont {Weigand}},\ }\bibfield  {title} {\bibinfo {title} {Single-mode displacement sensor},\ }\href {https://doi.org/10.1103/PhysRevA.95.012305} {\bibfield  {journal} {\bibinfo  {journal} {Physical Review A}\ }\textbf {\bibinfo {volume} {95}},\ \bibinfo {pages} {012305} (\bibinfo {year} {2017})}\BibitemShut {NoStop}%
\bibitem [{\citenamefont {Weedbrook}\ \emph {et~al.}(2012)\citenamefont {Weedbrook}, \citenamefont {Pirandola}, \citenamefont {{Garc{\'i}a-Patr{\'o}n}}, \citenamefont {Cerf}, \citenamefont {Ralph}, \citenamefont {Shapiro},\ and\ \citenamefont {Lloyd}}]{weedbrook_gaussian_2012}%
  \BibitemOpen
  \bibfield  {author} {\bibinfo {author} {\bibfnamefont {C.}~\bibnamefont {Weedbrook}}, \bibinfo {author} {\bibfnamefont {S.}~\bibnamefont {Pirandola}}, \bibinfo {author} {\bibfnamefont {R.}~\bibnamefont {{Garc{\'i}a-Patr{\'o}n}}}, \bibinfo {author} {\bibfnamefont {N.~J.}\ \bibnamefont {Cerf}}, \bibinfo {author} {\bibfnamefont {T.~C.}\ \bibnamefont {Ralph}}, \bibinfo {author} {\bibfnamefont {J.~H.}\ \bibnamefont {Shapiro}},\ and\ \bibinfo {author} {\bibfnamefont {S.}~\bibnamefont {Lloyd}},\ }\bibfield  {title} {\bibinfo {title} {Gaussian quantum information},\ }\href {https://doi.org/10.1103/RevModPhys.84.621} {\bibfield  {journal} {\bibinfo  {journal} {Reviews of Modern Physics}\ }\textbf {\bibinfo {volume} {84}},\ \bibinfo {pages} {621} (\bibinfo {year} {2012})}\BibitemShut {NoStop}%
\bibitem [{\citenamefont {Serafini}(2017)}]{serafini_quantum_2017}%
  \BibitemOpen
  \bibfield  {author} {\bibinfo {author} {\bibfnamefont {A.}~\bibnamefont {Serafini}},\ }\href@noop {} {\emph {\bibinfo {title} {Quantum Continuous Variables: A Primer of Theoretical Methods}}}\ (\bibinfo  {publisher} {CRC Press, Taylor \& Francis Group},\ \bibinfo {address} {Boca Raton},\ \bibinfo {year} {2017})\BibitemShut {NoStop}%
\bibitem [{\citenamefont {Brask}(2022)}]{brask_gaussian_2022}%
  \BibitemOpen
  \bibfield  {author} {\bibinfo {author} {\bibfnamefont {J.~B.}\ \bibnamefont {Brask}},\ }\href@noop {} {\bibinfo {title} {Gaussian states and operations -- a quick reference}} (\bibinfo {year} {2022}),\ \Eprint {https://arxiv.org/abs/2102.05748} {arXiv:2102.05748 [quant-ph]} \BibitemShut {NoStop}%
\bibitem [{\citenamefont {Hastrup}\ \emph {et~al.}(2021)\citenamefont {Hastrup}, \citenamefont {Park}, \citenamefont {Brask}, \citenamefont {Filip},\ and\ \citenamefont {Andersen}}]{hastrup_measurement-free_2021}%
  \BibitemOpen
  \bibfield  {author} {\bibinfo {author} {\bibfnamefont {J.}~\bibnamefont {Hastrup}}, \bibinfo {author} {\bibfnamefont {K.}~\bibnamefont {Park}}, \bibinfo {author} {\bibfnamefont {J.~B.}\ \bibnamefont {Brask}}, \bibinfo {author} {\bibfnamefont {R.}~\bibnamefont {Filip}},\ and\ \bibinfo {author} {\bibfnamefont {U.~L.}\ \bibnamefont {Andersen}},\ }\bibfield  {title} {\bibinfo {title} {Measurement-free preparation of grid states},\ }\href {https://doi.org/10.1038/s41534-020-00353-3} {\bibfield  {journal} {\bibinfo  {journal} {npj Quantum Information}\ }\textbf {\bibinfo {volume} {7}},\ \bibinfo {pages} {17} (\bibinfo {year} {2021})}\BibitemShut {NoStop}%
\bibitem [{\citenamefont {Grimsmo}\ and\ \citenamefont {Puri}(2021)}]{grimsmo_quantum_2021}%
  \BibitemOpen
  \bibfield  {author} {\bibinfo {author} {\bibfnamefont {A.~L.}\ \bibnamefont {Grimsmo}}\ and\ \bibinfo {author} {\bibfnamefont {S.}~\bibnamefont {Puri}},\ }\bibfield  {title} {\bibinfo {title} {Quantum {{Error Correction}} with the {{Gottesman-Kitaev-Preskill Code}}},\ }\href {https://doi.org/10.1103/PRXQuantum.2.020101} {\bibfield  {journal} {\bibinfo  {journal} {PRX Quantum}\ }\textbf {\bibinfo {volume} {2}},\ \bibinfo {pages} {020101} (\bibinfo {year} {2021})}\BibitemShut {NoStop}%
\bibitem [{\citenamefont {Mensen}\ \emph {et~al.}(2021)\citenamefont {Mensen}, \citenamefont {Baragiola},\ and\ \citenamefont {Menicucci}}]{mensen_phase-space_2021}%
  \BibitemOpen
  \bibfield  {author} {\bibinfo {author} {\bibfnamefont {L.~J.}\ \bibnamefont {Mensen}}, \bibinfo {author} {\bibfnamefont {B.~Q.}\ \bibnamefont {Baragiola}},\ and\ \bibinfo {author} {\bibfnamefont {N.~C.}\ \bibnamefont {Menicucci}},\ }\bibfield  {title} {\bibinfo {title} {Phase-space methods for representing, manipulating, and correcting {{Gottesman-Kitaev-Preskill}} qubits},\ }\href {https://doi.org/10.1103/PhysRevA.104.022408} {\bibfield  {journal} {\bibinfo  {journal} {Physical Review A}\ }\textbf {\bibinfo {volume} {104}},\ \bibinfo {pages} {022408} (\bibinfo {year} {2021})}\BibitemShut {NoStop}%
\bibitem [{\citenamefont {Walshe}\ \emph {et~al.}(2020)\citenamefont {Walshe}, \citenamefont {Baragiola}, \citenamefont {Alexander},\ and\ \citenamefont {Menicucci}}]{walshe_continuous-variable_2020}%
  \BibitemOpen
  \bibfield  {author} {\bibinfo {author} {\bibfnamefont {B.~W.}\ \bibnamefont {Walshe}}, \bibinfo {author} {\bibfnamefont {B.~Q.}\ \bibnamefont {Baragiola}}, \bibinfo {author} {\bibfnamefont {R.~N.}\ \bibnamefont {Alexander}},\ and\ \bibinfo {author} {\bibfnamefont {N.~C.}\ \bibnamefont {Menicucci}},\ }\bibfield  {title} {\bibinfo {title} {Continuous-variable gate teleportation and bosonic-code error correction},\ }\href {https://doi.org/10.1103/PhysRevA.102.062411} {\bibfield  {journal} {\bibinfo  {journal} {Physical Review A}\ }\textbf {\bibinfo {volume} {102}},\ \bibinfo {pages} {062411} (\bibinfo {year} {2020})}\BibitemShut {NoStop}%
\bibitem [{\citenamefont {Hastrup}\ and\ \citenamefont {Andersen}(2023)}]{hastrup_analysis_2023}%
  \BibitemOpen
  \bibfield  {author} {\bibinfo {author} {\bibfnamefont {J.}~\bibnamefont {Hastrup}}\ and\ \bibinfo {author} {\bibfnamefont {U.~L.}\ \bibnamefont {Andersen}},\ }\bibfield  {title} {\bibinfo {title} {Analysis of loss correction with the {{Gottesman-Kitaev-Preskill}} code},\ }\href {https://doi.org/10.1103/PhysRevA.108.052413} {\bibfield  {journal} {\bibinfo  {journal} {Physical Review A}\ }\textbf {\bibinfo {volume} {108}},\ \bibinfo {pages} {052413} (\bibinfo {year} {2023})}\BibitemShut {NoStop}%
\bibitem [{\citenamefont {Tzitrin}\ \emph {et~al.}(2021)\citenamefont {Tzitrin}, \citenamefont {Matsuura}, \citenamefont {Alexander}, \citenamefont {Dauphinais}, \citenamefont {Bourassa}, \citenamefont {Sabapathy}, \citenamefont {Menicucci},\ and\ \citenamefont {Dhand}}]{tzitrin_fault-tolerant_2021}%
  \BibitemOpen
  \bibfield  {author} {\bibinfo {author} {\bibfnamefont {I.}~\bibnamefont {Tzitrin}}, \bibinfo {author} {\bibfnamefont {T.}~\bibnamefont {Matsuura}}, \bibinfo {author} {\bibfnamefont {R.~N.}\ \bibnamefont {Alexander}}, \bibinfo {author} {\bibfnamefont {G.}~\bibnamefont {Dauphinais}}, \bibinfo {author} {\bibfnamefont {J.~E.}\ \bibnamefont {Bourassa}}, \bibinfo {author} {\bibfnamefont {K.~K.}\ \bibnamefont {Sabapathy}}, \bibinfo {author} {\bibfnamefont {N.~C.}\ \bibnamefont {Menicucci}},\ and\ \bibinfo {author} {\bibfnamefont {I.}~\bibnamefont {Dhand}},\ }\bibfield  {title} {\bibinfo {title} {Fault-{{Tolerant Quantum Computation}} with {{Static Linear Optics}}},\ }\href {https://doi.org/10.1103/PRXQuantum.2.040353} {\bibfield  {journal} {\bibinfo  {journal} {PRX Quantum}\ }\textbf {\bibinfo {volume} {2}},\ \bibinfo {pages} {040353} (\bibinfo {year} {2021})}\BibitemShut {NoStop}%
\bibitem [{\citenamefont {Marek}(2024)}]{marek_ground_2024}%
  \BibitemOpen
  \bibfield  {author} {\bibinfo {author} {\bibfnamefont {P.}~\bibnamefont {Marek}},\ }\bibfield  {title} {\bibinfo {title} {Ground {{State Nature}} and {{Nonlinear Squeezing}} of {{Gottesman-Kitaev-Preskill States}}},\ }\href {https://doi.org/10.1103/PhysRevLett.132.210601} {\bibfield  {journal} {\bibinfo  {journal} {Physical Review Letters}\ }\textbf {\bibinfo {volume} {132}},\ \bibinfo {pages} {210601} (\bibinfo {year} {2024})}\BibitemShut {NoStop}%
\bibitem [{\citenamefont {Hahn}\ \emph {et~al.}(2022)\citenamefont {Hahn}, \citenamefont {Holmvall}, \citenamefont {Stadler}, \citenamefont {Ferrini},\ and\ \citenamefont {Ferraro}}]{hahn_deterministic_2022}%
  \BibitemOpen
  \bibfield  {author} {\bibinfo {author} {\bibfnamefont {O.}~\bibnamefont {Hahn}}, \bibinfo {author} {\bibfnamefont {P.}~\bibnamefont {Holmvall}}, \bibinfo {author} {\bibfnamefont {P.}~\bibnamefont {Stadler}}, \bibinfo {author} {\bibfnamefont {G.}~\bibnamefont {Ferrini}},\ and\ \bibinfo {author} {\bibfnamefont {A.}~\bibnamefont {Ferraro}},\ }\bibfield  {title} {\bibinfo {title} {Deterministic {{Gaussian}} conversion protocols for non-{{Gaussian}} single-mode resources},\ }\href {https://doi.org/10.1103/PhysRevA.105.062446} {\bibfield  {journal} {\bibinfo  {journal} {Physical Review A}\ }\textbf {\bibinfo {volume} {105}},\ \bibinfo {pages} {062446} (\bibinfo {year} {2022})}\BibitemShut {NoStop}%
\bibitem [{\citenamefont {Laghaout}\ \emph {et~al.}(2013)\citenamefont {Laghaout}, \citenamefont {{Neergaard-Nielsen}}, \citenamefont {Rigas}, \citenamefont {Kragh}, \citenamefont {Tipsmark},\ and\ \citenamefont {Andersen}}]{laghaout_amplification_2013}%
  \BibitemOpen
  \bibfield  {author} {\bibinfo {author} {\bibfnamefont {A.}~\bibnamefont {Laghaout}}, \bibinfo {author} {\bibfnamefont {J.~S.}\ \bibnamefont {{Neergaard-Nielsen}}}, \bibinfo {author} {\bibfnamefont {I.}~\bibnamefont {Rigas}}, \bibinfo {author} {\bibfnamefont {C.}~\bibnamefont {Kragh}}, \bibinfo {author} {\bibfnamefont {A.}~\bibnamefont {Tipsmark}},\ and\ \bibinfo {author} {\bibfnamefont {U.~L.}\ \bibnamefont {Andersen}},\ }\bibfield  {title} {\bibinfo {title} {Amplification of realistic {{Schr{\"o}dinger-cat-state-like}} states by homodyne heralding},\ }\href {https://doi.org/10.1103/PhysRevA.87.043826} {\bibfield  {journal} {\bibinfo  {journal} {Physical Review A}\ }\textbf {\bibinfo {volume} {87}},\ \bibinfo {pages} {043826} (\bibinfo {year} {2013})}\BibitemShut {NoStop}%
\bibitem [{\citenamefont {Sychev}\ \emph {et~al.}(2017)\citenamefont {Sychev}, \citenamefont {Ulanov}, \citenamefont {Pushkina}, \citenamefont {Richards}, \citenamefont {Fedorov},\ and\ \citenamefont {Lvovsky}}]{sychev_enlargement_2017}%
  \BibitemOpen
  \bibfield  {author} {\bibinfo {author} {\bibfnamefont {D.~V.}\ \bibnamefont {Sychev}}, \bibinfo {author} {\bibfnamefont {A.~E.}\ \bibnamefont {Ulanov}}, \bibinfo {author} {\bibfnamefont {A.~A.}\ \bibnamefont {Pushkina}}, \bibinfo {author} {\bibfnamefont {M.~W.}\ \bibnamefont {Richards}}, \bibinfo {author} {\bibfnamefont {I.~A.}\ \bibnamefont {Fedorov}},\ and\ \bibinfo {author} {\bibfnamefont {A.~I.}\ \bibnamefont {Lvovsky}},\ }\bibfield  {title} {\bibinfo {title} {Enlargement of optical {{Schr{\"o}dinger}}'s cat states},\ }\href {https://doi.org/10.1038/nphoton.2017.57} {\bibfield  {journal} {\bibinfo  {journal} {Nature Photonics}\ }\textbf {\bibinfo {volume} {11}},\ \bibinfo {pages} {379} (\bibinfo {year} {2017})}\BibitemShut {NoStop}%
\bibitem [{\citenamefont {Vahlbruch}\ \emph {et~al.}(2016)\citenamefont {Vahlbruch}, \citenamefont {Mehmet}, \citenamefont {Danzmann},\ and\ \citenamefont {Schnabel}}]{vahlbruch_detection_2016}%
  \BibitemOpen
  \bibfield  {author} {\bibinfo {author} {\bibfnamefont {H.}~\bibnamefont {Vahlbruch}}, \bibinfo {author} {\bibfnamefont {M.}~\bibnamefont {Mehmet}}, \bibinfo {author} {\bibfnamefont {K.}~\bibnamefont {Danzmann}},\ and\ \bibinfo {author} {\bibfnamefont {R.}~\bibnamefont {Schnabel}},\ }\bibfield  {title} {\bibinfo {title} {Detection of 15 {{dB Squeezed States}} of {{Light}} and their {{Application}} for the {{Absolute Calibration}} of {{Photoelectric Quantum Efficiency}}},\ }\href {https://doi.org/10.1103/PhysRevLett.117.110801} {\bibfield  {journal} {\bibinfo  {journal} {Physical Review Letters}\ }\textbf {\bibinfo {volume} {117}},\ \bibinfo {pages} {110801} (\bibinfo {year} {2016})}\BibitemShut {NoStop}%
\bibitem [{\citenamefont {Virtanen}\ \emph {et~al.}(2020)\citenamefont {Virtanen}, \citenamefont {Gommers}, \citenamefont {Oliphant}, \citenamefont {Haberland}, \citenamefont {Reddy}, \citenamefont {Cournapeau}, \citenamefont {Burovski}, \citenamefont {Peterson}, \citenamefont {Weckesser}, \citenamefont {Bright}, \citenamefont {Van Der~Walt}, \citenamefont {Brett}, \citenamefont {Wilson}, \citenamefont {Millman}, \citenamefont {Mayorov}, \citenamefont {Nelson}, \citenamefont {Jones}, \citenamefont {Kern}, \citenamefont {Larson}, \citenamefont {Carey}, \citenamefont {Polat}, \citenamefont {Feng}, \citenamefont {Moore}, \citenamefont {VanderPlas}, \citenamefont {Laxalde}, \citenamefont {Perktold}, \citenamefont {Cimrman}, \citenamefont {Henriksen}, \citenamefont {Quintero}, \citenamefont {Harris}, \citenamefont {Archibald}, \citenamefont {Ribeiro}, \citenamefont {Pedregosa}, \citenamefont {Van~Mulbregt}, \citenamefont {{SciPy 1.0 Contributors}}, \citenamefont {Vijaykumar}, \citenamefont {Bardelli},
  \citenamefont {Rothberg}, \citenamefont {Hilboll}, \citenamefont {Kloeckner}, \citenamefont {Scopatz}, \citenamefont {Lee}, \citenamefont {Rokem}, \citenamefont {Woods}, \citenamefont {Fulton}, \citenamefont {Masson}, \citenamefont {H{\"a}ggstr{\"o}m}, \citenamefont {Fitzgerald}, \citenamefont {Nicholson}, \citenamefont {Hagen}, \citenamefont {Pasechnik}, \citenamefont {Olivetti}, \citenamefont {Martin}, \citenamefont {Wieser}, \citenamefont {Silva}, \citenamefont {Lenders}, \citenamefont {Wilhelm}, \citenamefont {Young}, \citenamefont {Price}, \citenamefont {Ingold}, \citenamefont {Allen}, \citenamefont {Lee}, \citenamefont {Audren}, \citenamefont {Probst}, \citenamefont {Dietrich}, \citenamefont {Silterra}, \citenamefont {Webber}, \citenamefont {Slavi{\v c}}, \citenamefont {Nothman}, \citenamefont {Buchner}, \citenamefont {Kulick}, \citenamefont {Sch{\"o}nberger}, \citenamefont {De~Miranda~Cardoso}, \citenamefont {Reimer}, \citenamefont {Harrington}, \citenamefont {Rodr{\'i}guez}, \citenamefont
  {{Nunez-Iglesias}}, \citenamefont {Kuczynski}, \citenamefont {Tritz}, \citenamefont {Thoma}, \citenamefont {Newville}, \citenamefont {K{\"u}mmerer}, \citenamefont {Bolingbroke}, \citenamefont {Tartre}, \citenamefont {Pak}, \citenamefont {Smith}, \citenamefont {Nowaczyk}, \citenamefont {Shebanov}, \citenamefont {Pavlyk}, \citenamefont {Brodtkorb}, \citenamefont {Lee}, \citenamefont {McGibbon}, \citenamefont {Feldbauer}, \citenamefont {Lewis}, \citenamefont {Tygier}, \citenamefont {Sievert}, \citenamefont {Vigna}, \citenamefont {Peterson}, \citenamefont {More}, \citenamefont {Pudlik}, \citenamefont {Oshima}, \citenamefont {Pingel}, \citenamefont {Robitaille}, \citenamefont {Spura}, \citenamefont {Jones}, \citenamefont {Cera}, \citenamefont {Leslie}, \citenamefont {Zito}, \citenamefont {Krauss}, \citenamefont {Upadhyay}, \citenamefont {Halchenko},\ and\ \citenamefont {{V{\'a}zquez-Baeza}}}]{virtanen_scipy_2020}%
  \BibitemOpen
  \bibfield  {author} {\bibinfo {author} {\bibfnamefont {P.}~\bibnamefont {Virtanen}}, \bibinfo {author} {\bibfnamefont {R.}~\bibnamefont {Gommers}}, \bibinfo {author} {\bibfnamefont {T.~E.}\ \bibnamefont {Oliphant}}, \bibinfo {author} {\bibfnamefont {M.}~\bibnamefont {Haberland}}, \bibinfo {author} {\bibfnamefont {T.}~\bibnamefont {Reddy}}, \bibinfo {author} {\bibfnamefont {D.}~\bibnamefont {Cournapeau}}, \bibinfo {author} {\bibfnamefont {E.}~\bibnamefont {Burovski}}, \bibinfo {author} {\bibfnamefont {P.}~\bibnamefont {Peterson}}, \bibinfo {author} {\bibfnamefont {W.}~\bibnamefont {Weckesser}}, \bibinfo {author} {\bibfnamefont {J.}~\bibnamefont {Bright}}, \bibinfo {author} {\bibfnamefont {S.~J.}\ \bibnamefont {Van Der~Walt}}, \bibinfo {author} {\bibfnamefont {M.}~\bibnamefont {Brett}}, \bibinfo {author} {\bibfnamefont {J.}~\bibnamefont {Wilson}}, \bibinfo {author} {\bibfnamefont {K.~J.}\ \bibnamefont {Millman}}, \bibinfo {author} {\bibfnamefont {N.}~\bibnamefont {Mayorov}}, \bibinfo {author} {\bibfnamefont
  {A.~R.~J.}\ \bibnamefont {Nelson}}, \bibinfo {author} {\bibfnamefont {E.}~\bibnamefont {Jones}}, \bibinfo {author} {\bibfnamefont {R.}~\bibnamefont {Kern}}, \bibinfo {author} {\bibfnamefont {E.}~\bibnamefont {Larson}}, \bibinfo {author} {\bibfnamefont {C.~J.}\ \bibnamefont {Carey}}, \bibinfo {author} {\bibfnamefont {{\.I}.}~\bibnamefont {Polat}}, \bibinfo {author} {\bibfnamefont {Y.}~\bibnamefont {Feng}}, \bibinfo {author} {\bibfnamefont {E.~W.}\ \bibnamefont {Moore}}, \bibinfo {author} {\bibfnamefont {J.}~\bibnamefont {VanderPlas}}, \bibinfo {author} {\bibfnamefont {D.}~\bibnamefont {Laxalde}}, \bibinfo {author} {\bibfnamefont {J.}~\bibnamefont {Perktold}}, \bibinfo {author} {\bibfnamefont {R.}~\bibnamefont {Cimrman}}, \bibinfo {author} {\bibfnamefont {I.}~\bibnamefont {Henriksen}}, \bibinfo {author} {\bibfnamefont {E.~A.}\ \bibnamefont {Quintero}}, \bibinfo {author} {\bibfnamefont {C.~R.}\ \bibnamefont {Harris}}, \bibinfo {author} {\bibfnamefont {A.~M.}\ \bibnamefont {Archibald}}, \bibinfo {author}
  {\bibfnamefont {A.~H.}\ \bibnamefont {Ribeiro}}, \bibinfo {author} {\bibfnamefont {F.}~\bibnamefont {Pedregosa}}, \bibinfo {author} {\bibfnamefont {P.}~\bibnamefont {Van~Mulbregt}}, \bibinfo {author} {\bibnamefont {{SciPy 1.0 Contributors}}}, \bibinfo {author} {\bibfnamefont {A.}~\bibnamefont {Vijaykumar}}, \bibinfo {author} {\bibfnamefont {A.~P.}\ \bibnamefont {Bardelli}}, \bibinfo {author} {\bibfnamefont {A.}~\bibnamefont {Rothberg}}, \bibinfo {author} {\bibfnamefont {A.}~\bibnamefont {Hilboll}}, \bibinfo {author} {\bibfnamefont {A.}~\bibnamefont {Kloeckner}}, \bibinfo {author} {\bibfnamefont {A.}~\bibnamefont {Scopatz}}, \bibinfo {author} {\bibfnamefont {A.}~\bibnamefont {Lee}}, \bibinfo {author} {\bibfnamefont {A.}~\bibnamefont {Rokem}}, \bibinfo {author} {\bibfnamefont {C.~N.}\ \bibnamefont {Woods}}, \bibinfo {author} {\bibfnamefont {C.}~\bibnamefont {Fulton}}, \bibinfo {author} {\bibfnamefont {C.}~\bibnamefont {Masson}}, \bibinfo {author} {\bibfnamefont {C.}~\bibnamefont {H{\"a}ggstr{\"o}m}}, \bibinfo
  {author} {\bibfnamefont {C.}~\bibnamefont {Fitzgerald}}, \bibinfo {author} {\bibfnamefont {D.~A.}\ \bibnamefont {Nicholson}}, \bibinfo {author} {\bibfnamefont {D.~R.}\ \bibnamefont {Hagen}}, \bibinfo {author} {\bibfnamefont {D.~V.}\ \bibnamefont {Pasechnik}}, \bibinfo {author} {\bibfnamefont {E.}~\bibnamefont {Olivetti}}, \bibinfo {author} {\bibfnamefont {E.}~\bibnamefont {Martin}}, \bibinfo {author} {\bibfnamefont {E.}~\bibnamefont {Wieser}}, \bibinfo {author} {\bibfnamefont {F.}~\bibnamefont {Silva}}, \bibinfo {author} {\bibfnamefont {F.}~\bibnamefont {Lenders}}, \bibinfo {author} {\bibfnamefont {F.}~\bibnamefont {Wilhelm}}, \bibinfo {author} {\bibfnamefont {G.}~\bibnamefont {Young}}, \bibinfo {author} {\bibfnamefont {G.~A.}\ \bibnamefont {Price}}, \bibinfo {author} {\bibfnamefont {G.-L.}\ \bibnamefont {Ingold}}, \bibinfo {author} {\bibfnamefont {G.~E.}\ \bibnamefont {Allen}}, \bibinfo {author} {\bibfnamefont {G.~R.}\ \bibnamefont {Lee}}, \bibinfo {author} {\bibfnamefont {H.}~\bibnamefont {Audren}},
  \bibinfo {author} {\bibfnamefont {I.}~\bibnamefont {Probst}}, \bibinfo {author} {\bibfnamefont {J.~P.}\ \bibnamefont {Dietrich}}, \bibinfo {author} {\bibfnamefont {J.}~\bibnamefont {Silterra}}, \bibinfo {author} {\bibfnamefont {J.~T.}\ \bibnamefont {Webber}}, \bibinfo {author} {\bibfnamefont {J.}~\bibnamefont {Slavi{\v c}}}, \bibinfo {author} {\bibfnamefont {J.}~\bibnamefont {Nothman}}, \bibinfo {author} {\bibfnamefont {J.}~\bibnamefont {Buchner}}, \bibinfo {author} {\bibfnamefont {J.}~\bibnamefont {Kulick}}, \bibinfo {author} {\bibfnamefont {J.~L.}\ \bibnamefont {Sch{\"o}nberger}}, \bibinfo {author} {\bibfnamefont {J.~V.}\ \bibnamefont {De~Miranda~Cardoso}}, \bibinfo {author} {\bibfnamefont {J.}~\bibnamefont {Reimer}}, \bibinfo {author} {\bibfnamefont {J.}~\bibnamefont {Harrington}}, \bibinfo {author} {\bibfnamefont {J.~L.~C.}\ \bibnamefont {Rodr{\'i}guez}}, \bibinfo {author} {\bibfnamefont {J.}~\bibnamefont {{Nunez-Iglesias}}}, \bibinfo {author} {\bibfnamefont {J.}~\bibnamefont {Kuczynski}}, \bibinfo
  {author} {\bibfnamefont {K.}~\bibnamefont {Tritz}}, \bibinfo {author} {\bibfnamefont {M.}~\bibnamefont {Thoma}}, \bibinfo {author} {\bibfnamefont {M.}~\bibnamefont {Newville}}, \bibinfo {author} {\bibfnamefont {M.}~\bibnamefont {K{\"u}mmerer}}, \bibinfo {author} {\bibfnamefont {M.}~\bibnamefont {Bolingbroke}}, \bibinfo {author} {\bibfnamefont {M.}~\bibnamefont {Tartre}}, \bibinfo {author} {\bibfnamefont {M.}~\bibnamefont {Pak}}, \bibinfo {author} {\bibfnamefont {N.~J.}\ \bibnamefont {Smith}}, \bibinfo {author} {\bibfnamefont {N.}~\bibnamefont {Nowaczyk}}, \bibinfo {author} {\bibfnamefont {N.}~\bibnamefont {Shebanov}}, \bibinfo {author} {\bibfnamefont {O.}~\bibnamefont {Pavlyk}}, \bibinfo {author} {\bibfnamefont {P.~A.}\ \bibnamefont {Brodtkorb}}, \bibinfo {author} {\bibfnamefont {P.}~\bibnamefont {Lee}}, \bibinfo {author} {\bibfnamefont {R.~T.}\ \bibnamefont {McGibbon}}, \bibinfo {author} {\bibfnamefont {R.}~\bibnamefont {Feldbauer}}, \bibinfo {author} {\bibfnamefont {S.}~\bibnamefont {Lewis}}, \bibinfo
  {author} {\bibfnamefont {S.}~\bibnamefont {Tygier}}, \bibinfo {author} {\bibfnamefont {S.}~\bibnamefont {Sievert}}, \bibinfo {author} {\bibfnamefont {S.}~\bibnamefont {Vigna}}, \bibinfo {author} {\bibfnamefont {S.}~\bibnamefont {Peterson}}, \bibinfo {author} {\bibfnamefont {S.}~\bibnamefont {More}}, \bibinfo {author} {\bibfnamefont {T.}~\bibnamefont {Pudlik}}, \bibinfo {author} {\bibfnamefont {T.}~\bibnamefont {Oshima}}, \bibinfo {author} {\bibfnamefont {T.~J.}\ \bibnamefont {Pingel}}, \bibinfo {author} {\bibfnamefont {T.~P.}\ \bibnamefont {Robitaille}}, \bibinfo {author} {\bibfnamefont {T.}~\bibnamefont {Spura}}, \bibinfo {author} {\bibfnamefont {T.~R.}\ \bibnamefont {Jones}}, \bibinfo {author} {\bibfnamefont {T.}~\bibnamefont {Cera}}, \bibinfo {author} {\bibfnamefont {T.}~\bibnamefont {Leslie}}, \bibinfo {author} {\bibfnamefont {T.}~\bibnamefont {Zito}}, \bibinfo {author} {\bibfnamefont {T.}~\bibnamefont {Krauss}}, \bibinfo {author} {\bibfnamefont {U.}~\bibnamefont {Upadhyay}}, \bibinfo {author}
  {\bibfnamefont {Y.~O.}\ \bibnamefont {Halchenko}},\ and\ \bibinfo {author} {\bibfnamefont {Y.}~\bibnamefont {{V{\'a}zquez-Baeza}}},\ }\bibfield  {title} {\bibinfo {title} {{{SciPy}} 1.0: Fundamental algorithms for scientific computing in {{Python}}},\ }\href {https://doi.org/10.1038/s41592-019-0686-2} {\bibfield  {journal} {\bibinfo  {journal} {Nature Methods}\ }\textbf {\bibinfo {volume} {17}},\ \bibinfo {pages} {261} (\bibinfo {year} {2020})}\BibitemShut {NoStop}%
\bibitem [{\citenamefont {Marshall}\ and\ \citenamefont {Anand}(2023)}]{marshall_simulation_2023}%
  \BibitemOpen
  \bibfield  {author} {\bibinfo {author} {\bibfnamefont {J.}~\bibnamefont {Marshall}}\ and\ \bibinfo {author} {\bibfnamefont {N.}~\bibnamefont {Anand}},\ }\bibfield  {title} {\bibinfo {title} {Simulation of quantum optics by coherent state decomposition},\ }\href {https://doi.org/10.1364/OPTICAQ.504311} {\bibfield  {journal} {\bibinfo  {journal} {Optica Quantum}\ }\textbf {\bibinfo {volume} {1}},\ \bibinfo {pages} {78} (\bibinfo {year} {2023})}\BibitemShut {NoStop}%
\bibitem [{\citenamefont {Zheng}\ \emph {et~al.}(2023)\citenamefont {Zheng}, \citenamefont {Ferraro}, \citenamefont {Kockum},\ and\ \citenamefont {Ferrini}}]{zheng_gaussian_2023}%
  \BibitemOpen
  \bibfield  {author} {\bibinfo {author} {\bibfnamefont {Y.}~\bibnamefont {Zheng}}, \bibinfo {author} {\bibfnamefont {A.}~\bibnamefont {Ferraro}}, \bibinfo {author} {\bibfnamefont {A.~F.}\ \bibnamefont {Kockum}},\ and\ \bibinfo {author} {\bibfnamefont {G.}~\bibnamefont {Ferrini}},\ }\bibfield  {title} {\bibinfo {title} {Gaussian conversion protocol for heralded generation of generalized {{Gottesman-Kitaev-Preskill}} states},\ }\href {https://doi.org/10.1103/PhysRevA.108.012603} {\bibfield  {journal} {\bibinfo  {journal} {Physical Review A}\ }\textbf {\bibinfo {volume} {108}},\ \bibinfo {pages} {012603} (\bibinfo {year} {2023})}\BibitemShut {NoStop}%
\bibitem [{\citenamefont {Chabaud}\ \emph {et~al.}(2020)\citenamefont {Chabaud}, \citenamefont {Markham},\ and\ \citenamefont {Grosshans}}]{chabaud_stellar_2020}%
  \BibitemOpen
  \bibfield  {author} {\bibinfo {author} {\bibfnamefont {U.}~\bibnamefont {Chabaud}}, \bibinfo {author} {\bibfnamefont {D.}~\bibnamefont {Markham}},\ and\ \bibinfo {author} {\bibfnamefont {F.}~\bibnamefont {Grosshans}},\ }\bibfield  {title} {\bibinfo {title} {Stellar {{Representation}} of {{Non-Gaussian Quantum States}}},\ }\href {https://doi.org/10.1103/PhysRevLett.124.063605} {\bibfield  {journal} {\bibinfo  {journal} {Physical Review Letters}\ }\textbf {\bibinfo {volume} {124}},\ \bibinfo {pages} {063605} (\bibinfo {year} {2020})}\BibitemShut {NoStop}%
\bibitem [{\citenamefont {Solodovnikova}\ \emph {et~al.}(2025)\citenamefont {Solodovnikova}, \citenamefont {Andersen},\ and\ \citenamefont {{Neergaard-Nielsen}}}]{solodovnikova_fast_2025}%
  \BibitemOpen
  \bibfield  {author} {\bibinfo {author} {\bibfnamefont {O.}~\bibnamefont {Solodovnikova}}, \bibinfo {author} {\bibfnamefont {U.~L.}\ \bibnamefont {Andersen}},\ and\ \bibinfo {author} {\bibfnamefont {J.~S.}\ \bibnamefont {{Neergaard-Nielsen}}},\ }\href@noop {} {\bibinfo {title} {Fast simulations of continuous-variable circuits using the coherent state decomposition}} (\bibinfo {year} {2025})\BibitemShut {NoStop}%
\bibitem [{\citenamefont {Br{\"a}uer}\ \emph {et~al.}(2025)\citenamefont {Br{\"a}uer}, \citenamefont {Provazn{\'i}k}, \citenamefont {Kala},\ and\ \citenamefont {Marek}}]{brauer_catability_2025}%
  \BibitemOpen
  \bibfield  {author} {\bibinfo {author} {\bibfnamefont {{\v S}.}~\bibnamefont {Br{\"a}uer}}, \bibinfo {author} {\bibfnamefont {J.}~\bibnamefont {Provazn{\'i}k}}, \bibinfo {author} {\bibfnamefont {V.}~\bibnamefont {Kala}},\ and\ \bibinfo {author} {\bibfnamefont {P.}~\bibnamefont {Marek}},\ }\href {https://doi.org/10.48550/arXiv.2505.19723} {\bibinfo {title} {Catability as a metric for evaluating superposed coherent states}} (\bibinfo {year} {2025}),\ \Eprint {https://arxiv.org/abs/2505.19723} {arXiv:2505.19723 [quant-ph]} \BibitemShut {NoStop}%
\bibitem [{\citenamefont {Harris}\ \emph {et~al.}(2020)\citenamefont {Harris}, \citenamefont {Millman}, \citenamefont {Van Der~Walt}, \citenamefont {Gommers}, \citenamefont {Virtanen}, \citenamefont {Cournapeau}, \citenamefont {Wieser}, \citenamefont {Taylor}, \citenamefont {Berg}, \citenamefont {Smith}, \citenamefont {Kern}, \citenamefont {Picus}, \citenamefont {Hoyer}, \citenamefont {Van~Kerkwijk}, \citenamefont {Brett}, \citenamefont {Haldane}, \citenamefont {Del~R{\'i}o}, \citenamefont {Wiebe}, \citenamefont {Peterson}, \citenamefont {{G{\'e}rard-Marchant}}, \citenamefont {Sheppard}, \citenamefont {Reddy}, \citenamefont {Weckesser}, \citenamefont {Abbasi}, \citenamefont {Gohlke},\ and\ \citenamefont {Oliphant}}]{harris_array_2020}%
  \BibitemOpen
  \bibfield  {author} {\bibinfo {author} {\bibfnamefont {C.~R.}\ \bibnamefont {Harris}}, \bibinfo {author} {\bibfnamefont {K.~J.}\ \bibnamefont {Millman}}, \bibinfo {author} {\bibfnamefont {S.~J.}\ \bibnamefont {Van Der~Walt}}, \bibinfo {author} {\bibfnamefont {R.}~\bibnamefont {Gommers}}, \bibinfo {author} {\bibfnamefont {P.}~\bibnamefont {Virtanen}}, \bibinfo {author} {\bibfnamefont {D.}~\bibnamefont {Cournapeau}}, \bibinfo {author} {\bibfnamefont {E.}~\bibnamefont {Wieser}}, \bibinfo {author} {\bibfnamefont {J.}~\bibnamefont {Taylor}}, \bibinfo {author} {\bibfnamefont {S.}~\bibnamefont {Berg}}, \bibinfo {author} {\bibfnamefont {N.~J.}\ \bibnamefont {Smith}}, \bibinfo {author} {\bibfnamefont {R.}~\bibnamefont {Kern}}, \bibinfo {author} {\bibfnamefont {M.}~\bibnamefont {Picus}}, \bibinfo {author} {\bibfnamefont {S.}~\bibnamefont {Hoyer}}, \bibinfo {author} {\bibfnamefont {M.~H.}\ \bibnamefont {Van~Kerkwijk}}, \bibinfo {author} {\bibfnamefont {M.}~\bibnamefont {Brett}}, \bibinfo {author} {\bibfnamefont
  {A.}~\bibnamefont {Haldane}}, \bibinfo {author} {\bibfnamefont {J.~F.}\ \bibnamefont {Del~R{\'i}o}}, \bibinfo {author} {\bibfnamefont {M.}~\bibnamefont {Wiebe}}, \bibinfo {author} {\bibfnamefont {P.}~\bibnamefont {Peterson}}, \bibinfo {author} {\bibfnamefont {P.}~\bibnamefont {{G{\'e}rard-Marchant}}}, \bibinfo {author} {\bibfnamefont {K.}~\bibnamefont {Sheppard}}, \bibinfo {author} {\bibfnamefont {T.}~\bibnamefont {Reddy}}, \bibinfo {author} {\bibfnamefont {W.}~\bibnamefont {Weckesser}}, \bibinfo {author} {\bibfnamefont {H.}~\bibnamefont {Abbasi}}, \bibinfo {author} {\bibfnamefont {C.}~\bibnamefont {Gohlke}},\ and\ \bibinfo {author} {\bibfnamefont {T.~E.}\ \bibnamefont {Oliphant}},\ }\bibfield  {title} {\bibinfo {title} {Array programming with {{NumPy}}},\ }\href {https://doi.org/10.1038/s41586-020-2649-2} {\bibfield  {journal} {\bibinfo  {journal} {Nature}\ }\textbf {\bibinfo {volume} {585}},\ \bibinfo {pages} {357} (\bibinfo {year} {2020})}\BibitemShut {NoStop}%
\bibitem [{\citenamefont {Hunter}(2007)}]{hunter_matplotlib_2007}%
  \BibitemOpen
  \bibfield  {author} {\bibinfo {author} {\bibfnamefont {J.~D.}\ \bibnamefont {Hunter}},\ }\bibfield  {title} {\bibinfo {title} {Matplotlib: {{A 2D Graphics Environment}}},\ }\href {https://doi.org/10.1109/mcse.2007.55} {\bibfield  {journal} {\bibinfo  {journal} {Computing in Science \& Engineering}\ }\textbf {\bibinfo {volume} {9}},\ \bibinfo {pages} {90} (\bibinfo {year} {2007})}\BibitemShut {NoStop}%
\bibitem [{\citenamefont {Kluyver}\ \emph {et~al.}(2016)\citenamefont {Kluyver}, \citenamefont {{Ragan-Kelley}}, \citenamefont {P{\'e}rez}, \citenamefont {Granger}, \citenamefont {Bussonnier}, \citenamefont {Frederic}, \citenamefont {Kelley}, \citenamefont {Hamrick}, \citenamefont {Grout}, \citenamefont {Corlay}, \citenamefont {P.Ivanov}, \citenamefont {Avila}, \citenamefont {{S. Abdalla}}, \citenamefont {Willing},\ and\ \citenamefont {{Jupyter Development Team}}}]{kluyver_jupyter_2016}%
  \BibitemOpen
  \bibfield  {author} {\bibinfo {author} {\bibfnamefont {T.}~\bibnamefont {Kluyver}}, \bibinfo {author} {\bibfnamefont {B.}~\bibnamefont {{Ragan-Kelley}}}, \bibinfo {author} {\bibfnamefont {F.}~\bibnamefont {P{\'e}rez}}, \bibinfo {author} {\bibfnamefont {B.}~\bibnamefont {Granger}}, \bibinfo {author} {\bibfnamefont {M.}~\bibnamefont {Bussonnier}}, \bibinfo {author} {\bibfnamefont {J.}~\bibnamefont {Frederic}}, \bibinfo {author} {\bibfnamefont {K.}~\bibnamefont {Kelley}}, \bibinfo {author} {\bibfnamefont {J.}~\bibnamefont {Hamrick}}, \bibinfo {author} {\bibfnamefont {J.}~\bibnamefont {Grout}}, \bibinfo {author} {\bibfnamefont {S.}~\bibnamefont {Corlay}}, \bibinfo {author} {\bibnamefont {P.Ivanov}}, \bibinfo {author} {\bibfnamefont {D.}~\bibnamefont {Avila}}, \bibinfo {author} {\bibnamefont {{S. Abdalla}}}, \bibinfo {author} {\bibfnamefont {C.}~\bibnamefont {Willing}},\ and\ \bibinfo {author} {\bibnamefont {{Jupyter Development Team}}},\ }\bibfield  {title} {\bibinfo {title} {Jupyter {{Notebooks}} - a
  publishing format for reproducible computational workflows},\ }in\ \href {https://doi.org/10.3233/978-1-61499-649-1-87} {\emph {\bibinfo {booktitle} {Positioning and {{Power}} in {{Academic Publishing}}: {{Players}}, {{Agents}} and {{Agendas}}}}},\ \bibinfo {editor} {edited by\ \bibinfo {editor} {\bibfnamefont {F.}~\bibnamefont {Loizides}}\ and\ \bibinfo {editor} {\bibfnamefont {B.}~\bibnamefont {Schmidt}}}\ (\bibinfo  {publisher} {IOS Press},\ \bibinfo {year} {2016})\ pp.\ \bibinfo {pages} {87--90}\BibitemShut {NoStop}%
\end{thebibliography}%
\newpage
\onecolumngrid
\appendix
\section{Symplectic matrix of the beam splitter cascade}\label{app:bs_cascade}
The $N$-mode beam splitter cascade in Fig. \ref{fig:cat_breeding_circuit} transforms the quadrature operators as $\hat{U}^{\dagger}\begin{pmatrix}
\hat{\vb*{x}} \\
\hat{\vb*{p}}
\end{pmatrix}\hat{U}=\begin{pmatrix}
\vb*{B} & \vb*{0} \\
\vb*{0} & \vb*{B}
\end{pmatrix}\begin{pmatrix}
\hat{\vb*{x}} \\
\hat{\vb*{p}}
\end{pmatrix}$
where $\vb*{B}\in \mathds{R}^{N\times N}$:
\[ \vb*{B} =
\begin{pmatrix}
\frac{1}{\sqrt{ 2 }} & -\frac{1}{\sqrt{ 2 }} & 0 & 0 & 0 & \dots & 0 \\
\frac{1}{\sqrt{ 2 }} \frac{1}{\sqrt{ 3 }} & \frac{1}{\sqrt{ 2 }} \frac{1}{\sqrt{ 3 }} & -\frac{\sqrt{ 2}}{\sqrt{ 3 }} & 0 & 0 & \dots & 0 \\
\frac{1}{\sqrt{ 3 }} \frac{1}{\sqrt{ 4 }} & \frac{1}{\sqrt{ 3 }} \frac{1}{\sqrt{ 4 }} & \frac{1}{\sqrt{ 3 }} \frac{1}{\sqrt{ 4 }} & -\frac{\sqrt{ 3 }}{\sqrt{ 4 }} & 0 & \dots & 0 \\
\frac{1}{\sqrt{ 4 }} \frac{1}{\sqrt{ 5 }} & \frac{1}{\sqrt{ 4 }} \frac{1}{\sqrt{ 5 }} & \frac{1}{\sqrt{ 4 }} \frac{1}{\sqrt{ 5 }} & \frac{1}{\sqrt{ 4 }} \frac{1}{\sqrt{ 5 }} & -\frac{\sqrt{4  }}{\sqrt{ 5 }} & \dots & 0 \\
\vdots & & & & \ddots &\ddots & \vdots \\
\frac{1}{\sqrt{ N-1 }} \frac{1}{\sqrt{ N }} & \dots & & & &\frac{1}{\sqrt{ N-1 }} \frac{1}{\sqrt{ N }}&  -\frac{\sqrt{ N-1 }}{\sqrt{ N }} \\
\frac{1}{\sqrt{ N }} & \dots  & & & & \dots & \frac{1}{\sqrt{ N }}
\end{pmatrix} .\label{eq:BS_cascade}
\]
Since we are using phase-less beam splitters, the transformation can be performed on the $x$ and $p$ quadrature separately. 

\section{Numerical stability}\label{app:numerical_stability}
Representing the Wigner function of even a single mode SCS can prove to be troublesome if the cat has a large amplitude. For example, when targeting a qunaught state, the initial squeezed cat must have amplitude $\alpha = \frac{1}{2}e^{r}\sqrt{N\pi}$. For parameters $r=-12$ dB, $N=9$, this corresponds to $\alpha = 10.58$. When evaluating Eq. \eqref{eq:Wigner_sqz_cat}, the two complex Gaussians, which form the interference fringes, have weights proportional to $e^{-2\alpha^2}$, i.e. $e^{-244}$ for our choice of parameters. Needless to say, we are in rounding error territory when performing any kind of summation with weights this small. One trick is to calculate the exponential argument of each element, and then take the sum over exponentials with the \texttt{logsumexp} function from SciPy \cite{virtanen_scipy_2020}. It employs shift rules in order to avoid rounding errors due to floating point precision. 
\begin{figure}[H]
    \centering
\includegraphics[width=0.5\linewidth]{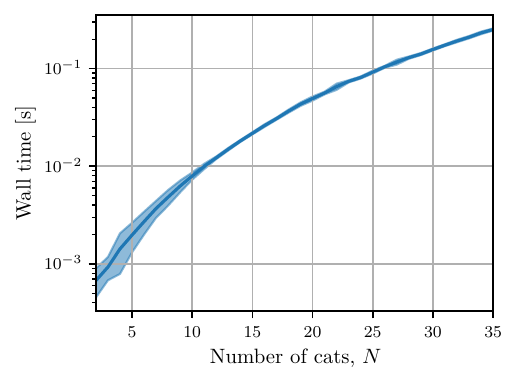}
    \caption{\textbf{Wall time of the code}. The runtime (averaged over 20 runs for each $N$) of the grid state preparation procedure when breeding $N$ $-20$ dB squeezed cats targeting a logical GKP state. The grid state is post-selected on $\vb*{p}=0$. Included in the runtime is the simplification of the output state, its normalisation and the computation of the effective squeezing $\Delta_x$ and $\Delta_p$ values.  }
    \label{fig:walltime}
\end{figure}

\section{S3: Supplementary figures}\label{app:sm_figures}
\begin{figure*}[h]
    \centering
    \includegraphics[width=0.95\linewidth]{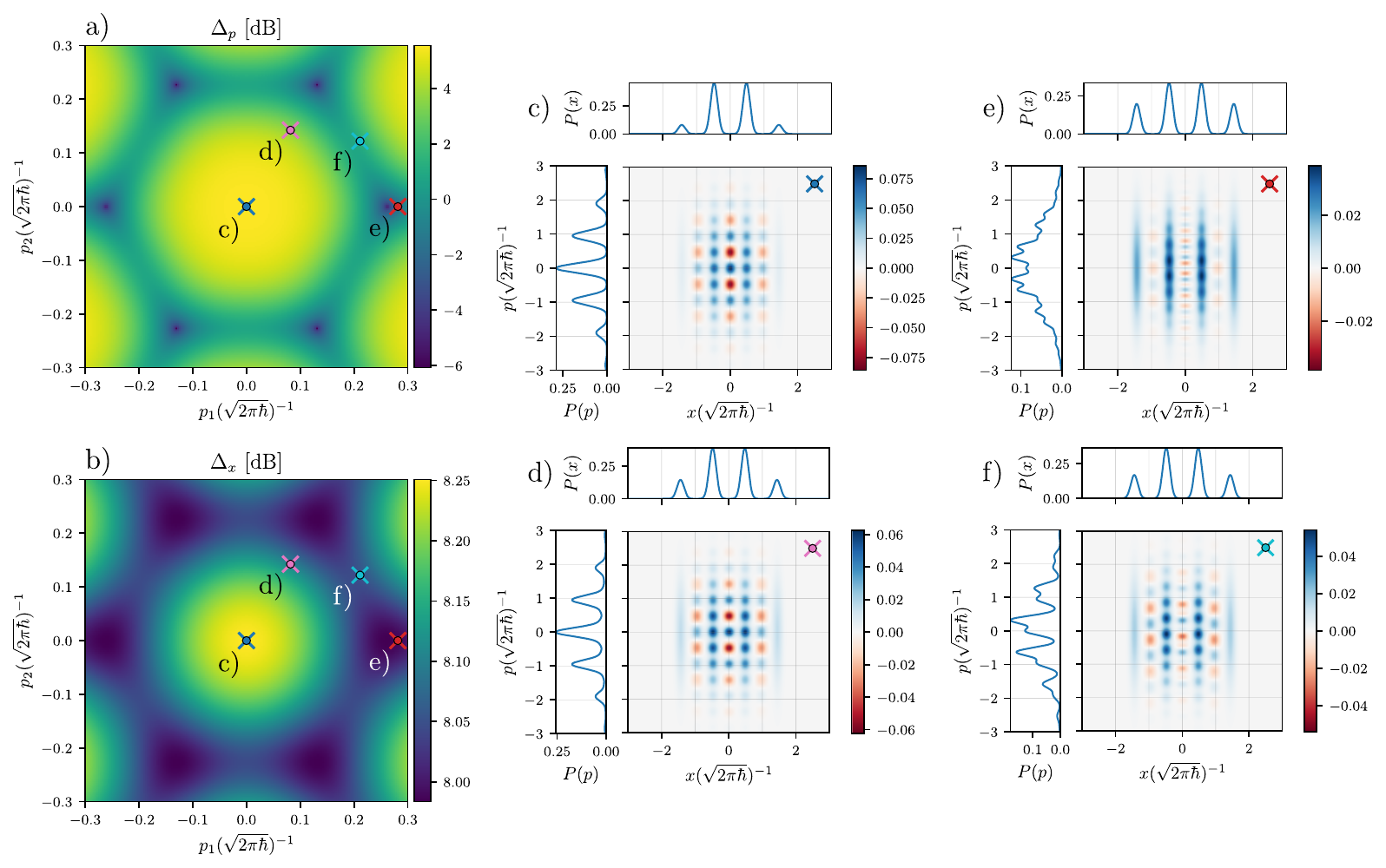}
    \caption{\textbf{The effect of loss on the post-selected output states}. A zoom in of the effective squeezing of the grid state produced by breeding three $-12$ dB lossy ($\eta=0.92$) squeezed cats a)-b). The Wigner functions of the generated grid state post selected on the colored points, which are the same points as in Fig. \ref{fig:N=3points} are shown in c)-f). The effective squeezing of the grid state is $(\Delta_x,\Delta_p)$ [dB]: c) $(8.25, 5.55)$, d) $(8.10, 3.57)$, e) $(7.99, -1.86)$, f) $(8.05, 1.21)$.}
    \label{fig:N=3_loss}
\end{figure*}
\begin{figure}[h]
    \centering
    \includegraphics[width=0.75\linewidth]{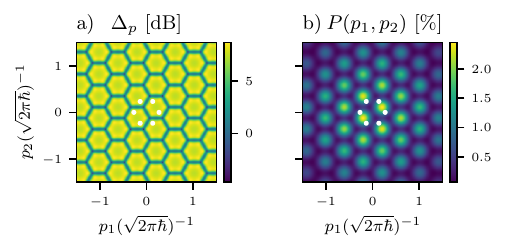}
    \caption{\textbf{{Breeding cats with different parities}}. Same as Fig. \ref{fig:N=3}, but where cats have parities $\vb*{k}=(0,1,0)$. The white points are the vertices of $(0,0,0)$ parity hexagons.  }
    \label{fig:N=3_parity_010}
\end{figure}
\begin{figure}[h]
    \centering
    \includegraphics[width=0.95\linewidth]{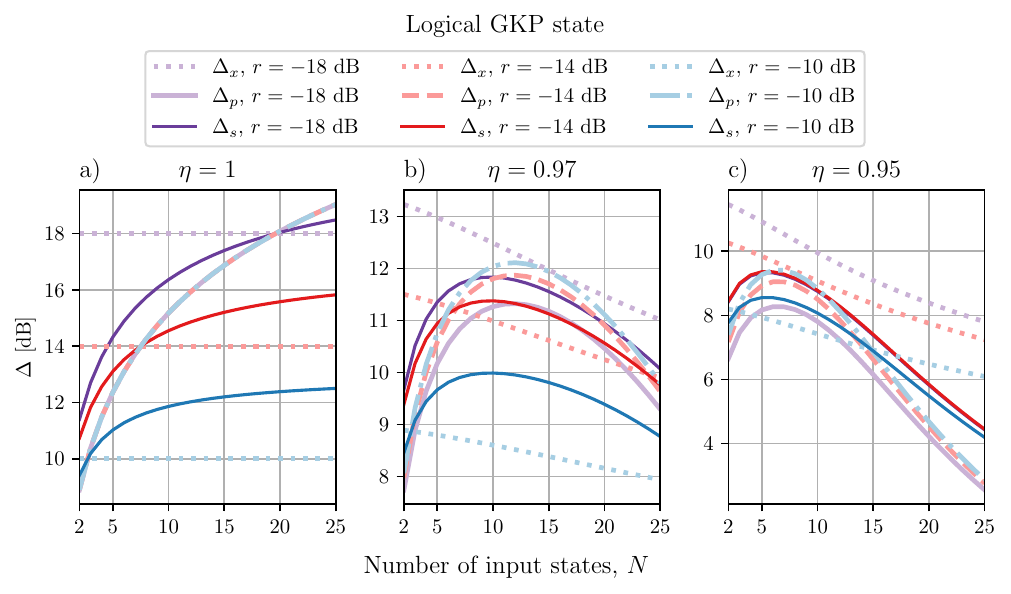}
    \caption{\textbf{Breeding a logical GKP state with loss}. The evolution of the effective squeezing in both quadratures $\Delta_x$ and $\Delta_p$ for the logical GKP states for 10, 14, 18 dB of cat squeezing as more input states are used for given loss channel transmissivities $\eta$ in the input state.  (projection on $p=0$).}
    \label{fig:effective_sqz_p=0_logical}
\end{figure}

\begin{figure*}[h]
    \centering
    \includegraphics[width=0.95\linewidth]{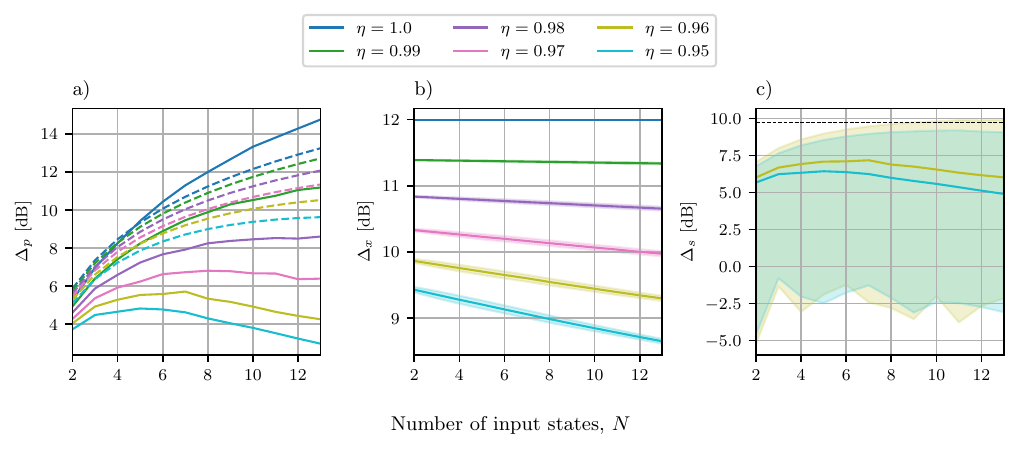}
    \caption{\textbf{The effect of loss on the effective squeezing, averaged over the homodyne measurement outcomes}.Same as Fig. \ref{fig:sampling_-15_dB}, but for $r=-12$ dB of squeezing. }    \label{fig:sampling_-12_dB}
\end{figure*}

\newpage

\end{document}